\documentclass[twocolumn]{aastex63}
\usepackage{amsmath}
\usepackage{xcolor}
\usepackage{soul}
\usepackage{multirow}
\usepackage{bm}
\usepackage{tabularx}
\usepackage{hyperref}

\definecolor{nsgreen}{rgb}{0.1,0.5,0.1}

\newcommand{\kms}{\,km\,s$^{-1}$} 

\shorttitle{USNO VLBI Spectroscopic Catalog}
\shortauthors{Sexton et al.}

\begin{document} 

\title{The U.S.\ Naval Observatory VLBI Spectroscopic Catalog}

\correspondingauthor{Remington O. Sexton}
\email{rsexton2@gmu.edu; remington.o.sexton.ctr@us.navy.mil}

\author[0000-0003-3432-2094]{Remington O. Sexton}
\affiliation{U.S. Naval Observatory, 3450 Massachusetts Ave NW, Washington, DC 20392-5420, USA}
\affiliation{Department of Physics and Astronomy, George Mason University, 4400 University Dr, Fairfax, VA 22030-4444, USA}

\author[0000-0002-4902-8077]{Nathan J. Secrest}
\affiliation{U.S. Naval Observatory, 3450 Massachusetts Ave NW, Washington, DC 20392-5420, USA}

\author[0000-0002-4146-1618]{Megan C. Johnson}
\affiliation{U.S. Naval Observatory, 3450 Massachusetts Ave NW, Washington, DC 20392-5420, USA}

\author[0000-0002-5604-5254]{Bryan N. Dorland}
\affiliation{U.S. Naval Observatory, 3450 Massachusetts Ave NW, Washington, DC 20392-5420, USA}



\begin{abstract}
Despite their importance for astrometry and navigation, the active galactic nuclei (AGNs) that comprise the International Celestial Reference Frame (ICRF) are relatively poorly understood, with key information such as their spectroscopic redshifts, AGN spectral type, and emission/absorption line properties generally missing from the literature. Using updated, publicly available, state-of-the-art spectroscopic fitting code optimized for the spectra of AGNs from low to high redshift, we present a catalog of emission line and spectral continuum parameters for 1,014 unique ICRF3 objects with single-fiber spectra from the Sloan Digital Sky Survey DR16. We additionally present black hole virial mass scaling relationships that use H$\alpha$-, H$\beta$-, \ion{Mg}{2}-, and \ion{C}{4}-based line widths, all consistent with each other, which can be used in studies of radio-loud objects across a wide range of redshifts, and we use these scaling relationships to provide derived properties such as black hole masses and bolometric luminosities for the catalog. We briefly comment on these properties for the ICRF objects, as well as their overall spectroscopic characteristics.\\
\end{abstract}

\section{Introduction} \label{section:introduction}

Since its adoption more than two decades ago, efforts to build the International Celestial Reference Frame \citep[ICRF;][]{1998AJ....116..516M} have led to cataloging the celestial coordinates of thousands of compact extragalactic radio sources across the sky, which are powered by active galactic nuclei (AGNs).  Decades of very long baseline interferometry (VLBI) observations and careful selection of sources for the ICRF has resulted in an unprecedented increase in accuracy and precision in the fields of deep space and terrestrial navigation, geodesy, and studies of various astrophysical phenomena.

The adoption of a reference frame based on extragalactic sources was first mandated by the International Astronomical Union (IAU) in 1991.  This reference frame, based on extragalactic radio observations using dual-frequency 2.3 and 8.4 GHz (S/X band) VLBI, was recommended to the IAU as the realization of the International Celestial Reference System \citep[ICRS;][]{1995A&A...303..604A}, and is defined using a set of sources carefully selected for precision and uniformity across the sky.
The IAU formally adopted the ICRF as part of the ICRS to replace the stellar-based Fifth Fundamental Catalog of Stars \citep[FK5;][]{1988VeARI..32....1F} on 1 January 1998.

With the advent of the National Radio Astronomy Observatory's  Very Long Baseline Array \citep[VLBA;][]{1994IEEEP..82..658N}, astrometric measurements of quasars became very efficient, resulting in a more-than five-fold increase in the number of VLBI sources by the mid 2000s, prompting the production of the second realization of the ICRF.  ICRF2 \citep{2015AJ....150...58F} was constructed at S/X bands using a total of 3,414 objects, 295 of which are classified as ``defining sources" and set the quasi-inertial barycenter frame of reference.  These defining sources were selected to have the optimal astrometric precision and minimal source structure.

The current ICRF, ICRF3, incorporates additional VLBI observations, and accounts for the galactocentric acceleration of the solar system \citep{2011A&A...529A..91T}, greatly improving accuracy. ICRF3 has 4,536 sources at S/X, with 303 defining sources chosen to prioritize isotropy, and includes 824 objects observed at 24 GHz (K band) and 678 objects observed with dual-frequency 8.4 and 32 GHz (X/Ka band), making ICRF3 the first multi-frequency celestial reference frame.



Despite their foundational importance to astronomy and geodesy, relatively little is known about the intrinsic properties of the sources that make up the ICRF, such as black hole (BH) masses, bolometric luminosities, Eddington ratios, host galaxy types, and stellar masses, other than properties derived directly from radio observations.  Many sources exhibit significant source structure at VLBI resolution, likely caused by the presence of relativistic jets commonly found in radio-selected AGNs and quasars \citep{2019ARA&A..57..467B}. Positional offsets between the VLBI radio and optical positions of ICRF sources, initially detected prior to Gaia by multiple groups \citep{2002AJ....124..612D,2013MNRAS.430.2797A,2013A&A...553A..13O,2014AJ....147...95Z}, have been a subject of ongoing study \citep{2016A&A...595A...5M, 2017A&A...598L...1K, 2017MNRAS.467L..71P,2019ApJ...873..132M}, and observing programs to understand the astrophysical nature of these offsets are underway \citep{2020jsrs.conf..179Z}.  To date, there is evidence that suggests that off-center optical AGN jets are a significant contributor to observed radio-optical offsets \citep{2019MNRAS.482.3023P,2019MNRAS.485.1822P}.

Moreover, AGNs and quasars can be some of the most persistently variable objects in the universe \citep[e.g.,][]{1997ARA&A..35..445U}, and exhibit an enormous range of morphologies and spectral energy distributions that change with time. In principle, reference frame objects that exhibit positional stability over one epoch may change significantly in the next epoch, so cataloging objects for the ICRF is not a ``one and done'' process: objects must be physically characterized and monitored if the ICRF is to become a truly multi-wavelength CRF that is stable and accurate across epochs. It is therefore necessary to have an in-depth understanding of the astrophysical properties of these objects, so as to minimize the impact of the peculiarities of AGNs on the accuracy of the ICRF. Understanding the relevant physics of AGNs and their environments is one of the guiding principles of the Fundamental Reference Frame AGN Monitoring Experiment \citep[FRAMEx:][]{2020jsrs.conf..165D, 2021ApJ...906...88F, 2022arXiv220105152F}.

\indent While there have been systematic programs to concatenate catalog photometry and redshifts of nearly all ICRF objects \citep[e.g.,][]{2018ApJS..239...20M}, relatively few ICRF objects have been spectroscopically characterized \citep[e.g.,][]{2017AJ....153..157T}, and efforts have mainly focused on determination of redshift \citep[e.g.,][]{2009A&A...506.1477T}. Optical spectra contain an abundant amount of information regarding the physical processes powering AGNs and quasars.  Most notable are the presence of permitted Balmer emission features broadened by the bulk motions of gas orbiting the BH, from which we can estimate BH mass \citep{2004IAUS..222...15P}, and the apparent flux ratio of permitted emission lines of a particular ionized species can give a direct measure of the line-of-sight (LOS) AGN obscuration.  Additionally, the narrow forbidden emission lines of AGNs and quasars can exhibit line profile asymmetries indicative of ionized gas outflows \citep{1996ApJ...465...96N}, which can provide insight into how the AGN may be interacting with its host galaxy.  Properties of the host galaxy can also be inferred from the presence of stellar absorption features to estimate stellar kinematics \citep{2006ApJ...641..117G}.  Most, if not all, of these intrinsic properties are unknown for ICRF sources but can be estimated via spectroscopy.

In this work, we produce a catalog of spectral properties for ICRF3 sources with spectra available from the Sloan Digital Sky Survey (SDSS) Data Release 16 \citep[DR16;][]{2017AJ....154...28B}, and provide value-added properties such as BH masses and bolometric luminosities. In producing this catalog, we have additionally derived new BH virial mass scaling relationships that will be of value for work on radio-loud AGNs, and we have updated the spectral-fitting code of \citet{2021MNRAS.500.2871S} to handle high-redshift and radio loud objects. In Section \ref{sec:data} we describe the SDSS DR16 spectroscopic data used in this work.  Section \ref{sec:methods} provides a detailed description of the spectral code updates, fitting algorithm, strategy, model implementation, and quality metrics used for our sample. Section \ref{sec:derived_properties} details the calculation of derived properties such as luminosities, broad emission line widths, and BH mass, as well as the relevant relations between these quantities.  We briefly discuss the general spectroscopic properties of ICRF3 objects in Section \ref{sec:general_properties}.  Finally, in Section \ref{sec:catalog_structure} we describe the structure, proper usage, caveats, and availability of the catalog for the astronomical community.

Where distances are employed, we use a flat $\Lambda{\textrm{CDM}}$ cosmology with $H_0=70$ km s$^{-1}$ Mpc$^{-1}$ and $\Omega_\mathrm{m}=0.3$.  Specific ICRF3 sources are referenced by their IERS designation.



\section{Data}
\label{sec:data}

Optical spectra for ICRF3 sources were obtained from the SDSS DR16 \citep{2017AJ....154...28B}, given its large availability of calibrated AGN and quasar spectra and large wavelength and sky coverage.  The original SDSS spectroscopic survey \citep{2000AJ....120.1579Y} employed spectral fiber apertures with $3\arcsec$ diameter and an observed wavelength coverage from 3800\AA\ to 9200\AA. The Baryon Oscillation Spectroscopic Survey \citep[BOSS;][]{2013AJ....145...10D} and extended BOSS \citep[eBOSS;][]{2016AJ....151...44D} introduced in SDSS-III \citep{2011AJ....142...72E}, employed a $2\arcsec$ diameter fiber with a slightly larger wavelength coverage from 3650\AA\ to 10,400\AA.  While the main SDSS, BOSS, and eBOSS programs focus primarily on extragalactic sources and avoid the galactic disk, the Sloan Extension for Galactic Understanding and Exploration \citep[SEGUE;][]{2009AJ....137.4377Y} program probes the Milky way disk and halo stars to obtain stellar parameters, and occasionally observes background extragalactic sources.  All SDSS spectra are logarithmically re-binned \citep{1979AJ.....84.1511T} such that each pixel corresponds to a constant velocity of $\sim69$\kms\ in the observed frame.

We cross-matched 4,536 ICRF3 source coordinates with the \texttt{specObj-dr16.fits} table, using a $1\farcs5$ match tolerance, resulting in 1,559 matches, of which 11 spectra contained no data and were removed from the sample. Of the remaining 1,548 spectra, 534 are repeat observations of the unique set of 1,014 sources, which will be the subject of a future multi-epoch study. To ensure that the best version of each duplicate spectrum was fit, we prioritized available wavelength coverage, which favored BOSS spectra over SDSS spectra.

The sample of unique ICRF3 sources consists of 1,014 SDSS spectra from both the SDSS and BOSS spectrographs, and were observed in various programs including the original SDSS (355), BOSS (383), eBOSS (263), SEGUE1 (7), and SEGUE2 (6). 

These individual survey programs cover an area of sky of approximately 14,555 square degrees, predominantly in the northern hemisphere and largely avoiding the Galactic plane, whereas the ICRF3 spans the full sky by design.  The area surveyed by the SDSS therefore allows only $\sim1600$ ICRF3 objects, or $35\%$ of all ICRF3 objects, to be potentially observed within the SDSS footprint.  The 1,014 available spectra therefore amounts to $63\%$ of observable ICRF3 sources, or $22\%$ of all ICRF3 sources, with SDSS.  The limiting magnitude for our sample is $i_{\rm{PSF}}=20.6$, which is consistent with the $i_{\rm{PSF}}=20.2$ limiting magnitude for quasar target selection for the SDSS spectrograph. Objects observed using the more-sensitive BOSS spectrograph are limited to $i_{\rm{PSF}}=22.1$.

\begin{figure*}[htb!]
\includegraphics[width=\textwidth]{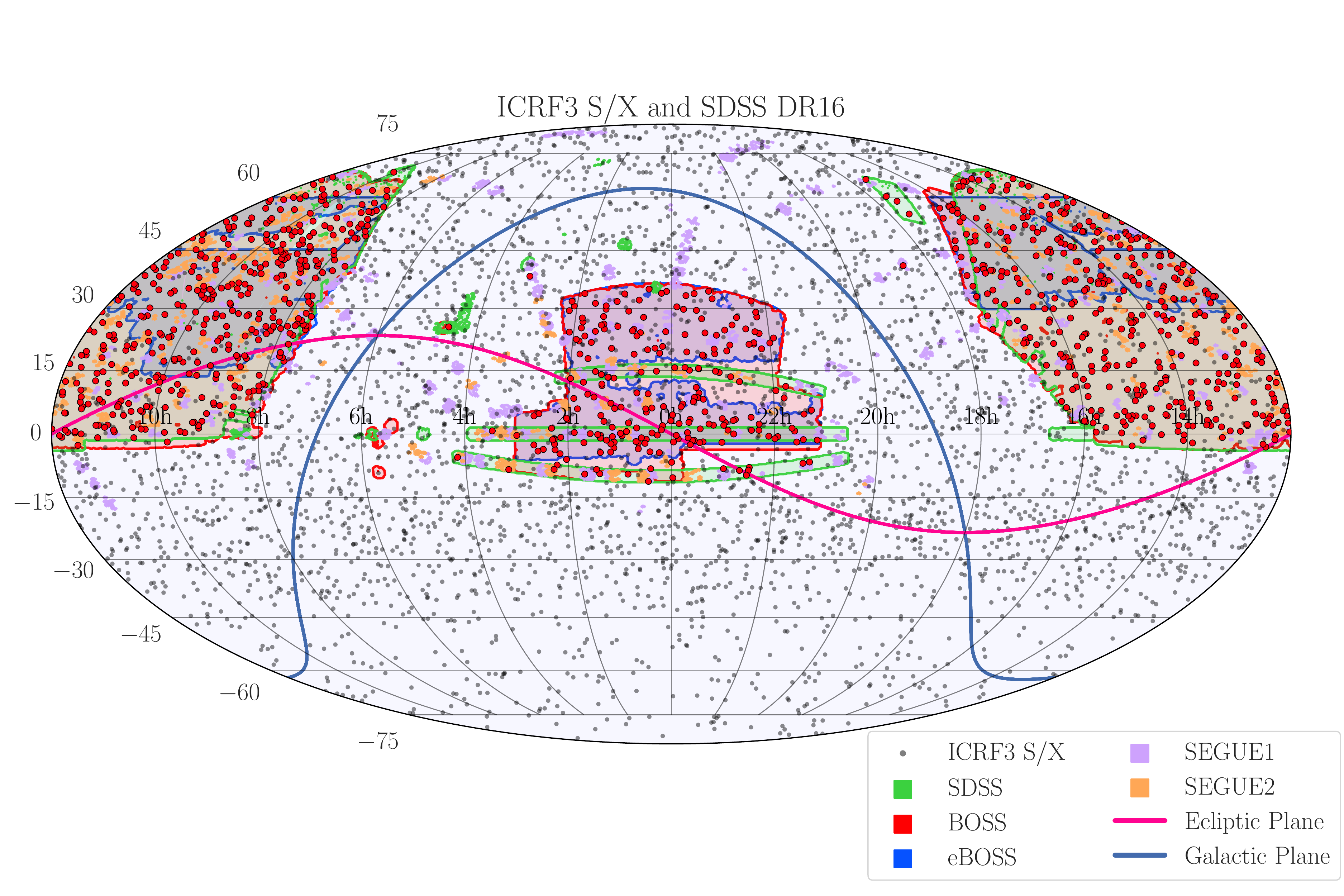}
\caption{The Mollweide projection of all 4,536 S/X ICRF3 sources from \citet{2020A&A...644A.159C}.  Outlined regions correspond to various SDSS programs from which SDSS spectra of ICRF3 sources are obtained.  The 1,014 ICRF3 sources with available spectra are shown as red points. }
\label{fig:sky_coverage}
\end{figure*}

Figure \ref{fig:sky_coverage} shows the Mollweide projection of all 4,536 S/X sources from ICRF3 from \citet{2020A&A...644A.159C} with the sky coverage of various SDSS programs from which ICRF3 spectra are obtained.  Of the 1,014 unique ICRF3 sources, 58 (5.7\%) are defining sources and the remaining 956 objects (94.3\%) are densifying sources.

Visual inspection of all objects with the \texttt{specObj} \texttt{ZWARNING} flag set to $>0$ (indicating a potential issue in the pipeline redshift determination) revealed 121 sources with no visible emission or absorption line features from which a redshift could be determined and therefore cannot be fit.  These featureless objects are flagged as ``featureless'' in the catalog.

Twenty-two spectra with $\texttt{ZWARNING}>0$ were determined to have redshifts suitable for fitting purposes.  Additionally, seven spectra were determined to have incorrect redshifts, which were corrected using visual inspection using any visible emission or absorption features.  Finally, one spectrum (IERS 1048+470) appears to be contaminated by an foreground star in SDSS imaging, producing a unidentifiable spectrum from which no redshift could be determined.

\begin{figure}[htb!]
\includegraphics[width=0.5\textwidth]{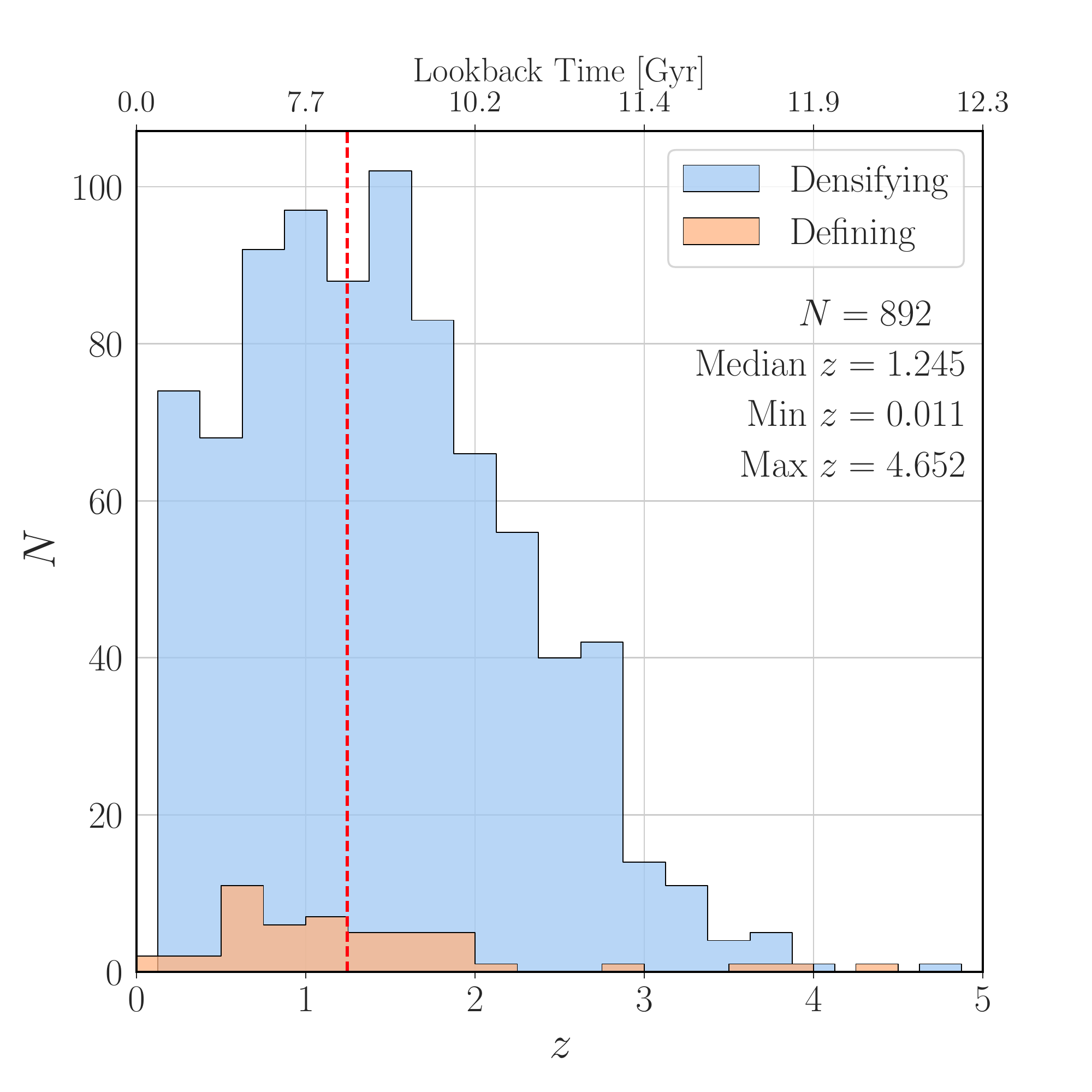}
\caption{Histogram of the redshift distribution of of ICRF3 SDSS spectra for which redshifts could be accurately determined ($N=892$).  The dashed red line marks the median redshift of $z=1.245$.}
\label{fig:redshift_dist}
\end{figure}

The final sample consists of 892 unique ICRF3 spectra with accurate redshift determinations suitable for fitting (19.7\%). This consists of 48 (5.4\%) defining and 844 (94.6\%) densifying sources with a median redshift of 1.245.  Figure \ref{fig:redshift_dist} shows the redshift distribution of the final fitted sample.

\section{Methods}
\label{sec:methods}

\subsection{Spectral Fitting}

The fitting of all spectra is performed using the Bayesian AGN Decomposition Analysis for SDSS Spectra \citep[BADASS\footnote{\url{https://github.com/remingtonsexton/BADASS3}};][]{2021MNRAS.500.2871S} software, which is written in Python and specifically designed for fitting AGNs and quasars.  The BADASS software has been recently upgraded to allow for the fitting and analysis of high-redshift spectra specific to the needs of the ICRF3 spectroscopic sample, and we provide a brief description of these relevant upgrades.

The original release of BADASS was designed for rest-frame optical SDSS spectra, specifically to model the regions spanning 3500 \AA~to 7000 \AA, and was limited by the available coverage of empirical stellar templates libraries used to simultaneously model the stellar line-of-sight velocity distribution (LOSVD).  In order to be able to model the spectra of ICRF3 objects, the majority of which have a $z>1$, significant modifications and additions to the fitting model were made.  These include models for the Balmer pseudo-continuum \citep{1982ApJ...255...25G,1985ApJ...288...94W} as well as an empirical UV iron template \citep{2001ApJS..134....1V}, both of which are discussed in further detail in Section \ref{sec:continuum_fit}.  Additional line profile shapes such as pseudo-Voigt and Gauss-Hermite were added for increased flexibility in modelling broad emission line profiles typically seen in \ion{Mg}{2} and \ion{C}{4}.  The largest structural change to the algorithm was a complete overhaul of the BADASS framework to allow for easy customization emission lines and the addition of optional soft and hard constraints on emission line parameters.  A complete description of all new additions and features to BADASS will be provided in a future manuscript.

The BADASS fitting algorithm relies on the same Bayesian principles outlined in \citet{2021MNRAS.500.2871S}, with posterior sampling performed using the affine-invariant Markov-Chain Monte Carlo (MCMC) sampler \texttt{emcee} \citep{2019JOSS....4.1864F}.  A preliminary maximum-likelihood fit to the data is performed to obtain initial parameter values using a normal (Gaussian) likelihood distribution.  Flat priors with liberal bounds are initialized for all fitted parameters.  A final fit to the data is performed using MCMC to obtain robust uncertainty estimates on all parameters.  We do not utilize the built-in autocorrelation analysis functionality of BADASS, since the majority of our parameters are those of emission lines, which nominally converge in under 1000 iterations depending on their signal-to-noise (S/N).  We instead perform a maximum of 2500 iterations of MCMC sampling with 100 walkers per parameter and use the final 1000 iterations (1500 burn-in) for the posterior distributions, from which the median (50\textsuperscript{th} percentile) of the distribution is considered the best-fit values to our parameters, and the 16\textsuperscript{th} and 84\textsuperscript{th} percentiles of the distributions are used as the lower- and upper-bound $1\sigma$ uncertainties on each parameter.

Prior to fitting, BADASS performs preliminary data preparation on SDSS spectra, which includes the correction for Galactic extinction. Conversion to rest frame is performed using the redshift determined by the SDSS pipeline except in the seven cases where the redshift was determined to be incorrect, for which we determined the correct redshift using visual inspection. We used the extinction map of \citet{2011ApJ...737..103S} and corrected using a \citet{1989ApJ...345..245C} extinction law assuming an $R_V = 3.1$. BADASS does not fit and subtract off any spectral components as a preliminary step to measuring other components such as emission lines, and instead fits all components simultaneously in order to properly constrain their relative contributions and covariances.

In the following subsections, we describe the fitting strategy implemented to fit the ICRF3 spectra, with particular attention given to regions that are important for deriving virial BH masses from broad emission line features and continuum luminosities.

\subsection{Fitting Regions}
To facilitate the detailed fitting of all ICRF3 spectra, we divide each spectrum into eight fitting regions that span the entire optical and near-UV wavelength coverage of the SDSS and BOSS.  The availability of these regions in each spectrum is dependent on the redshift of each object, and therefore not all spectra have data in each region.  A description of these fitting regions and their wavelength ranges is shown in Table \ref{tab:fitting_regions}.

Fitting regions are chosen based on available spectral features as a means to limit the number of free parameters in the fit to reduce computation time.  Fitting regions, as opposed to full-spectrum fitting, also reduces the chance that a poorly fit region of the spectrum not described by our model (such as poorly subtracted skylines) adversely affects the overall fit.  We make an exception for the fitting of continuum quantities (described in Section \ref{sec:continuum_fit}), since continuum components (such as power-law, Balmer, and host galaxy continuum) can span multiple regions, and thus we use the entire available spectrum to properly constrain their respective parameters.

To avoid excessive noise at the edges of SDSS spectra, we require that the spectrum have a minimum of 100~\AA\ overlap in a given region for that region to be fit, and require that any fit emission lines are at least 100~\AA\ from the edges of a given spectrum.  The model components, such as emission lines, \ion{Fe}{2} templates, UV iron, etc., that are fit in each region is determined automatically by BADASS, based on available wavelength coverage and a predefined line list given in Table \ref{tab:emline}.  We do not set a minimum S/N requirement to fit any emission lines. Instead, the detection confidence of all lines is determined after fitting by performing a two-tailed hypothesis test of the likelihood distributions, which takes into account the spectral noise.  The number of detections of each line with a confidence $\geq95\%$ is also given in Table \ref{tab:emline}.

\begin{deluxetable}{ccc}
\tablecaption{Fitting regions used for the USNO VLBI Spectroscopic Catalog}
\tablehead{
\colhead{Region} &  \colhead{Range (\AA)} & \colhead{$N$}
}
\colnumbers
\startdata
1 & $> 8000$      & 56  \\
2 & $6800 - 8000$ & 125 \\
3 & $6200 - 6800$ & 175\\
4 & $5500 - 6200$ & 254 \\
5 & $4400 - 5500$ & 431\\
6 & $3500 - 4400$ & 629\\
7 & $2000 - 3500$ & 856 \\
8 & $< 2000$      & 566 \\
\enddata
\tablecomments{
Column 1: Region number.  Column 2: rest-frame wavelength range spanned by region.  Column 3: number of objects $N$ with at least 100~\AA\ of data in the corresponding region.
}
\label{tab:fitting_regions}
\end{deluxetable}

\subsection{Continuum Modelling}
\label{sec:continuum_fit}

Since the ICRF3 sample contains various types of objects, each with varying contributions from the host galaxy and/or the AGN depending on the spectral type, we do not limit the fitting of continuum quantities to the regions described in Table \ref{tab:fitting_regions}.  While smaller fitting regions are useful for detailed fitting of features such as emission lines, the host galaxy and AGN continuum require the full available wavelength range in order to constrain their relative contributions to the overall spectrum.  This is especially crucial for Seyfert~1 (Sy~1) type objects, for which there are varying contributions from both the host galaxy and AGN.

In order to limit the number of free fit parameters introduced by including emission lines, we utilize the automated emission and absorption masking feature built into BADASS, which implements an iterative moving median filter to identify features of considerable variance to mask from the fit.  The masking of features only removes these pixels from contributing to the Gaussian log-likelihood calculated during the fit.  The model components included in the continuum fit for all objects with determined redshifts ($N=892$) are a composite host galaxy template, power-law continuum, Balmer pseudo-continuum, optical broad and narrow \ion{Fe}{2} template, and UV iron template.

The composite host galaxy template consists of three single stellar population (SSP) models from the E-MILES SSP library \citep{2016MNRAS.463.3409V} of ages 0.1, 1.0, and 10.0 Gyr.  The relative contributions of the individual models are determined using a non-negative least squares minimization during the fitting process.  The velocity shift and dispersion are free parameters in the host galaxy model.

The AGN contribution of the continuum is assumed to be the sum of contributions from a power-law continuum and a Balmer pseudo-continuum \citep{1982ApJ...255...25G,1985ApJ...288...94W}, the latter modeled using similar methods to \citet{2002ApJ...564..581D} and \citet{2017MNRAS.472.4051C}.  The exact relative contributions of each the power-law and Balmer continua are difficult to constrain despite using the full available spectral range in the fit, as the only defining feature of the Balmer pseudo-continuum is the Balmer edge at 3640\AA, which can be obscured by strong optical \ion{Fe}{2} emission.  The amplitude, width, velocity offset, and ratio between the higher-order Balmer lines and the blackbody component of the Balmer pseudo-continuum are allowed to be free parameters in the fit to the Balmer continuum, while the temperature and optical depth at the Balmer edge are fixed following \citet{2002ApJ...564..581D} since these parameters are often degenerate.

Broad and narrow optical \ion{Fe}{2} emission was modeled using the template from \citet{2004A&A...417..515V}, which covers the spectral range from 3400\AA\ to 7200\AA.  Iron emission in the near UV is modeled using the ``B'' template from \citet{2001ApJS..134....1V}, covering the spectral range from 1250\AA\ to 3090\AA. To account for any iron emission in the gap between the optical and UV templates, two broad lines were added between 3100\AA\ and 3400\AA\ for the sole purpose of improving the overall fit to the region.  Amplitude, velocity offset, and width are free parameters for both the optical and UV templates.  Both the optical and UV iron templates are based on the spectrum of I~Zw~1, which is a limitation in our ability to accurately model iron emission.

\begin{figure*}
\gridline{
            \fig{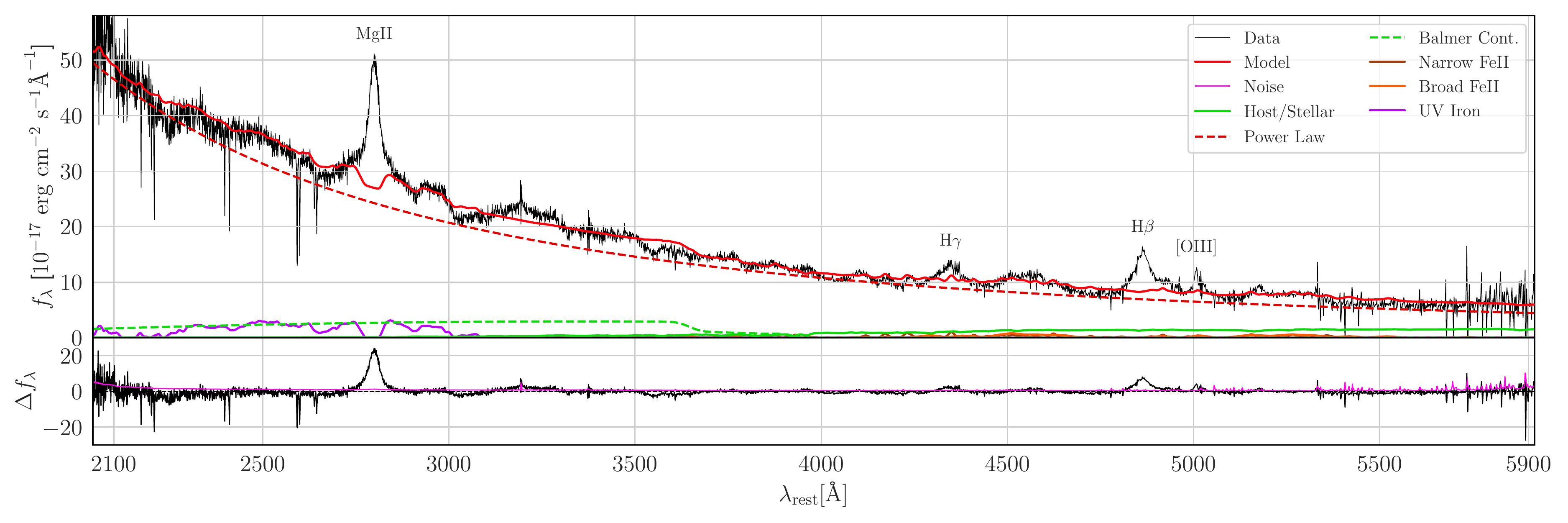}{1.0\textwidth}{(a)}
          }
\gridline{
            \fig{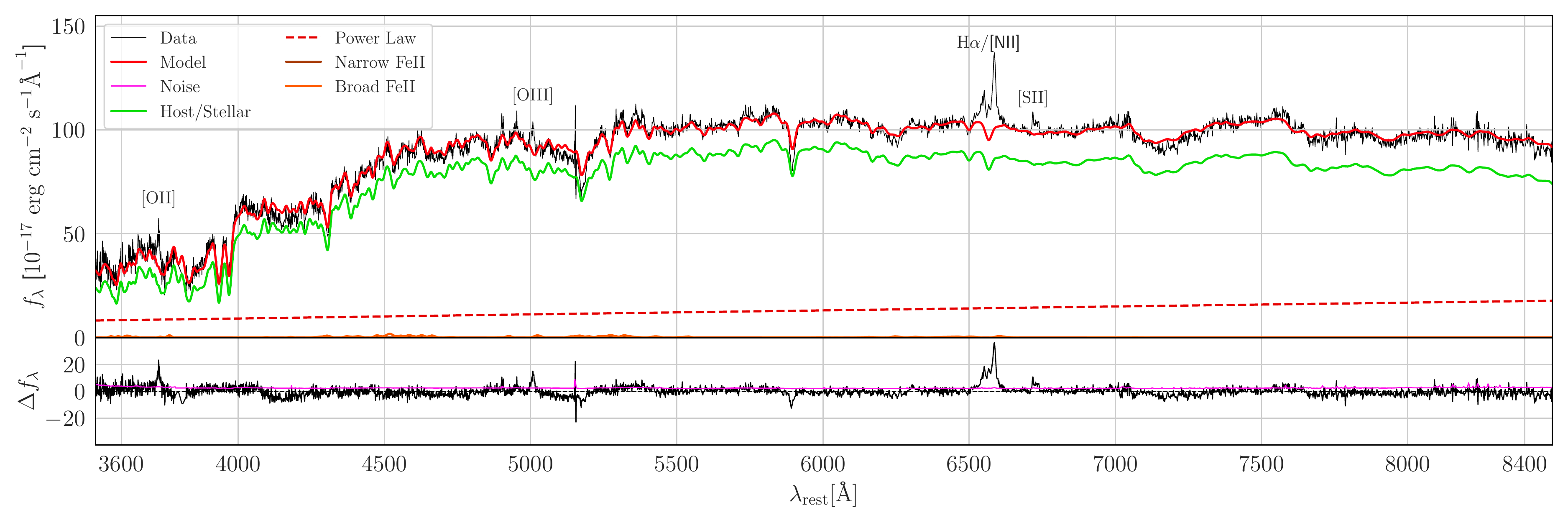}{1.0\textwidth}{(b)}
          }
\caption{Examples of the full-spectrum continuum fitting performed to constrain large-scale parameters such as the host/stellar and AGN fractions, and power-law index ($\alpha_\lambda$).  Prominent emission features that appear in the residuals are labeled. (a) Example continuum fit typical of Sy~1 and quasar-type objects (IERS~0043+246), exhibiting strong power-law and Balmer continuum, and weaker host/stellar contributions.  (b) Example continuum fit of a typical Sy~2 (IERS~0026-015) exhibiting a strong host/stellar and weaker AGN power-law contributions.
\label{fig:cont_fit}}
\end{figure*}

Figure \ref{fig:cont_fit} shows two example continuum fits that represent the extremes of spectral types encountered in the ICRF3 sample, namely (a) quasars or Sy~1 objects, and (b) host-dominated Seyfert~2 (Sy~2) objects.  Note that the primary purpose of these fits is to properly model the AGN component (power-law $+$ Balmer continuum) of the spectrum in order to accurately obtain continuum luminosities and estimate virial BH mass. Additionally, we use the continuum fits to estimate power-law slope ($\alpha_\lambda$) and ``Eigenvector 1'' parameters inferred from principal component analysis-based studies \citet{1992ApJS...80..109B}.  Other continuum quantities such as the stellar LOSVD are measured from individual regions.

\subsection{Emission \& Absorption Line Modelling} \label{subsec: emission and absorption line modelling}
Emission lines listed in Table~\ref{tab:emline} that are not in boldface are modeled as simple Gaussians characterized by an amplitude, FWHM, and velocity.  Lines of the same ion within a given region are tied in width and velocity in order to decrease the number of free parameters and constrain their respective fits.  Special consideration is given to lines of more importance (boldface text in Table~\ref{tab:emline}), such as Baldwin-Phillips-Terlevich \citep[BPT;][]{1981PASP...93....5B} and broad lines, require more careful treatment to properly model their profiles, and we discuss these cases in the subsequent sections.

Narrow emission line features are parametrically distinguished from broad emission lines by the allowed parameter spaces of the FWHM and velocity offset.  The allowed parameter spaces for individual narrow emission lines is [0,800] (\kms) for FWHM and [-1000,1000] (\kms) for velocity offset.  For individual broad emission lines, these intervals are [500,15000] (\kms) and [-2000,2000] (\kms) for width and velocity offset, respectively.  Outflow components, such as those used to model ionized gas outflows in [\ion{O}{3}], have allowed parameter spaces [0,5000] (\kms) and [-1000,1000] (\kms) for width and velocity offset, respectively.  Emission lines with fit parameters close (within $\sim 1\sigma$) to the limits of these parameter spaces should be used with caution.

For lines that require multiple components, such as broad lines and lines with ionized gas outflows, integrated quantities such as flux, luminosity, equivalent width, dispersion, and velocity must be re-derived using the full model consisting of the sum of individual emission line components.  Since the parameter chains of multiple components belonging to a single emission line are necessarily correlated, the MCMC chains must be used to reconstruct the line profile for each iteration of the fit.  This yields new distributions of parameters from which multi-component best-fit values and uncertainties can be obtained.  Expressions for flux, luminosity, equivalent width, and FWHM are derived in the usual way. The expression adopted for the integrated flux-weighted velocity and velocity dispersion reported for all emission lines is given by

\begin{equation}
    v = \frac{\int x~f_x~dx}{\int~f_x~dx}
\end{equation}

\begin{equation}
    \sigma^2 = \frac{\int x^2~f_x~dx}{\int~f_x~dx} - v^2
\end{equation}
where $f_x$ is the normalized flux density at each pixel $x$ expressed in units of velocity (Section~\ref{sec:data}). The distributions for the majority of parameters are typically close to Gaussian, and uncertainties reported for all parameters are the averaged 16th and 84th percentiles of each distribution.

\subsubsection{Ionized Gas Outflows}
The presence of asymmetric and kinematically broadened forbidden (narrow) line profiles, indicative of ionized gas outflows \citep{1980A&A....87..152H,1996ApJ...465...96N}, are commonplace within the ICRF3 sample.  Asymmetric narrow-line profiles are most apparent in the H$\beta$/[\ion{O}{3}] and [\ion{O}{1}]/[\ion{N}{2}]/H$\alpha$/[\ion{S}{2}] line groups, but can occasionally appear in [\ion{Ne}{5}] in higher-$z$ spectra.  To model these cases, we take advantage of the built-in outflow fitting capabilities for which BADASS was originally designed.

To determine if any of the aforementioned lines show evidence of asymmetry or broadening, we utilize the line detection capabilities of BADASS to obtain a probability that a two-component fit to the line performs significantly better than a single-component fit.  This is most-easily performed with the [\ion{O}{3}]$\lambda5007$ line because it is relatively unaffected by broad H$\beta$.

If an object that shows at least 50\% confidence that an outflow is detected is fit with a two-component model, comprised of a narrow ``core'' and broader ``outflow'' component, BADASS constrains the width of the core components to be narrower than the outflow component by default.  BADASS also ties the respective widths and velocities of the core and outflow components for all lines for which the multi-component model is applied.  These constraints are justified by the limited resolution and S/N of SDSS spectra.  For the vast majority of cases, a double component model achieves a satisfactory fit.  A much smaller fraction of objects cannot be accurately modeled using a double component and show extreme outflow profiles, and are therefore fit with up to four components to accurately model the line profiles until a satisfactory fit is achieved.

Reported catalog quantities, such as velocity, dispersion, flux, etc., for all multi-component lines reflect the sum of the individual line components.  A more detailed study of the kinematics of ionized gas outflows in the ICRF3 sample will be left for a future investigation.

\subsubsection{Broad Emission Lines}
For the majority of the sample, a pseudo-Voigt profile adequately describes the broad line emission profile in ICRF3 type~1 objects.  Normally, the implementation of a true Voigt profile requires the convolution of a Gaussian and Lorentzian function.  To reduce the computational resources required for convolution, BADASS implements a pseudo-Voigt profile as the linear combination of a Gaussian and Lorentzian functions.  This allows us to simply model broad lines that may exhibit extended wings such as in narrow-line Sy~1 (NLS1) galaxies \citep{1999ApJS..125..297L,2001A&A...372..730V,2020A&A...636A..64B}, or more relevantly, radio-loud quasars that may exhibit similar extended wings \citep{1989ApJ...343...54S,1992ApJS...80..109B,1996ApJS..102....1B}.  The pseudo-Voigt profile requires only one additional free parameter, a ``shape'' variable, that determines how close the profile is to either Gaussian or Lorentzian.  Since the extended wings of the Lorentzian profile can extend to infinity, introducing increased bias in the measurement of the flux-weighted velocity and velocity dispersion, BADASS truncates the pseudo-Voigt function to zero below the median noise level during the fitting process, setting an effective limit on the extent of the profile wings.

For a fraction of cases in which the symmetric pseudo-Voigt profile does not satisfactorily fit a broad emission line, BADASS instead implements either a 4- or 6-moment Gauss-Hermite profile, or a multi-component (up to 3) Gaussian broad line profile, until a satisfactory fit is achieved.  We purposefully choose not to implement Gauss-Hermite profiles for all broad lines because Gauss-Hermite parameterization is not strictly unimodal nor positive-definite, and the systematics involving such possibly ill-behaved functions are not well understood in the context of broad emission line studies, despite their frequent use in the literature.  For example, we find that Gauss-Hermite profiles alone cannot adequately model the peaks attributed to the narrow component of broad lines, especially in broad \ion{C}{4}, as shown in Figure \ref{fig:region_fits}h.  Gauss-Hermite profiles are also prone to fit excess emission in the wings of the profile which can be attributed to optical or UV iron in type~1 AGNs.  For these reasons, the use of Gauss-Hermite profiles are used only on a case-by-case basis when the quality of the fit can be visually assessed.

For both H$\alpha$ and H$\beta$, the broad component is modeled separately from the narrow component, and reported catalog measurements are for the broad components alone.  Narrow line velocities and widths for these regions (3 and 5) are tied in order to help constrain the fits for narrow H$\beta$ and H$\alpha$, which are obscured by broad lines.  Example fits of the H$\alpha$ and H$\beta$ broad lines are shown in Figures \ref{fig:region_fits}c and \ref{fig:region_fits}e, respectively.

Regions 7 and 8, which respectively host the \ion{Mg}{2} and \ion{C}{4} lines, can show a significant number of metal absorption features, which can bias the fitting of broad lines if the two coincide.  To avoid this, BADASS uses a moving median filter of various bandwidths to determine the location of possible metal absorption lines and attempts to interpolate over them using a simple linear interpolation.  We find this technique only marginally better than masking, since in some cases, masking leads to poor constraints on the fits of narrow line components if the mask covers a number of pixels comparable to the narrow line width, which occurs frequently in \ion{C}{4} emission.

Following previous studies that performed fitting of the broad \ion{Mg}{2} line \citep{2011ApJS..194...45S,2018ApJ...859..138W,2020ApJ...901...35L}, we treat the \ion{Mg}{2} emission line as a single feature.  Although the \ion{Mg}{2} feature is actually a doublet, the limited spectral resolution of SDSS coupled with the velocity broadening of the complex prevents us from modeling the narrow line component as an independent kinematic component.  We find that the best method for fitting broad \ion{Mg}{2} is to implement a multi-component (up to 2) pseudo-Voigt fit with an additional single Gaussian narrow line component, similar to the methods of \citet{2011ApJS..194...45S}. Other more recent studies \citep{2018ApJ...859..138W,2020ApJ...901...35L} used 6\textsuperscript{th}-order Gauss-Hermite profiles to fit broad \ion{Mg}{2}, however we found that this parameterization can result in unexpected or multi-modal behavior in the wings of the profile, which can be affected by UV iron emission either side of \ion{Mg}{2}.  To prevent the possibility of over-fitting or inadvertently fitting UV iron emission, we only fit two broad line components if it is justified by the residuals of the fit.  Figure \ref{fig:region_fits}g shows an example of a fit to Region 7 which includes the \ion{Mg}{2} line.

Finally, broad \ion{C}{4} is fit in a similar way to H$\beta$ and H$\alpha$, with a single pseudo-Voigt broad and narrow component.  Again, we found that a single broad pseudo-Voigt profile with a single Gaussian narrow line profile is favorable to a single Gauss-Hermite profile, for reasons that the Gauss-Hermite model does not model the narrower peaks of some \ion{C}{4} profiles.  \citet{2017ApJ...839...93P} similarly fit a 6\textsuperscript{th}-order Gauss-Hermite with an additional Gaussian to account for the narrower peaks of some \ion{C}{4} profiles.  Figure \ref{fig:region_fits}h shows an example fit of Region 8, which contains the \ion{C}{4} line.  Due to the combined effects of metal absorption, UV iron contamination, and neighboring broad lines, \ion{C}{4} remains the most problematic line for fitting.

\subsubsection{Stellar Absorption Features}
For Regions 1 through 6, the stellar continuum is modeled using 50 template stars of various stellar type from the Indo-US Library of Coud{\'e} Feed Stellar spectra \citep{2004ApJS..152..251V}. BADASS utilizes the penalized pixel-fitting algorithm \citep[pPXF;][]{2017MNRAS.466..798C} to determine the best-fitting linear combination of stellar templates, convolved with a Gauss-Hermite series, to model the LOSVD.  Stellar kinematics are reported for objects with spectral coverage that includes the \ion{Ca}{0}K+H ($\sim 3950$ \AA), \ion{Mg}{1}b ($\sim 5175$ \AA), and CaT ($\sim 8545$ \AA) features, as is commonly found in the literature.

Despite the fact that the majority (59\%) of the ICRF3 objects are quasars, in which the host stellar continuum is too diluted by the AGN to be properly modeled \citep{2006ApJ...641..117G}, we nonetheless perform a full fit of the LOSVD in Regions 1 through 6 to ensure that the general shape of the stellar continuum is modeled accurately, if present.  For quasars, the host component is typically suppressed to zero or negligible levels. In Sy~1 objects, there is enough host contribution that the general shape of the host stellar continuum can be constrained by the data, even if individual absorption features are not resolved and the LOSVD is unconstrained. The fit to the LOSVD is only well-constrained in Sy~2 objects, where the continuum is dominated by the host.

At wavelengths below 3500\AA\ (Regions 7 and 8) we can no longer utilize stellar templates due to the limited wavelength range of the Indo-US stellar library (3460\AA\ to 9464\AA).  We instead use the E-MILES SSP templates \citep{2016MNRAS.463.3409V}, which extend down to 1680\AA. For the vast majority of objects with coverage at wavelengths $\lambda_{\rm{rest}}<3500$\AA, which are typically higher-$z$ type~1 objects with strong power-law contributions, the host galaxy contribution is small, or in the case of quasars, the host galaxy contribution is zero.  The catalog includes the relative contribution of the AGN component at 4000\AA\ and 7000\AA, which can serve as an indication of the relative strength of the host galaxy contribution.

\begin{figure*}
\gridline{
            \fig{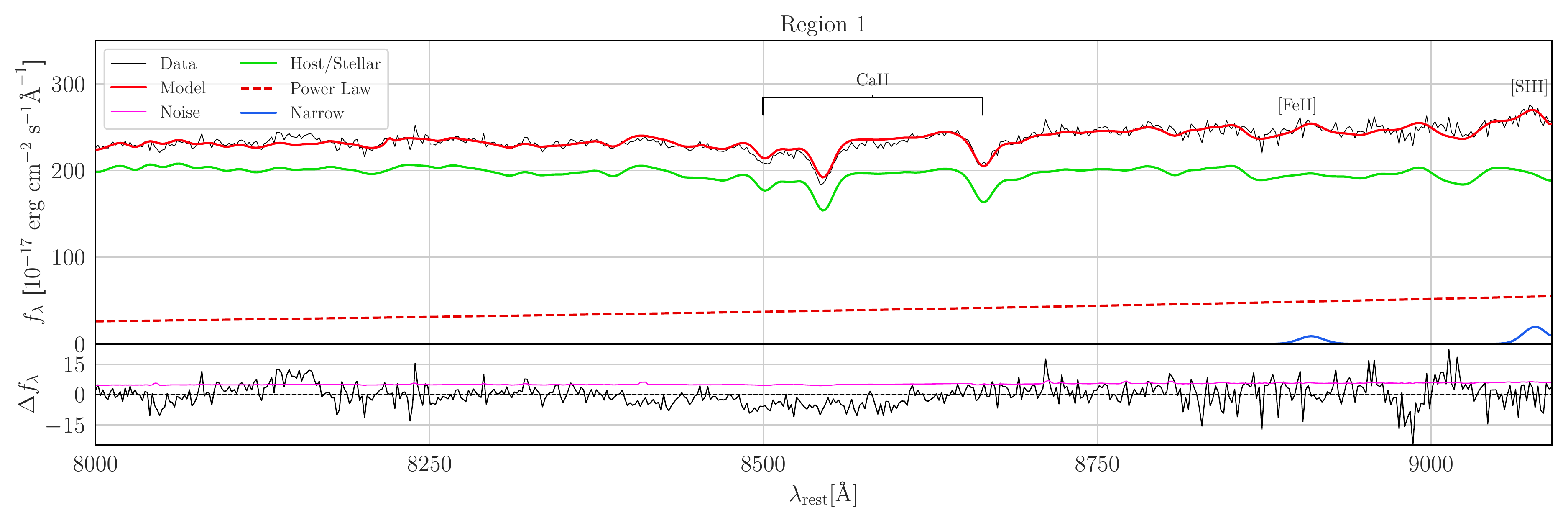}{0.80\textwidth}{(a)}
          }
\gridline{
            \fig{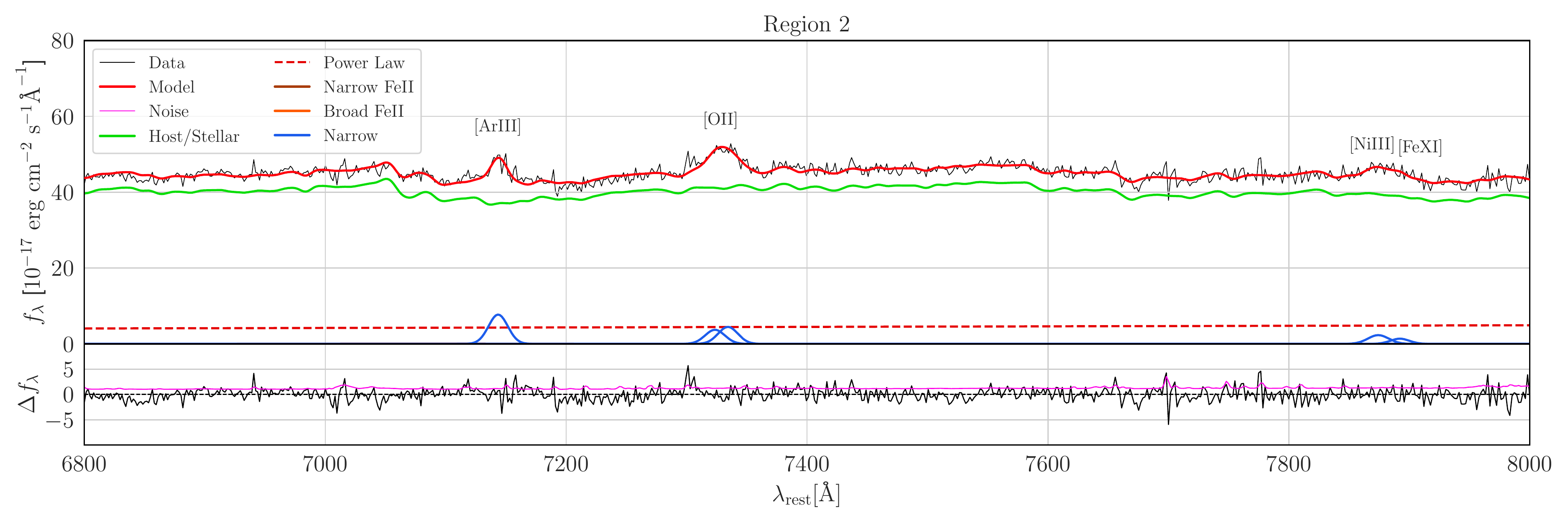}{0.80\textwidth}{(b)}
          }
\gridline{
            \fig{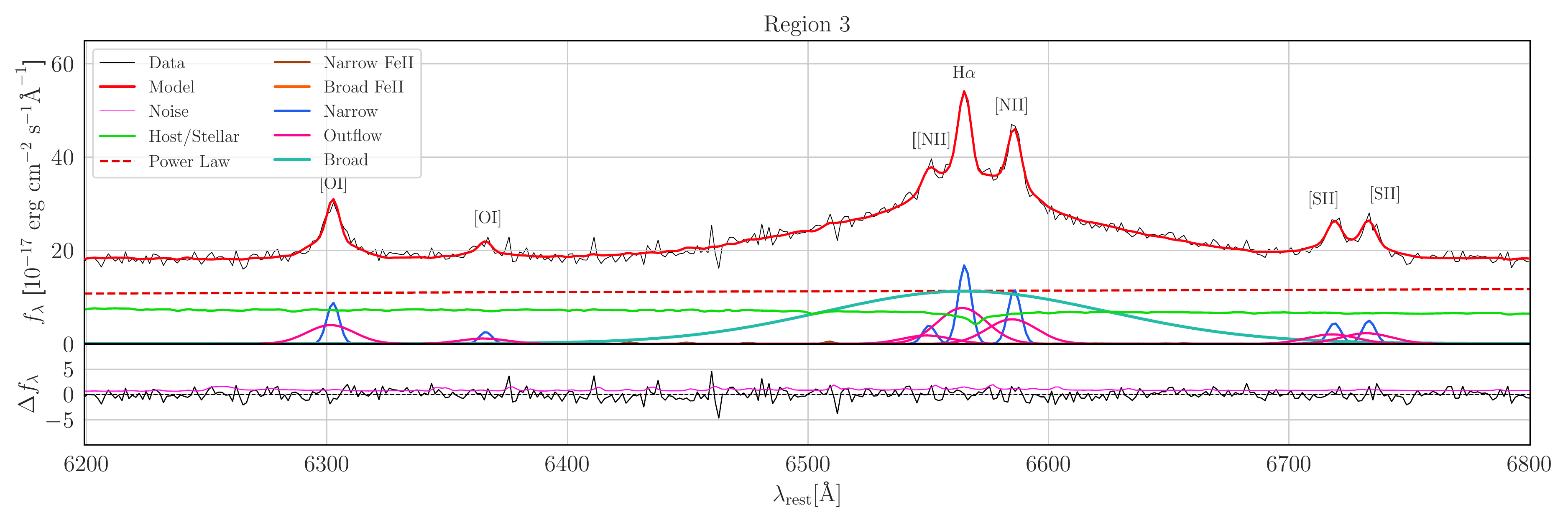}{0.80\textwidth}{(c)}
          }
\gridline{
            \fig{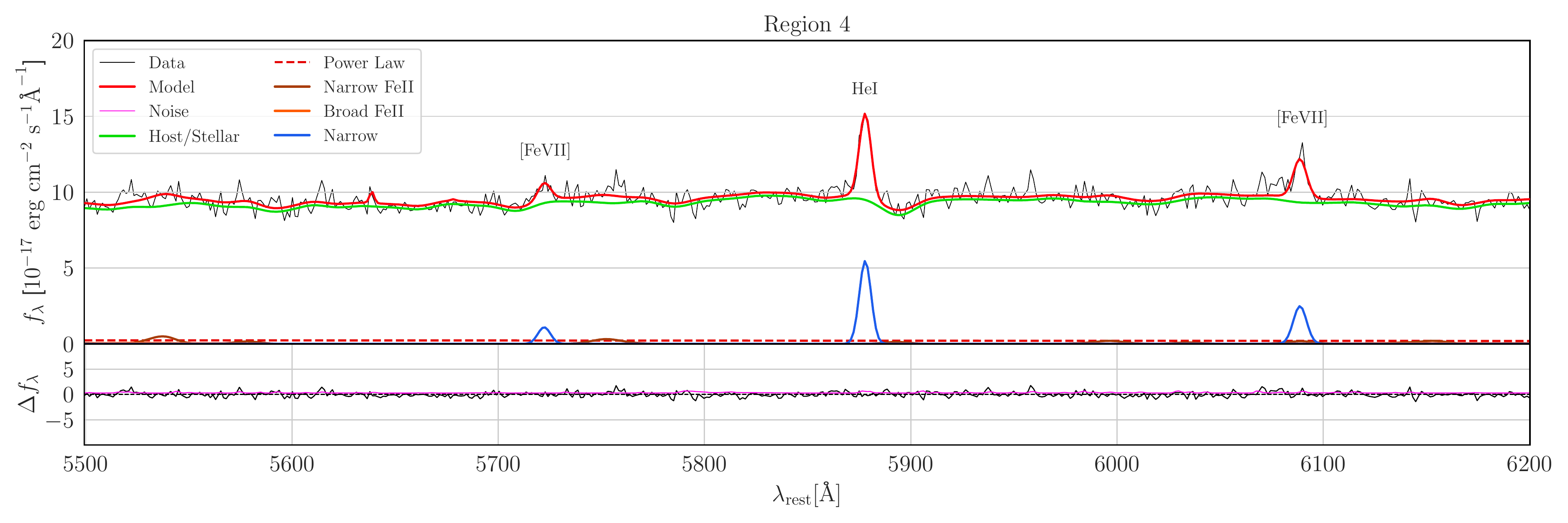}{0.80\textwidth}{(d)}
          }          
\caption{ Example fits to regions in the order defined in Table \ref{tab:fitting_regions}.  (a) Region 1 (IERS 1645+365); (b) Region 2 (IERS 1509+054); (c) Region 3 (IERS 1330+022); (d) Region 4 (IERS 2305+187); (e) Region 5 (IERS 1415+385); (f) Region 6 (IERS 0037+011); (g) Region 7 (IERS 0043+246); (h) Region 8 (IERS 0955+476)
\label{fig:region_fits}}
\end{figure*}

\setcounter{figure}{3}
\begin{figure*}
\gridline{
            \fig{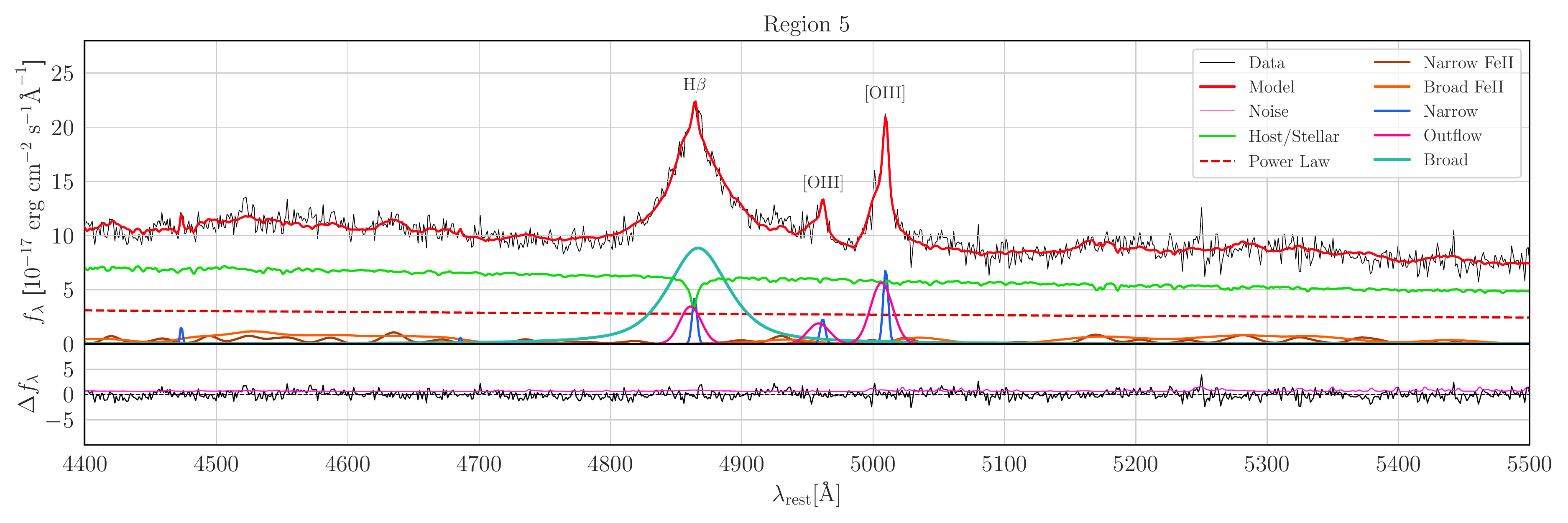}{0.80\textwidth}{(e)}
          }
\gridline{
            \fig{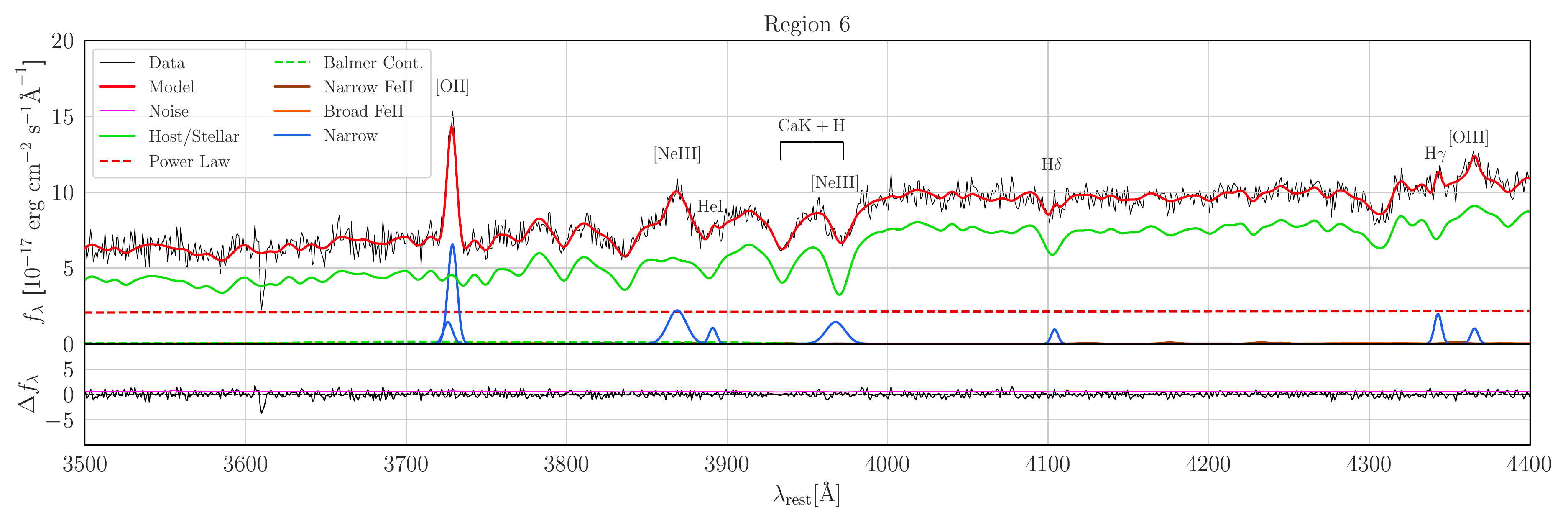}{0.80\textwidth}{(f)}
          }
\gridline{
            \fig{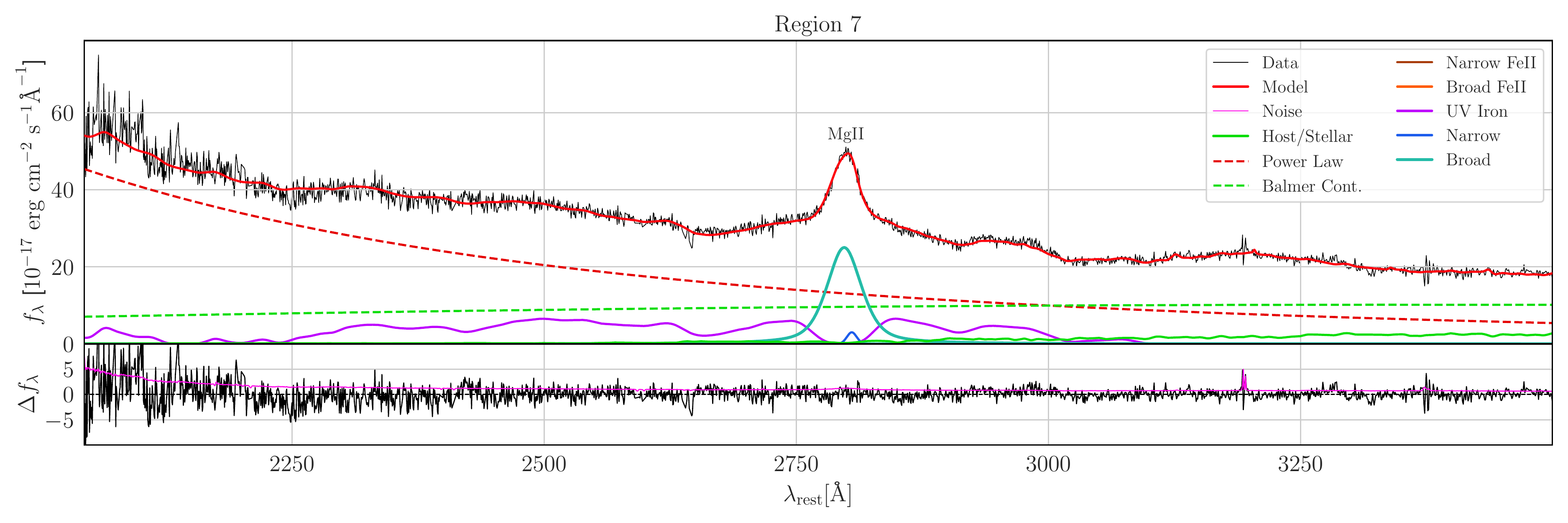}{0.80\textwidth}{(g)}
          }
\gridline{   
            \fig{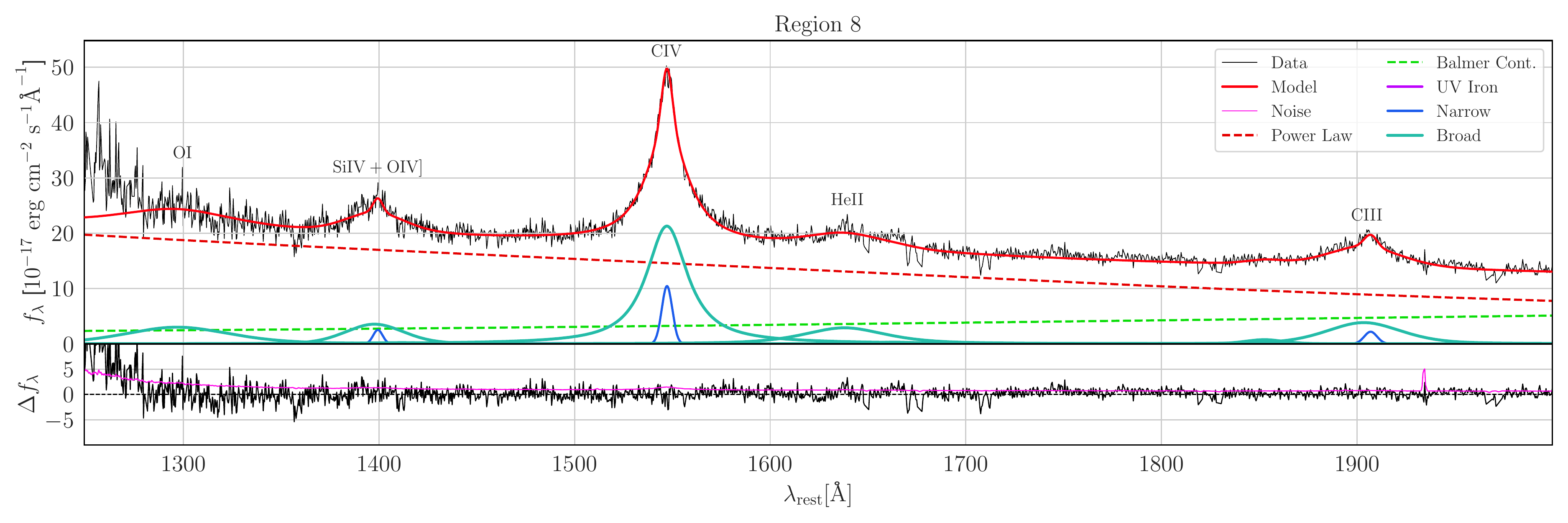}{0.80\textwidth}{(h)}
          }       
\caption{(\emph{continued})
}
\end{figure*}



\subsection{Fit Quality \& Validation}
\begin{figure}[htb!]
\includegraphics[width=\columnwidth]{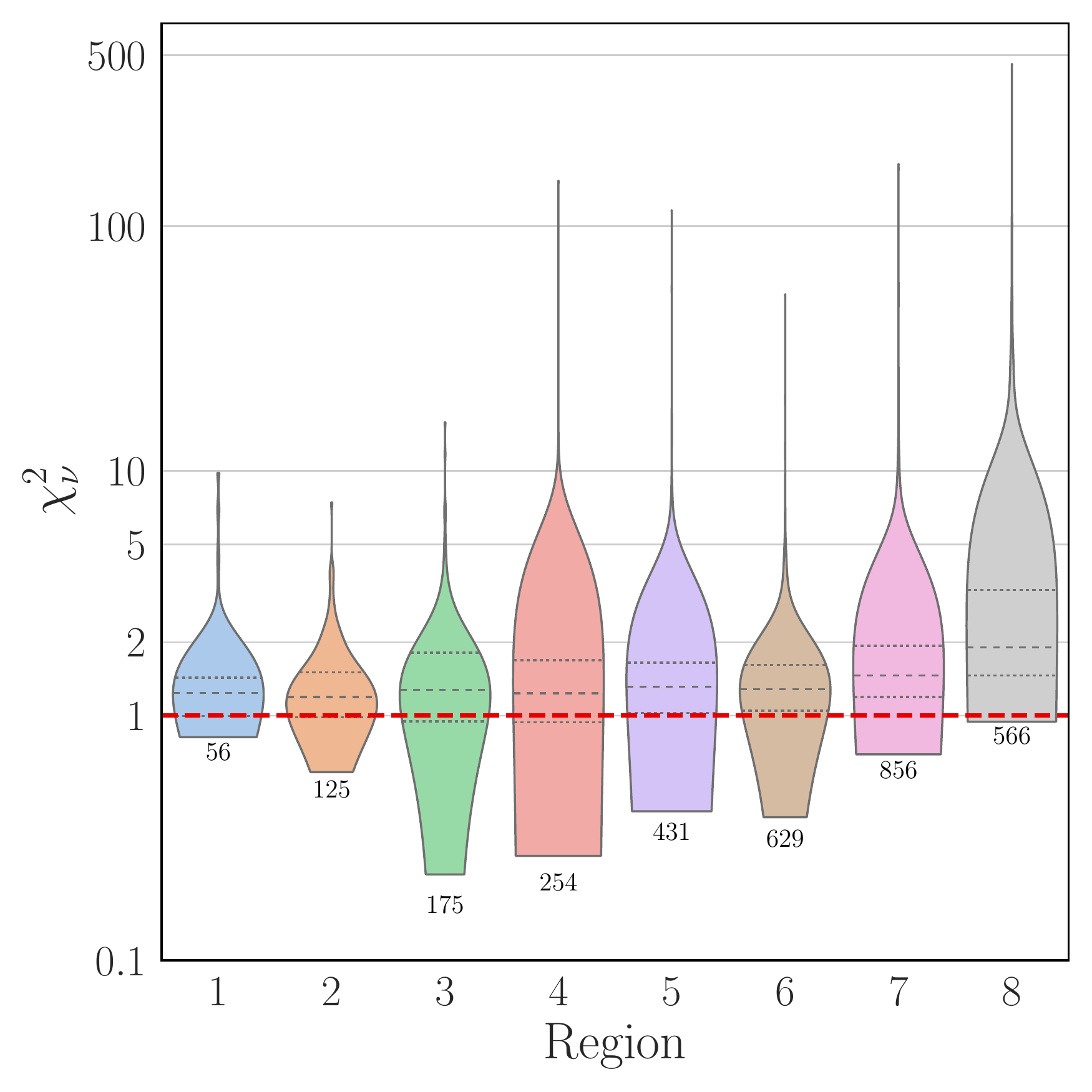}
\caption{
The reduced $\chi^2$ distributions for each fitting region.  The width of each distribution represents the relative fraction of objects of a given $\chi_\nu^2$ value in each region, with the total number of objects in each region given below.  The interquartile range (IQR) is shown inside of each distribution by dashed lines.  The red dashed line represents a $\chi_\nu^2=1$.  
}
\label{fig:rchi2_regions}
\end{figure}

We quantify the quality of our fits in each region using the standard definition of the reduced chi-squared statistic $\chi_\nu^2$.  The boxplot distributions of $\chi_\nu^2$ for each region is shown in Figure \ref{fig:rchi2_regions}.

The median $\chi_\nu^2$ value for all fits to all regions is 1.28, however the median value and scatter of $\chi_\nu^2$ can vary significantly by region for specific reasons.  For instance, the reduced scatter for $\chi_\nu^2<1$ in Regions 1 and 2 can be attributed to type~2 objects with relatively high S/N and spectra well-described by the stellar template fitted to that region.  However, for Regions 3 -- 6, the stellar continuum becomes increasingly diluted by the AGN, which can cause the stellar template fitting algorithm to inadvertently confuse stellar absorption features with noise, resulting in over-fitting of the continuum and a $\chi_\nu^2<1$.  The number of objects with $\chi_\nu^2<1$ decreases again for Regions 7 and 8 where the stellar template fitting algorithm is turned off (due to the wavelength coverage limits of the chosen stellar template library).

For objects with $\chi_\nu^2>1$, we see a linear increase in the number of outliers with number of objects in each region; however, we do see a significant shift in the median $\chi_\nu^2$ values in Regions 7 and 8, likely due to the increased number of metal absorption features in the near-UV and their imperfect interpolation by our algorithm.  This effect is especially strong for region 8 where metal lines can be extremely broad ($>500$~\kms) and strong. In general, the quality of the fit to each region is very good, with $\sim95\%$ of fits within one standard deviation of $\chi_\nu^2 = 1$.

\begin{figure}[htb!]
\includegraphics[width=\columnwidth]{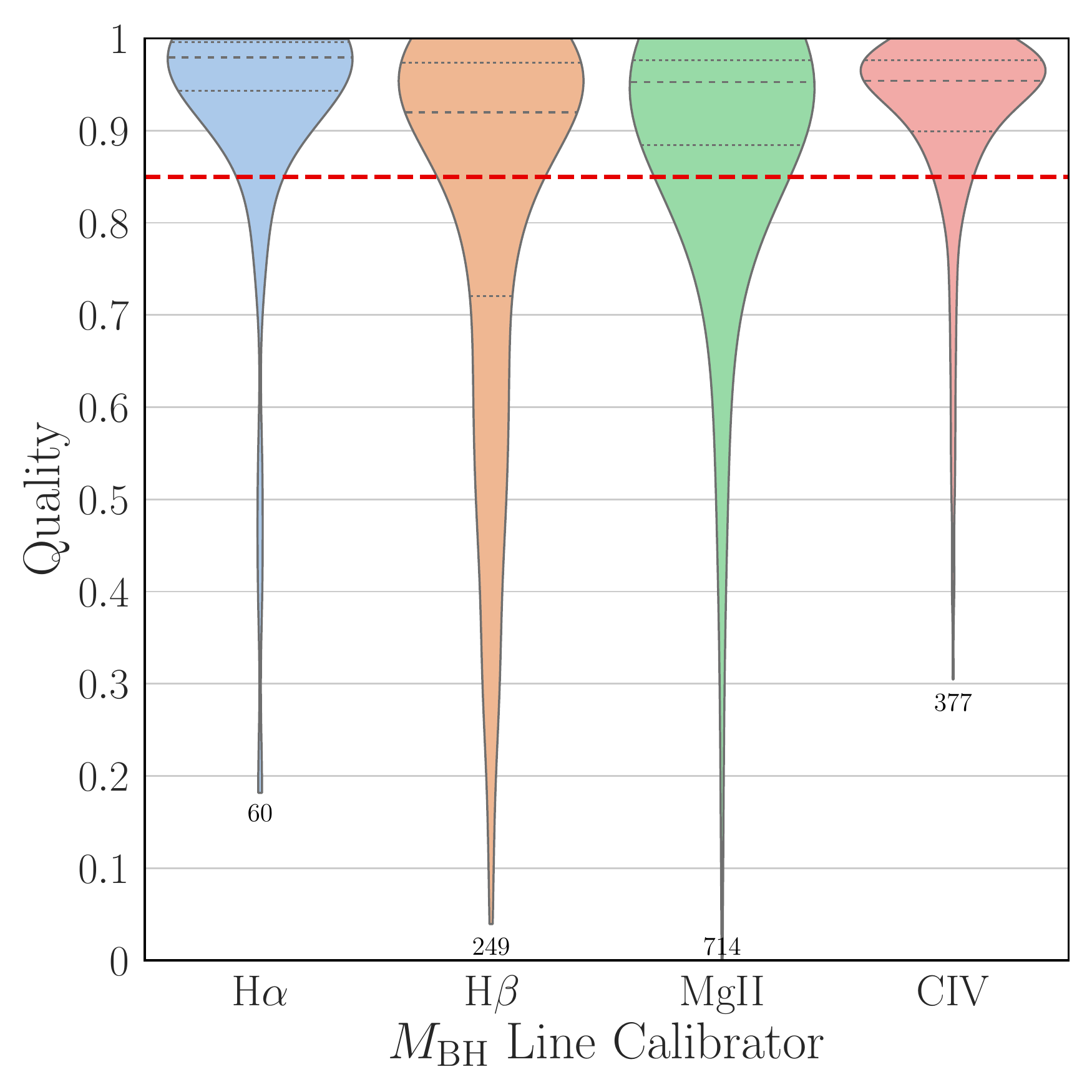}
\caption{
The fit quality distributions for BH mass line calibrators.  The width of each distribution represents the relative fraction of objects with a given fit quality for each line calibrator, with the total number of objects in each region posted below each distribution.  The IQR is shown inside of each distribution by dashed lines.  The red dashed line represents fit quality of 0.85 considered here to be excellent.  
}
\label{fig:quality_lines}
\end{figure}

For broad lines used to estimate virial BH mass, we provide a $\chi_\nu^2$ statistic as well as a quality metric analogous to $R^2$, which is used to asses how well the emission line model explains line flux variance that is not attributable to noise.  Figure \ref{fig:quality_lines} shows violin plots of the distributions of the quality metric for each BH mass line calibrator.  We consider a quality of 0.85 excellent for the use of calibration purposes for the relations derived in Sections \ref{sec:agn_luminosity}, \ref{sec:blr_emission}, and \ref{sec:mbh_calibration}.  Fits to broad H$\alpha$ are generally good, with 90\% of cases with quality metrics $\geq0.85$.  This is juxtaposed by a much lower number of fits with quality metrics $\geq0.85$ for H$\beta$, for which only 63\% meet the criterion.  Factors such as stellar continuum, \ion{Fe}{2} contamination, and asymmetries that can be attributed to ionized gas outflows all contribute to the non-trivialities of fitting broad H$\beta$, and we surmise that these factors play a role in a significantly lower fit quality.  The fits to \ion{Mg}{2} and \ion{C}{4} are comparatively better, with 80\% and 86\% of fits with quality metrics $\geq0.85$, respectively.  This could be attributed to the fact that features in Regions 7 and 8, which contain \ion{Mg}{2} and \ion{C}{4}, are generally more broad, and the continuum more featureless, allowing one to achieve a better quality fit by our metric, albeit with larger parameter uncertainties.

\begin{figure*}[htb!]
\includegraphics[width=\textwidth,trim={0cm 6cm 0 0},clip]{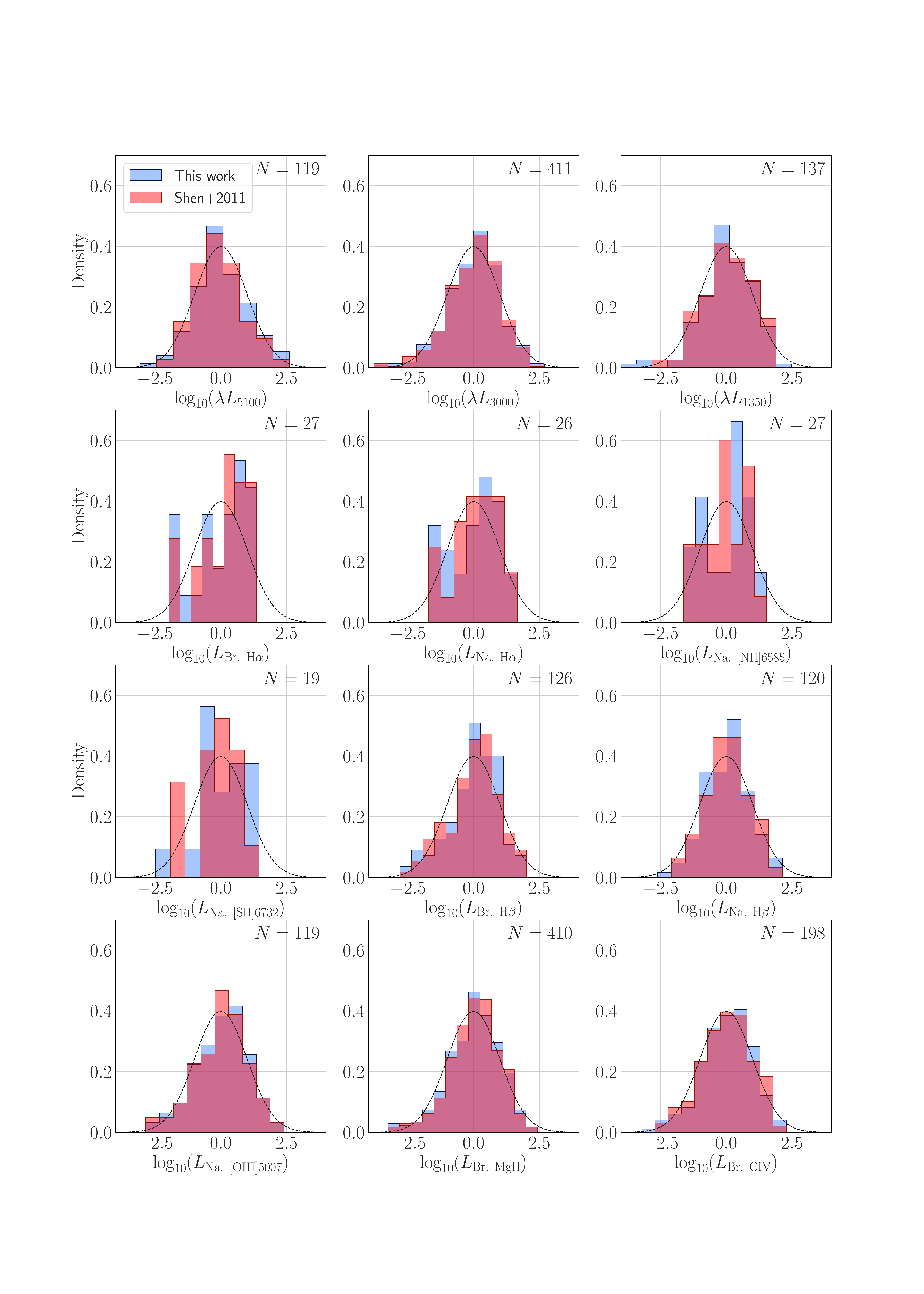}
\caption{
Comparison of error-normalized distributions of select parameters from our catalog and the \citet{2011ApJS..194...45S} QSO catalog. The dashed line represents the normal distribution $\mathcal{N}(0,1)$.
}
\label{fig:Shen2011_valid}
\end{figure*}

To validate our fitting results, we compare common fitted parameters against previously published catalogs of similar objects.  The \citet{2011ApJS..194...45S} QSO catalog includes 38 parameters comparable to our catalog measurements for 506 broad line (QSO and Sy 1) objects in our sample (63\%).  Considering the differences in fitting models (e.g., line profile shapes, stellar/host continuum, multiple line components, etc.), we do expect some systematic offset in the distribution of parameter values.  We report that the vast majority of systematic offsets in parameters between catalogs are significantly less than their respective scatter ($<1\sigma$ on average), and once corrected for these offsets, more than 85\% of variance between parameter distributions can be accounted for on average.  Figure \ref{fig:Shen2011_valid} shows the error-normalized distributions of select parameters between the two catalogs.

\begin{figure*}[htb!]
\includegraphics[width=\textwidth,trim={0cm 6cm 0 0},clip]{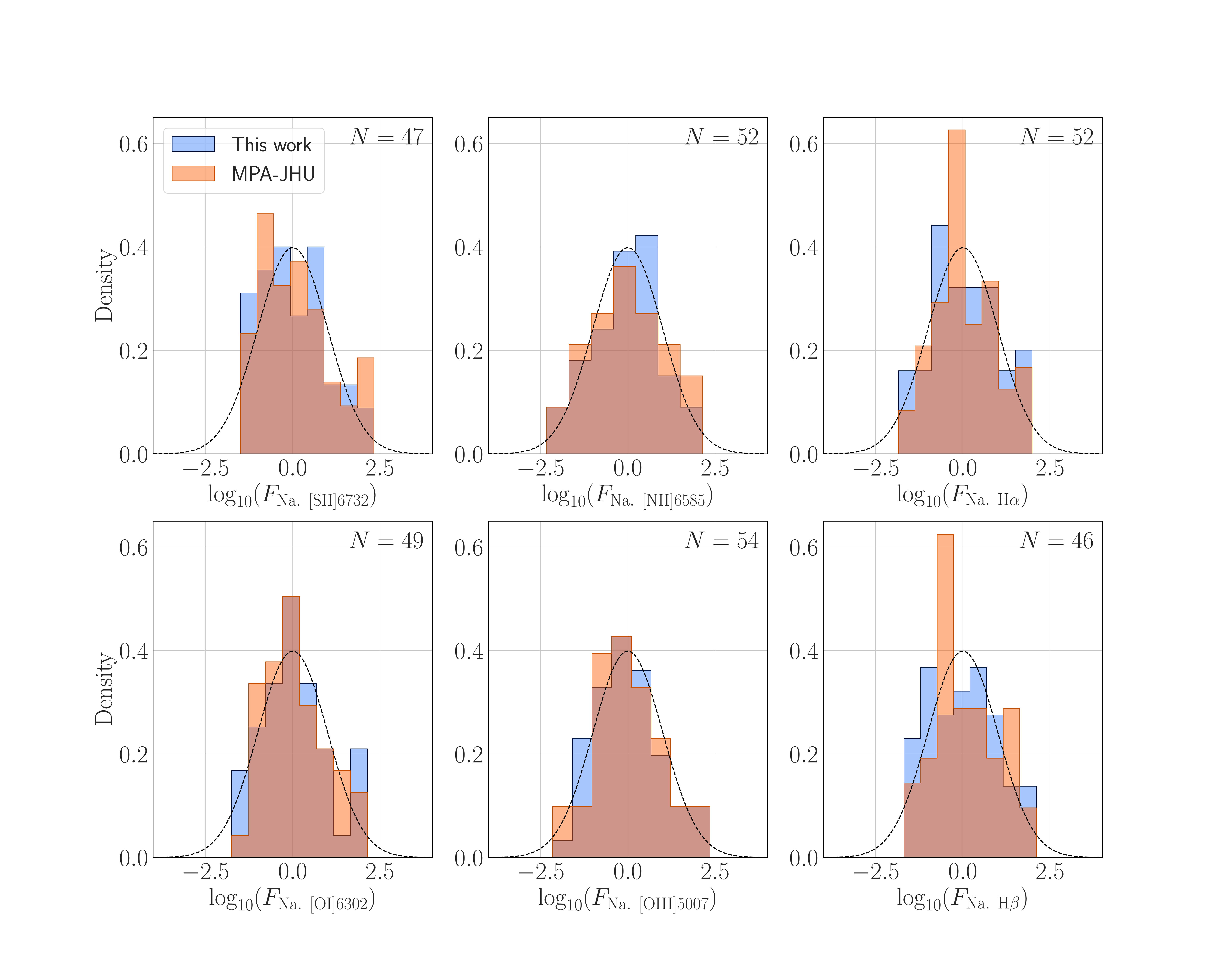}
\caption{
Comparison of error-normalized distributions of select parameters from our catalog and the SDSS DR8 MPA-JHU catalog.  The dashed line represents the normal distribution $\mathcal{N}(0,1)$.
}
\label{fig:MPAJHU_valid}
\end{figure*}

For the 92 objects in our catalog that are classified as Sy~2, we compare our parameter values to the SDSS DR8 MPA-JHU catalog of galaxy emission lines\footnote{\url{http://www.mpa-garching.mpg.de/SDSS/}}, from which there are 55 object matches (60\%) and 12 directly comparable parameters.  We prefer the MPA-JHU catalog over the more-recent Portsmouth catalog \citep{2013MNRAS.431.1383T}, although the general quality of the two catalogs similar in regards to emission line parameters.   For Sy~2 types, we report that the systematic offsets between the MPA-JHU and our catalog is also significantly less than the scatter, and 91\% of the variance in the parameter distributions is accounted for.  A comparison of the error-normalized distributions of select parameters between our catalog and the MPA-JHU catalog is shown in Figure \ref{fig:MPAJHU_valid}.

Given the excellent agreement between $\sim 63\%$ of our catalog with previously published catalogs, we conclude that our fitting method and results are sufficiently validated. 

\section{Derived Properties}
\label{sec:derived_properties}

Whereas the previous section detailed our method for fitting the continuum and line features of the ICRF3 objects, in this section we produce value-added properties derived from these fits. Our analysis includes quantities of relevance to virial BH mass estimation, including proxies of AGN luminosity, broad line width, and BH mass calibration between different estimators, and compare these to those in the literature.

\subsection{AGN Luminosity}
\label{sec:agn_luminosity}

First, we investigate typical proxies of AGN luminosity, including those luminosities measured directly from the AGN component of the spectral continuum, as well as luminosities from the broad line profile of relevant estimators.  Ideally, one would obtain an accurate estimate of the AGN luminosity from high-resolution imaging and performing surface-brightness decomposition of the AGN PSF from the host galaxy \citep{2012ApJ...760...38C,2018ApJ...864..146B,2019ApJ...878..101S}, and then use the $R_{\rm{BLR}}-L_{\rm{AGN}}$ relation from \citet{2013ApJ...767..149B} to obtain a proxy for the broad line region size.  However, a strong correlation between continuum and broad line luminosities and BH mass makes estimating AGN luminosity directly from single-epoch spectra a viable alternative, provided that the host stellar component of the continuum can be accurately subtracted from the AGN continuum component.

When possible, we prefer the continuum over the broad line luminosity, since the modelling of the continuum using the full spectrum is well-constrained, and not subject to uncertainties introduced by the shape of the broad line profile.  Black hole masses estimated using the H$\alpha$ or H$\beta$ width estimator use the continuum luminosity at 5100\AA, the \ion{Mg}{2} width estimator uses the continuum luminosity at 3000\AA, and \ion{C}{4} width estimator uses the continuum luminosity at 1350\AA.  When the continuum luminosity is not available for the respective broad line width estimator, we use the broad line luminosity as a proxy for the continuum luminosity to estimate BH mass.

\begin{figure*}[htb!]
\includegraphics[width=\textwidth]{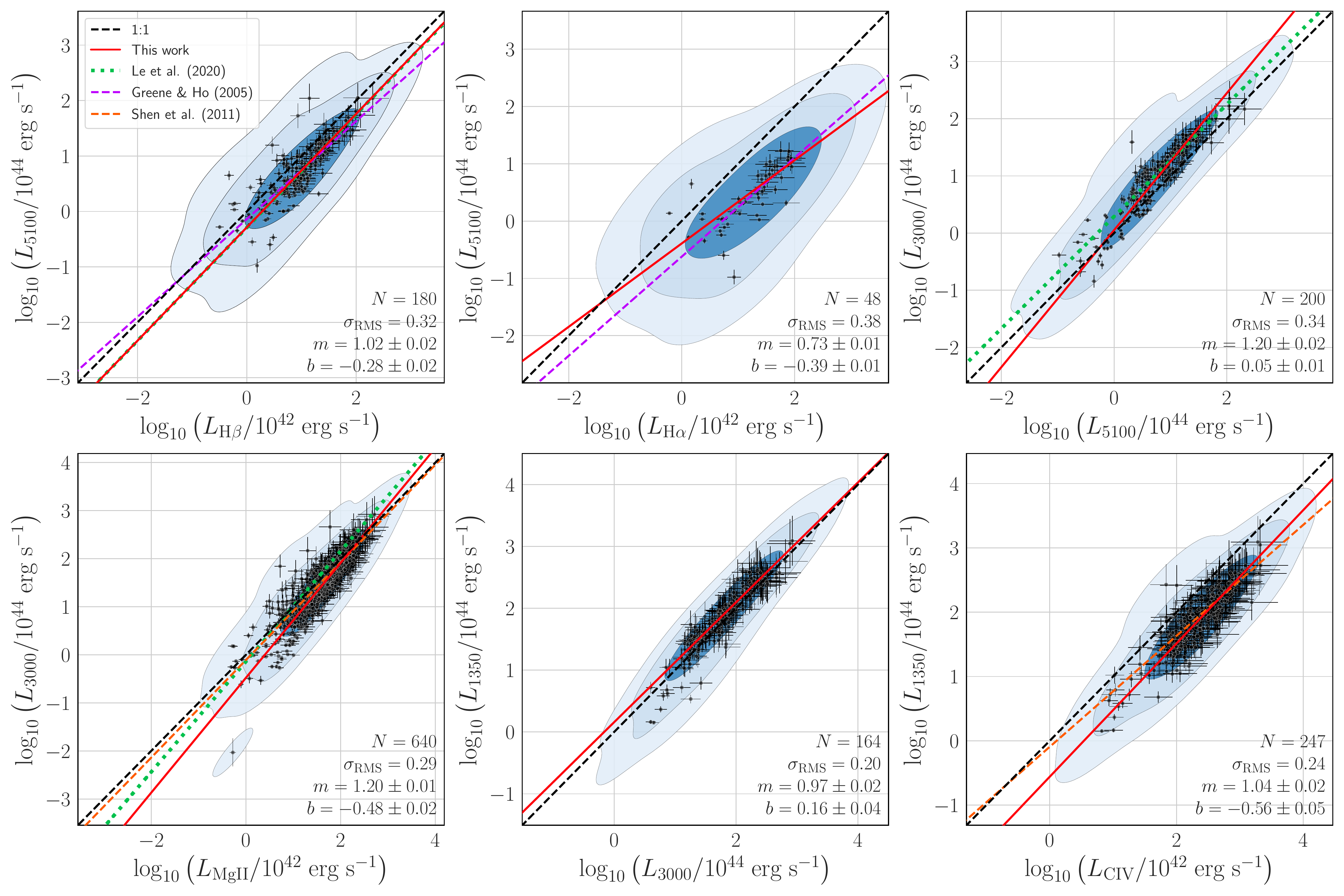}
\caption{KDE plots of various continuum and broad line luminosity relations with comparisons from the literature.  Contour levels represent constant density that include 68\%, 95\% and 99.7\% of $N$ objects.  The intrinsic RMS scatter is given with respect to our best-fit relation (solid red line).  Error bars incorporate the 13\% uncertainty in single epoch luminosities from \citet{2011nlsg.confE..39W}. With the exception of H$\alpha$, all other relations have slopes close to unity, however all relations have slopes similar to their respective empirical relations found in the literature. }
\label{fig:luminosity_relations}
\end{figure*}

To verify that our measurements are valid proxies for AGN luminosity, we measure the log-linear relationship between various luminosity estimators and compare them to those found in the literature.  We perform linear regression using maximum likelihood techniques, taking into account uncertainties in both variables, and estimate uncertainties in the slope and intercept using the affine-invariant Markov chain Monte Carlo (MCMC) sampler \texttt{emcee} \citep{2019JOSS....4.1864F}.  Figure \ref{fig:luminosity_relations} shows the kernel density estimation (KDE) distributions and fits to the various luminosity relations under consideration, as well as comparison fits from literature.  The results of these fits are provided in Table \ref{tab:luminosity_relations}.

In general, the intrinsic AGN continuum and broad emission line luminosities are strongly correlated, which is very important as BH virial mass estimation relies on accurate measurement of the intrinsic AGN continuum luminosity. Of these correlated luminosities, the relationship between H$\alpha$ and $\lambda L_{\rm{5100}}$ has the shallowest slope and greatest intrinsic RMS scatter; however, our relation agrees well with that of \citet{2005ApJ...630..122G}, which has a slope of 0.86.  Relations with \ion{Mg}{2} and \ion{C}{4} are also in good agreement with those from \citet{2011ApJS..194...45S} and \citet{2020ApJ...901...35L}.

From these relations, we can surmise that the relationships between broad line and continuum luminosities for exclusively radio-loud objects, such as those in the ICRF3 sample, appear to be statistically indistinguishable from the same relations for samples that may contain both radio-loud and/or radio-quiet objects.  Previous studies \citep{1992ApJS...80..109B,1996ApJS..102....1B} using correlation analysis and principal component analysis found significant differences between radio-loud and radio-quiet objects in equivalent widths of [\ion{O}{3}], H$\beta$, and \ion{Fe}{2}, however continuum or broad line luminosities were not among these significant differences.

\begin{deluxetable}{lllrc}
\tablecaption{Results of log-linear fits to various continuum and broad line luminosity relations, as shown in Figure \ref{fig:luminosity_relations}.
}
\tablehead{
\colhead{Y} & \colhead{X} & \colhead{$m$} & \colhead{$b$} & \colhead{$\sigma_{\rm{RMS}}$}
}
\colnumbers
\startdata
 $\lambda L_{5100}$ & $L_{\rm{H}\beta}$       & $1.02\pm0.02$ & $-0.28\pm0.02$ & $0.32$ \\
 $\lambda L_{5100}$ & $L_{\rm{H}\alpha}$      & $0.73\pm0.01$ & $-0.39\pm0.01$ & $0.38$ \\
 $\lambda L_{3000}$ & $\lambda L_{5100}$      & $1.20\pm0.02$ & $0.05\pm0.01$  & $0.34$ \\
 $\lambda L_{3000}$ & $\lambda L_{\rm{MgII}}$ & $1.20\pm0.01$ & $-0.48\pm0.02$ & $0.29$ \\
 $\lambda L_{1350}$ & $\lambda L_{3000}$      & $0.97\pm0.02$ & $0.16\pm0.04$  & $0.20$ \\
 $\lambda L_{1350}$ & $\lambda L_{\rm{CIV}}$  & $1.04\pm0.02$ & $-0.56\pm0.05$ & $0.24$ 
\enddata
\tablecomments{Column 1: dependent variable.  Column 2: independent variable.  Column 3: best fit slope. Column 4: best fit $y$-intercept.  Column 5:  Intrinsic RMS scatter (dex).  Column 6: number of objects included in fit.}
\label{tab:luminosity_relations}
\end{deluxetable}

\subsection{Broad Line Width}
\label{sec:blr_emission}

\begin{figure*}[htb!]
\includegraphics[width=\textwidth]{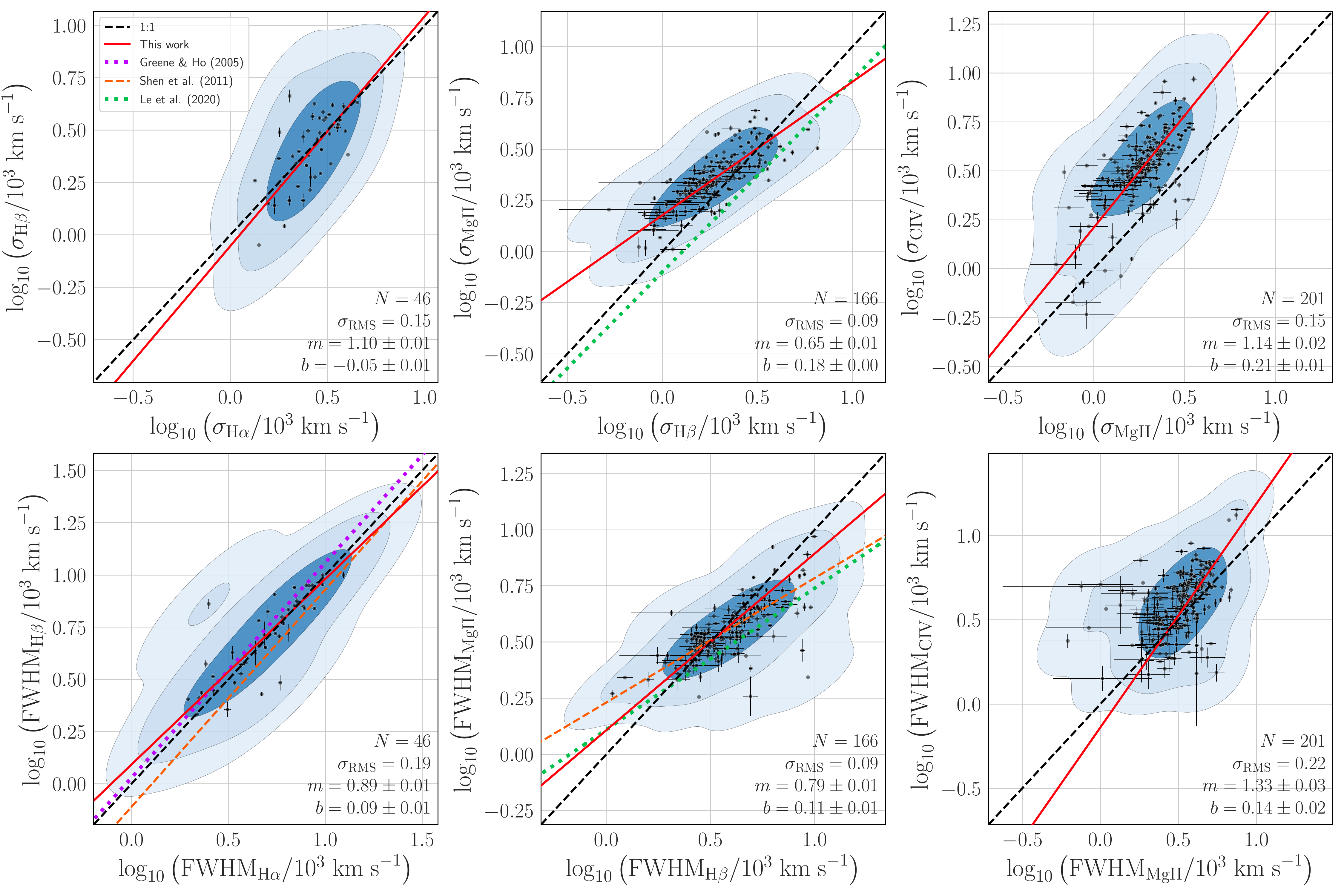}
\caption{KDE plots of various relations with broad line width calibrators with comparisons from the literature.  As with the luminosity relations, contour levels represent constant density that include 68\%, 95\% and 99.7\% of $N$ objects.  RMS scatter is given with respect to our best-fit relation (solid red line). }
\label{fig:width_relations}
\end{figure*}

As with luminosity, we investigate the relations between width calibrators used in the calculation of BH mass to ensure the consistency and validity of our measurements.  Figure~\ref{fig:width_relations} shows the various relationships between broad line width calibrators for dispersion and FWHM.  Results of the fits to these relationships are provided in Table \ref{tab:width_relations}.  We find that there is good agreement between different broad lines and their respective dispersions or FWHMs, on average.

We recover a comparable slope to \citet{2005ApJ...630..122G} and \citet{2011ApJ...737..103S} when comparing the FWHMs of H$\beta$ and H$\alpha$, with the dispersions showing a similar slope, indicating that the ratio of the widths $\rm{FWHM}/\sigma$ converges to a constant value as expected by design of the symmetric line profile (pseudo-Voigt) chosen for this catalog. 

When comparing the \ion{Mg}{2} and H$\beta$ widths, we find a mean offset of $+0.13$~dex for $\sigma_{\rm{MgII}}$ relative to $\sigma_{\rm{H}\beta}$, whereas the mean offset for FWHM is only $-0.03$~dex.  We suspect that the excess dispersion relative to H$\beta$ is due to the fact that strong UV iron contamination in the wings of the line profile can bias the dispersion to larger values, and in part that \ion{Mg}{2} is a doublet.  Excess UV iron emission can push the pseudo-Voigt profile towards a Lorentzian shape with extended symmetric wings, biasing the dispersions toward higher values while leaving the FWHM unaffected. By comparison, the relation from \citet{2020ApJ...901...35L} for both dispersion and FWHM have a negative offset by a comparable amount, while our offset for FWHM is closer to unity and more comparable to those of \citet{2011ApJS..194...45S}. For \ion{Mg}{2}, the difference in BH mass by choice of width estimator ($\sigma$ versus FWHM) is 0.1 dex, with dispersion resulting in lower BH mass on average.

The \ion{C}{4} line profile is fit in a similar way to the \ion{Mg}{2} profile, by considering the full line profile instead of decomposing the line into separate narrow and broad kinematic components.  We fit a single broad pseudo-Voigt profile with a single narrow Gaussian profile.  There appears to be a dichotomy of profile shapes for \ion{C}{4}, which differ by the presence of a peaked narrow component and thus requiring the single narrow Gaussian component in the model.  We suspect that the difference in line profile shape is caused by truncation of the line profile due to narrow and broad metal absorption features typically observed in rest frame near-UV spectra.  The \ion{C}{4} line is fit within Region 8, which is defined as $\lambda_{\rm{rest}}<2000$\AA, which is necessary to constrain an apparently weaker UV iron continuum, as well as other broad emission line features in the region.

We again observe a positive offset of 0.28~dex in dispersion when comparing \ion{C}{4} to \ion{Mg}{2}, compared to an only 0.09~dex for FWHM, as shown in Figure \ref{fig:width_relations}.  We suspect this is also caused by the biasing of the pseudo-Voigt line profile towards a Lorentzian shape, which increases the value of dispersion due to excess flux in the wings of the profile, despite the noise floor truncation we employ in the calculating of dispersion.  It is uncertain if a multi-Gaussian \citep{2011ApJS..194...45S} or Gauss-Hermite \citep{2017ApJ...839...93P} model would yield different results, since both implementations could allow for excess flux in the wings of the profile to be included in the model unless otherwise constrained. The difference in choice of width calibrator ($\sigma$ versus FWHM) for the \ion{C}{4} is 0.17~dex, with dispersion resulting in a lower BH mass than FWHM on average.

Broad \ion{C}{4} is particularly difficult to obtain consistently reliable fits due to the prevalence of  narrow and/or broad metal absorption features in the near-UV.  We use an iterative moving median filter, which varies in size, to detect and interpolate over $3\sigma$ negative outliers.  This method has a high success rate for narrow ($<100$~\kms) features but fails to fully mask broader features ($>1000$~\kms) which cover much larger portions of the fitting region. \citet{2011ApJS..194...45S} used a similar 20-pixel moving boxcar method to detect and mask absorption features at the $3\sigma$ level.  We find that interpolation is preferable to masking, since the size of the mask is often such that a significant portion of the \ion{C}{4} line is masked, causing the fit to be unconstrained within the mask, whereas interpolation preserves these pixels as constraints for the fit.  However, this method is only marginally better for cases in which large absorption features affect the \ion{C}{4} line profile, since we cannot fully reconstruct the original line shape from either method.  The affect of strong absorption features on the \ion{C}{4} line profile can truncate the full amplitude of the line and bias the quantities derived from the model such as width, flux, and equivalent width.  A larger study of the systematics regarding the fitting of \ion{C}{4} is necessary to address these  modelling issues. 

In summary, it would appear that when comparing across different line estimators (H$\alpha$, H$\beta$, \ion{Mg}{2}, \ion{C}{4}), the FWHM width estimator is more consistent; that is, the FWHM measurements of different lines scale closer to unity.  Lines such as \ion{Mg}{2} and \ion{C}{4}, which can experience varying amounts of UV iron contamination, make choice of line profile shape non-trivial, and can lead to biased estimates in velocity dispersion.  While dispersion has been considered the width estimator of choice for BH studies, in part due to the decreased scatter as shown in Figure \ref{fig:width_relations}, it is possible that FWHM may be a better and unbiased estimator of width across different line estimators.  A systematic study of the relationship between line profile shapes across different line estimators is necessary to determine which model implementation is ideal for large samples. 

\begin{deluxetable}{lllrc}
\tablecaption{Results of log-linear fits to various broad line width relations, as shown in Figure \ref{fig:width_relations}.
}
\tablehead{
\colhead{Y} & \colhead{X} & \colhead{$m$} & \colhead{$b$} & \colhead{$\sigma_{\rm{RMS}}$}
}
\colnumbers
\startdata
 $\sigma_{\rm{H}\beta}$    & $\sigma_{\rm{H}\alpha}$    & $1.10\pm0.01$ & $-0.05\pm0.01$ & $0.15$ \\
 $\sigma_{\rm{MgII}}$      & $\sigma_{\rm{H}\beta}$     & $0.65\pm0.01$ & $0.18\pm0.00$ & $0.09$  \\
 $\sigma_{\rm{CIV}}$       & $\sigma_{\rm{MgII}}$       & $1.14\pm0.02$ & $0.21\pm0.01$ & $0.15$  \\
 $\rm{FWHM}_{\rm{H}\beta}$ & $\rm{FWHM}_{\rm{H}\alpha}$ & $0.89\pm0.01$ & $0.09\pm0.01$ & $0.19$  \\
 $\rm{FWHM}_{\rm{MgII}}$   & $\rm{FWHM}_{\rm{H}\beta}$  & $0.79\pm0.01$ & $0.11\pm0.01$ & $0.09$  \\
 $\rm{FWHM}_{\rm{CIV}}$    & $\rm{FWHM}_{\rm{MgII}}$    & $1.33\pm0.03$ & $0.14\pm0.02$ & $0.22$  
\enddata
\tablecomments{Column 1: dependent variable.  Column 2: independent variable.  Column 3: best fit slope. Column 4: best fit $y$-intercept.  Column 5:  Intrinsic RMS scatter (dex).  Column 6: number of objects included in fit.}
\label{tab:width_relations}
\end{deluxetable}

\subsection{Black Hole Mass}
\label{sec:mbh_calibration}

Estimates of BH mass are inferred from the size and kinematics of gas assumed to be virialized within the gravitational potential of the BH \citep{2004IAUS..222...15P}, given by the equation

\begin{equation}
    M_{\rm{BH}}=f\frac{\Delta V^2 R_{\rm{BLR}}}{G}
\end{equation}

\noindent where $f$ is the virial coefficient, $\Delta V$ is the gas velocity dispersion, $R_{\rm{BLR}}$ is the effective radius of the broad line region, and $G=5.124~\rm{km}^2~\rm{s}^{-2}~\rm{ld}~\rm{M}_\odot^{-1}$ is the gravitational constant in the relevant units.  The virial coefficient $f$ is used to scale single-epoch BH estimates such that they are consistent with those of well-determined reverberation-mapped and quiescent galaxies using the broad H$\beta$ emission line.  The gas velocity $\Delta V$ is estimated via the gas dispersion or FWHM of broad emission lines. The radius of the broad line region $R_{\rm{BLR}}$ is empirically determined from a strong correlation with the AGN luminosity at 5100 $\textrm{\AA}$ from \citet{2013ApJ...767..149B}, given by

\begin{align}
    \log_{10}\left(\frac{R_{\rm{BLR}}}{\rm{ld}}\right)&=(1.527\pm0.031)+(0.533\pm0.035)\nonumber\\
    &\times\log_{10}\left(\frac{\lambda L_{5100}}{10^{44}\rm{~erg~s}^{-1}}\right)
\end{align}

\noindent \citet{2015ApJ...801...38W} determined the virial factor by re-calibrating the reverberation-mapped and quiescent galaxies, finding $\log_{10}(f) = 0.05\pm 0.12$ and $\log_{10}(f) = 0.65 \pm 0.12$ for dispersion- and FWHM-based BH masses, respectively.

We use the above relations and values to determine BH mass for H$\beta$ line for both the dispersion and FWHM width estimators, and use H$\beta$ as the baseline for determining the appropriate calibration for H$\alpha$ and \ion{Mg}{2}.  We then use the \ion{Mg}{2} BH mass to calibrate \ion{C}{4} BH masses. 


The redshift range spanned by ICRF3 objects, as well as the available wavelength coverage in SDSS, ensures that the number and type of available broad lines and continuum luminosities required for single-epoch virial BH mass estimation varies from object to object.  As a result, it is necessary to re-derive BH mass calibrators for \ion{Mg}{2} and \ion{C}{4} in order to ensure that BH masses are self-consistent for all objects in the catalog.  We note that while these calibrations have been performed in the literature in the past, calibrations of these estimators have not been performed for a sample of exclusively radio-loud objects.

The functional form of the BH mass estimator we adopt is given by

\begin{align}
    \log_{10}\left(\frac{M_{\rm{BH}}}{\rm{M}_\odot}\right) =~& \alpha + \beta\log_{10}\left(\frac{\Delta V}{10^3\rm{\;km\;s}^{-1}}\right)  \\
    & + \gamma\log_{10}\left(\frac{\lambda L_\lambda}{10^{44}\rm{\;erg\;s}^{-1}}\right)
\label{eq:BH_mass_1}
\end{align}

\noindent where $\Delta V$ is estimated using either the broad line dispersion ($\sigma$) or FWHM, and $\lambda L_\lambda$ is the monochromatic luminosity of the AGN measured from the AGN continuum (power-law component) of the best-fit model.  The coefficients $\beta=2.0$ and $\gamma=0.5$ were fixed constants in the calculation of $\alpha$ for both the H$\alpha$, \ion{Mg}{2}, and \ion{C}{4} calibrations. 

If instead the integrated luminosity of the broad line is used to as a proxy for the monochromatic AGN luminosity, the functional form of the BH mass estimator used to estimate $\alpha$ becomes:

\begin{align}
    \log_{10}\left(\frac{M_{\rm{BH}}}{\rm{M}_\odot}\right) =~& \alpha + \beta\log_{10}\left(\frac{\Delta V}{10^3\rm{\;km\;s}^{-1}}\right) + \nonumber \\ & \gamma\log_{10}\left(\frac{L_{\rm{br.}}}{10^{42}\rm{\;erg\;s}^{-1}}\right)
\label{eq:BH_mass_2}
\end{align}

\noindent where $L_{\rm{br.}}$ is the integrated broad line luminosity, and the luminosity scaling factor becomes $10^{42}$ erg s$^{-1}$, following the conventions frequently used in the literature.  We calibrate H$\alpha$ and \ion{Mg}{2} BH masses using H$\beta$ BH masses to determine the best fitting $\alpha$.  Due to the wavelength range of the SDSS, we cannot use H$\beta$ to determine the calibration for \ion{C}{4}, and therefore \ion{Mg}{2} is used for the calibration of \ion{C}{4} instead, with uncertainties for these calibrations propagated appropriately.

The results of the calibrations for H$\alpha$, \ion{Mg}{2}, and \ion{C}{4} are provided in Table \ref{tab:mbh_calibration}.  Objects used for calibrations are those with line detections of at least 95\% and quality flags of at least 0.85 to ensure that poor quality fits do not adversely bias the calibration.  The dispersion width calibrator consistently results in the lowest scatter in each resulting calibration, while the FWHM width calibrator consistently produces the most amount of scatter. The scatter is largest at $\sim0.36$~dex for the \ion{C}{4}, likely due to increased scatter in both FWHM and dispersion, combined with the stronger dependence on width in Equations~\ref{eq:BH_mass_1} and \ref{eq:BH_mass_2}.  The resulting values of $\alpha$ in Table \ref{tab:mbh_calibration} are used to calculate BH mass for H$\alpha$, \ion{Mg}{2}, and \ion{C}{4}.

\begin{figure}[htb!]
\includegraphics[width=\columnwidth,scale=0.8]{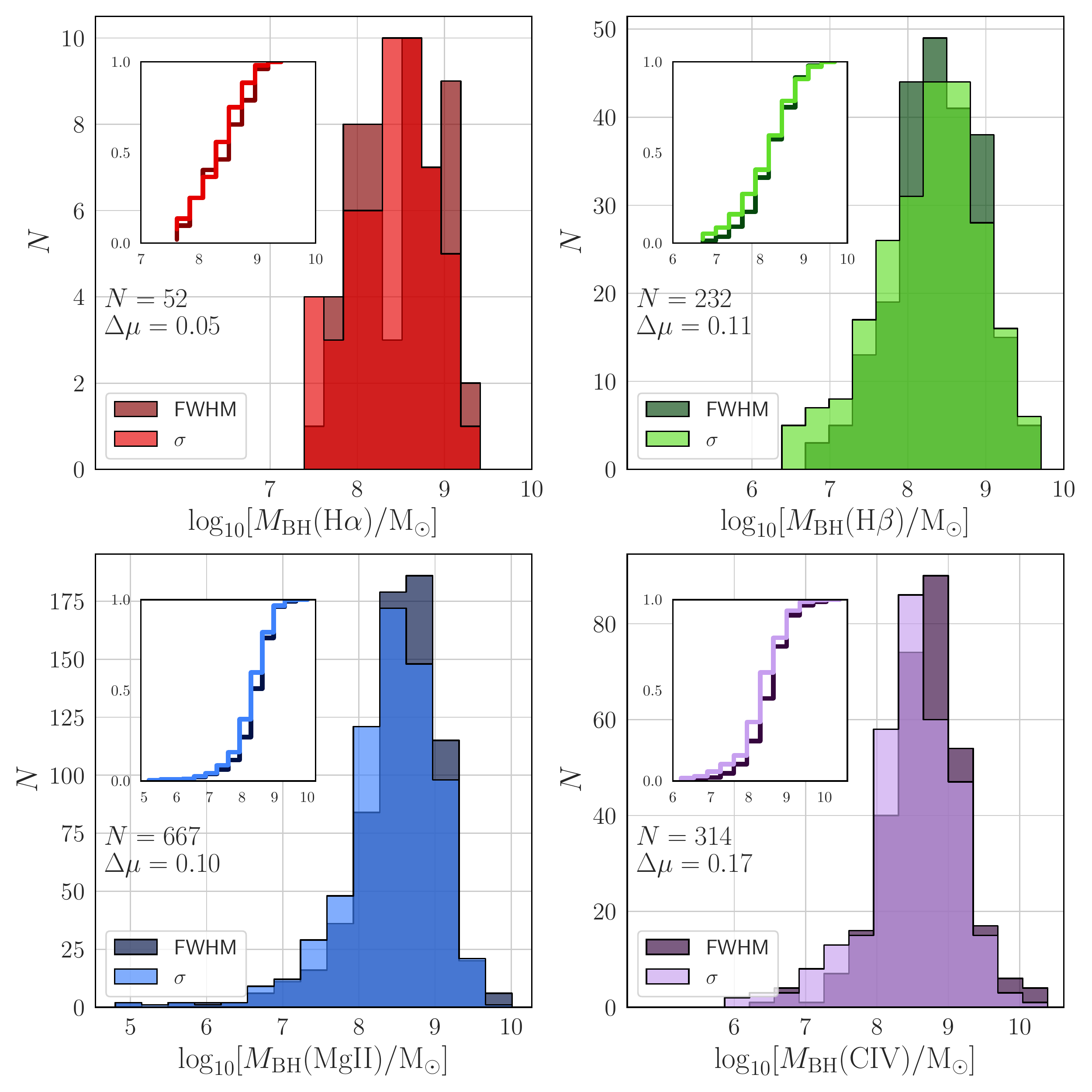}
\caption{
Distributions of BH mass for H$\alpha$, H$\beta$, \ion{Mg}{2}, and \ion{C}{4} for both dispersion and FWHM width calibrators.  The cumulative distribution functions for each distribution are inset in the top left of every panel to show the systematic overestimation of dispersion-based BH mass relative to those based on FWHM. 
}
\label{fig:mbh_hist}
\end{figure}

Considering the availability of the four broad line BH estimators for each spectrum, there are a total of 795 objects with unique BH masses, of which 467 (59\%) have at least two estimates of BH mass from available broad lines. The BH mass distributions for H$\alpha$, H$\beta$, \ion{Mg}{2}, and \ion{C}{4} calibrators, for both dispersion and FWHM-based BH mass, are shown in Figure~\ref{fig:mbh_hist}. Average uncertainty for H$\beta$ masses derived from empirical relations is 0.46~dex, based on the scatter derived from the calibration of the virial factor from \citet{2015ApJ...801...38W} as well as uncertainties in the measurements of width and luminosity. Average uncertainties for H$\alpha$, \ion{Mg}{2}, and \ion{C}{4} are $\sim0.51$~dex for dispersion and $\sim 0.55$ for FWHM.

We find that dispersion systematically overestimates BH mass relative to masses estimated using FWHM, with the largest differences found in \ion{C}{4}, with a mean offset of $+0.17$ dex.  The H$\beta$ and \ion{Mg}{2} BH masses show nearly identical mean offsets at $+0.11$ dex, and H$\alpha$ shows the least difference at $+0.05$ dex.  We suspect that the difference in BH mass estimated with dispersion is due to the bias towards a Lorentzian profile of our pseudo-Voigt implementation, caused by optical \ion{Fe}{2} and UV iron emission in the regions including H$\beta$, \ion{Mg}{2}, and \ion{C}{4}.  Optical \ion{Fe}{2} emission generally extends up to 7000\AA; however, the relative intensities of \ion{Fe}{2} for our model \citep{2004A&A...417..515V} are significantly weaker than they are near H$\beta$.

\begin{deluxetable}{llcrc}
\tablecaption{Calibration of H$\alpha$, \ion{Mg}{2}, and \ion{C}{4} BH mass for the ICRF3 sample.
}
\tablehead{
\colhead{$\Delta V$} & \colhead{$L_\lambda$} & \colhead{$\alpha$} & \colhead{$\mu_{\rm{offset}}$} & \colhead{$\sigma_{\rm{RMS}}$} \\
& & & (dex) & (dex)
}
\colnumbers
\startdata
$\sigma_{\rm{H}\alpha}$     & $\lambda L_{\rm{5100}}$    & $7.46\pm0.07$    & $0.00$     & $0.25$   \\
$\sigma_{\rm{H}\alpha}$     & $L_{\rm{H}\alpha}$         & $7.06\pm0.07$    & $+0.03$    & $0.26$   \\
$\rm{FWHM}_{\rm{H}\alpha}$  & $\lambda L_{\rm{5100}}$    & $6.91\pm0.07$    & $-0.03$    & $0.31$   \\
$\rm{FWHM}_{\rm{H}\alpha}$  & $L_{\rm{H}\alpha}$         & $6.50\pm0.07$    & $-0.06$    & $0.31$   \\
\hline
$\sigma_{\rm{MgII}}$        & $\lambda L_{\rm{3000}}$    & $7.20\pm0.04$    & $-0.03$    & $0.27$   \\
$\sigma_{\rm{MgII}}$        & $L_{\rm{MgII}}$            & $7.07\pm0.04$    & $-0.01$    & $0.32$   \\
$\rm{FWHM}_{\rm{MgII}}$     & $\lambda L_{\rm{3000}}$    & $6.83\pm0.04$    & $+0.01$    & $0.32$   \\
$\rm{FWHM}_{\rm{MgII}}$     & $L_{\rm{MgII}}$            & $6.69\pm0.04$    & $-0.02$    & $0.36$   \\
\hline
$\sigma_{\rm{CIV}}$        & $\lambda L_{\rm{1350}}$     & $6.60\pm0.06$    & $-0.03$    & $0.26$    \\
$\sigma_{\rm{CIV}}$        & $L_{\rm{CIV}}$              & $6.39\pm0.06$    & $-0.03$    & $0.30$    \\
$\rm{FWHM}_{\rm{CIV}}$     & $\lambda L_{\rm{1350}}$     & $6.63\pm0.06$    & $-0.13$    & $0.36$    \\
$\rm{FWHM}_{\rm{CIV}}$     & $L_{\rm{CIV}}$              & $6.42\pm0.06$    & $-0.16$    & $0.35$    \\
\enddata
\tablecomments{Column 1: velocity calibrator.  Column 2: luminosity calibrator.  Column 3: best-fit $\alpha$. Column 4: mean offset of residuals from unity.  Column 5:  Intrinsic RMS scatter. }
\label{tab:mbh_calibration}
\end{deluxetable}

\section{General Properties of ICRF3 Objects}
\label{sec:general_properties}

\subsection{AGN Spectral Type}
We classify each ICRF3 object in our sample by its AGN spectral type, of which there are four that can be distinguished by general spectroscopic characteristics.  These four AGN spectral types are ``QSO'', ``Sy 1'', ``Sy 2'', and ``featureless'', and are provided in the catalog for each object by the ``\texttt{AGN\_TYPE}'' column.

\begin{figure*}[htb!]
\includegraphics[width=1.0\textwidth]{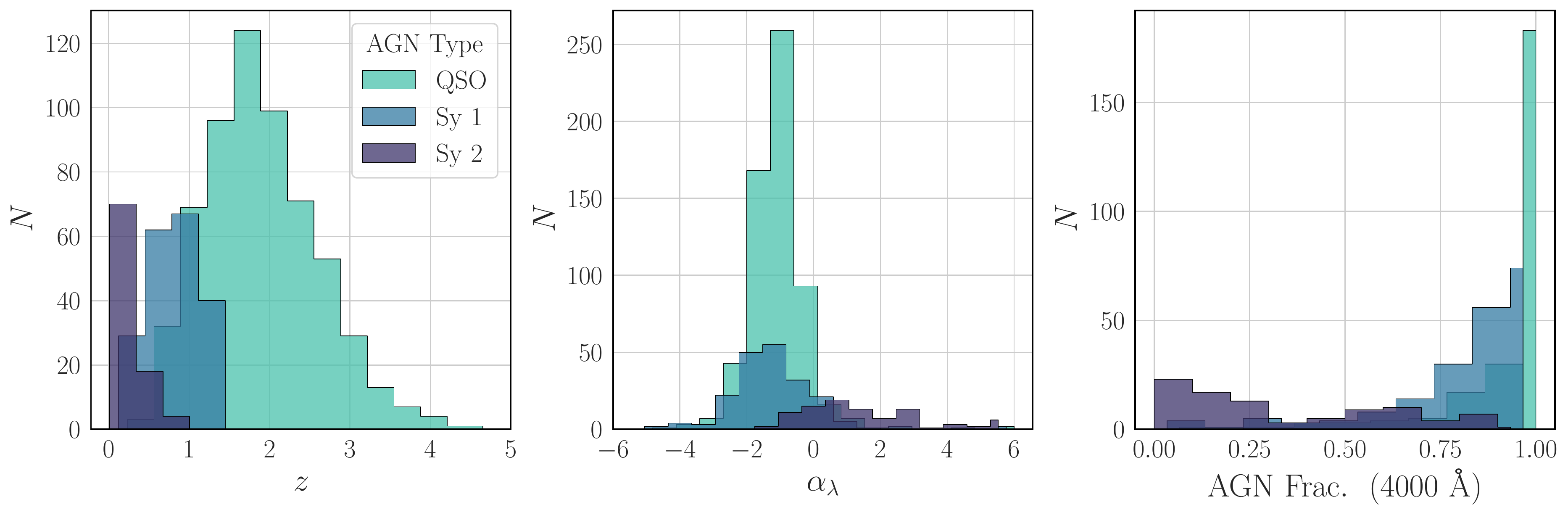}
\caption{General properties of ICRF3 objects that are indicative of AGN types.  From left to right, distributions of redshift $z$, power-law index (slope) $\alpha_\lambda$, and AGN fraction of the continuum at 4000 $\rm{\AA}$ as a function of AGN type in our catalog.}
\label{fig:prop_agn_type}
\end{figure*}

The vast majority ($N=602$) of ICRF3 objects are quasars (QSOs), which are typically at the highest redshifts in the sample.  Quasars are typically distinguished by their luminosities, which are comparatively higher than those of Sy~1 objects.  The AGN power-law continuum, parameterized by index $\alpha_\lambda$, is nearly identical between Sy~1 objects and QSOs; QSOs in our sample have a mean $\alpha_\lambda = -1.58$, while Sy~1 objects have a mean $\alpha_\lambda=-1.56$.  To further distinguish these two classes, we consider the fraction of the total continuum at 4000\AA\ attributed solely to the AGN continuum.  For our sample, we classify an object as a QSO if the continuum fraction at 4000 \AA\ is consistent with unity, within its $1\sigma$ uncertainty, that is, there is no detectable host galaxy component in the spectrum at 4000 \AA. The distributions of QSOs with redshift, $\alpha_\lambda$, and AGN flux fraction at 4000~\AA\ is shown in Figure \ref{fig:prop_agn_type}.


Sy~1 objects ($N=198$) occupy an intermediate space in redshift between QSOs and type~2 objects, and are therefore distinguished by a strong AGN fraction at 4000\AA\ and strong host galaxy fractions at 7000\AA.  The majority of Sy~1 objects have AGN continuum fractions at 4000\AA\ greater than 50\%.

Finally, Sy~2 objects ($N=92$), which have host galaxy contributions that dominate the entire optical spectrum, presumably due to obscuration of the AGN, occupy the lowest redshifts in the ICRF3 sample.  The shape of the AGN power-law continuum for Sy~2 objects is closer to flat with a mean $\alpha_\lambda=1.01$.  The AGN fraction at 4000\AA\ is negligible or is is consistent with zero, and the lack of dilution from an AGN continuum coupled with lower redshifts can allow for accurate recovery of the stellar LOSVD due to higher S/N.

A fourth class of 122 objects, for which the only spectral feature is a featureless power-law continuum, we classify as ``featureless'' in the catalog. The lack of any spectral features, coupled with the fact that these are radio-loud objects, may indicate that these are AGNs with featureless radio continua beamed into the LOS.  Because they have no distinguishing emission or absorption line features, we cannot determine the redshifts of these objects.  We do however provide values and uncertainties for the power-law continuum slope for these objects to characterize the general shape of the spectrum. 

\subsection{BPT Diagrams of Type~2 Objects}
While we have assumed that accretion onto a SMBH is the primary engine powering radio-loud emission in ICRF3 objects, it is not obvious from the spectra of Sy~2 objects that this is the case, due to a lack of distinguishing AGN features such as broad lines or a steep power-law continuum.  We can further classify Sy~2 objects in our sample using BPT diagrams \citep{1981PASP...93....5B} to determine if other radiative mechanisms contribute to the optical emission we observe.

\begin{figure*}[htb!]
\includegraphics[width=1.0\textwidth]{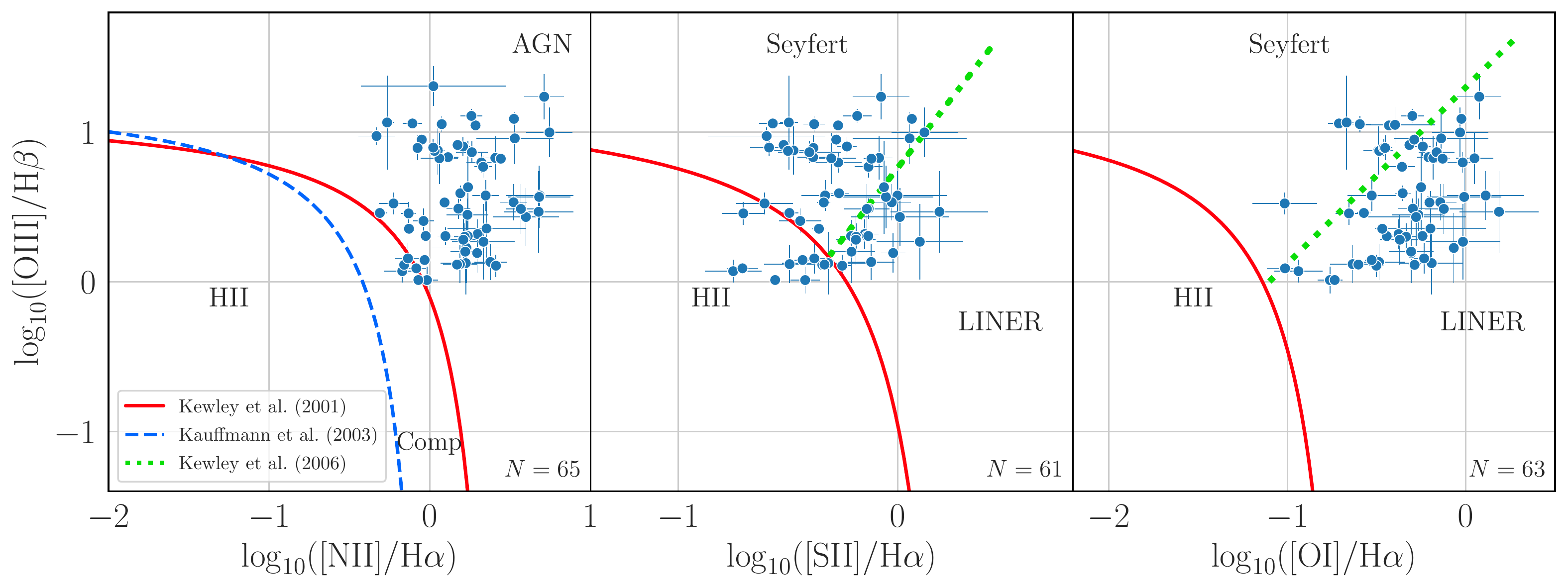}
\caption{BPT diagrams of ICRF3 objects classified as Sy~2 objects.  The theoretical and empirical demarcations between star forming \ion{H}{2} galaxies and AGNs from \citet{2001ApJ...556..121K} and \citet{2003MNRAS.346.1055K} are shown by the solid red and dashed blue lines, respectively. Objects in the region between these two demarcations are often referred to as being of composite AGN/star-forming type.  The demarcations between Seyfert and LINER type galaxies from \citet{2006MNRAS.372..961K} is shown by the green dotted line.}
\label{fig:bpt_diagram}
\end{figure*}

Figure~\ref{fig:bpt_diagram} shows the three BPT diagrams for objects classified as Sy~2 in our sample. The H$\alpha$ BPT diagram (left panel of Figure \ref{fig:bpt_diagram}) classifies the majority of our objects as AGNs, with only a handful of objects in the composite category and none in the star-forming \ion{H}{2} region category.  The [\ion{S}{2}] and [\ion{O}{1}] BPT diagrams (middle and right panels of Figure \ref{fig:bpt_diagram}, respectively) show slightly differing distributions, with a handful of objects classified as \ion{H}{2} or low-ionization nuclear emission-line region (LINER) types for [\ion{S}{2}], and the majority as LINER for [\ion{O}{1}].  However, given the relative strengths of [\ion{S}{2}] and [\ion{O}{1}] compared to H$\alpha$, these BPT classifications are likely less reliable compared to the H$\alpha$ BPT diagram.  Nevertheless, all three BPT diagrams show clear preference toward larger flux ratios inconsistent with \ion{H}{2} regions.  We therefore conclude that the Sy~2 classified objects within our sample are likely obscured AGNs.

Finally, while it is possible that the emission from type 1 objects may not be completely dominated by the emission from the AGN, as in the case of Sy 1 types, we do not attempt to classify Sy 1 or QSOs based on BPT classification, as the presence of broad line emission can result in poor constraints on the fitting of narrow H$\alpha$ and H$\beta$, resulting in large uncertainties in line ratios and/or questionable results.

\subsection{BH Mass \& Accretion Rate}
\begin{figure}[htb!]
\includegraphics[width=\columnwidth]{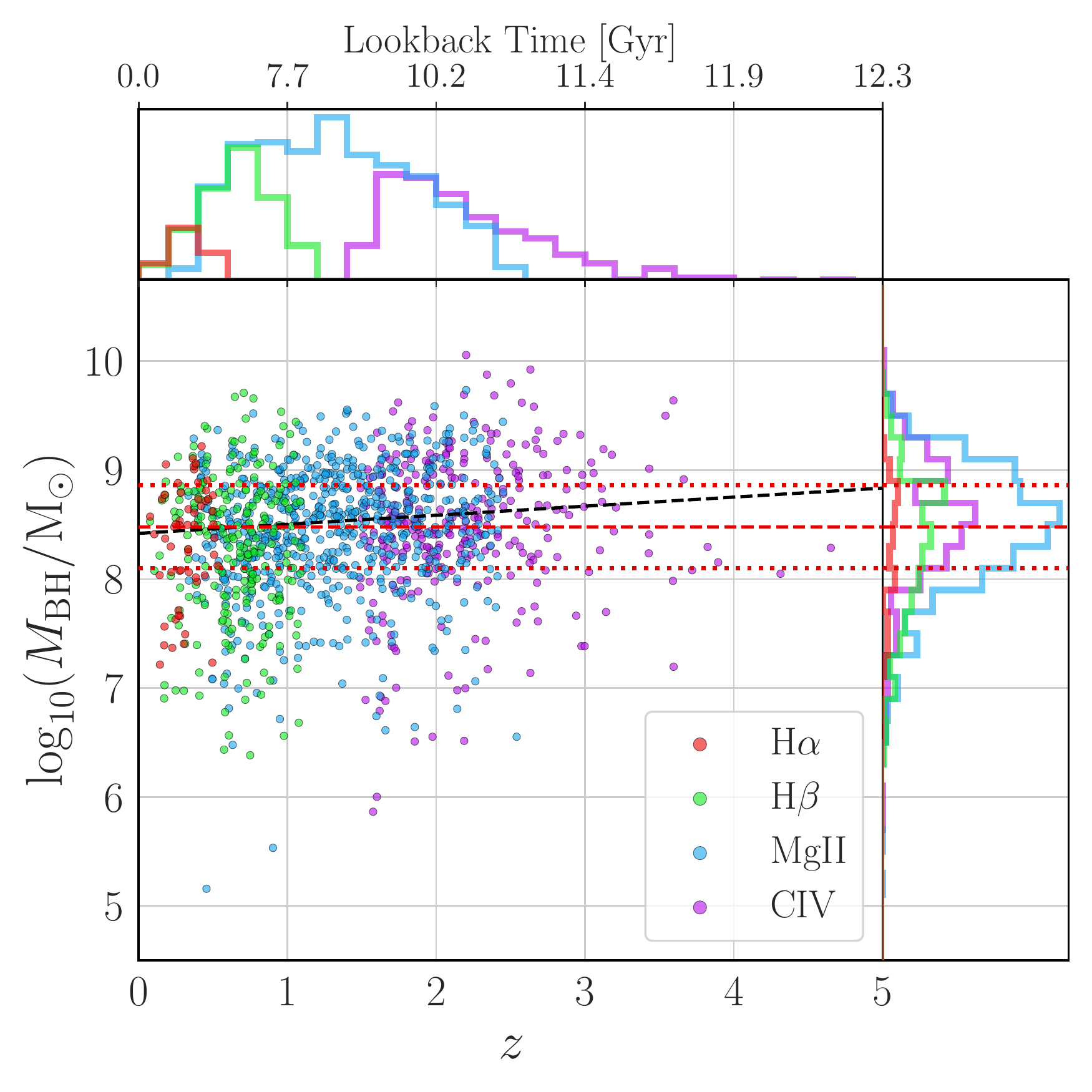}
\caption{The distribution of ICRF3 BH masses over cosmic time.  Marginalized distribution of $M_{\rm{BH}}$ and $z$/lookback time are shown in the right and top histograms, respectively.  The black dashed line represents a linear regression of slope $0.08\pm0.03$.  The red dashed line indicates the median BH mass of $8.48$, and the red dotted lines represent the median absolute deviation of $0.38$ dex.}
\label{fig:mbh_vs_z}
\end{figure}

We plot the distribution of BH mass for all line calibrators as a function of cosmic time in Figure \ref{fig:mbh_vs_z}.  The BH masses as a function of $z$ have a slope of $0.08\pm 0.03$, significant at the $2.7\sigma$ level. The relation is dominated by a total scatter of $\sigma_{\rm{RMS}}=0.68$~dex and and intrinsic scatter of $\sigma_{\rm{intr}}=0.30$~dex.  Despite the fact that our objects are radio selected, the number of spectra from which we can accurately measure BH mass is greatest for those at $z>1$, and we are likely sampling from the highest-luminosity-- and therefore highest-mass --objects in the BH mass distribution at high $z$ \citep{2007ApJ...670..249L,2008ApJ...680..169S}. Notwithstanding any effects of optical selection bias, the distribution of BH masses at all $z$ is generally flat over the last 12.3 Gyr, with a median value of $8.48$ in $\log_{10}(M_{\rm{BH}}/\rm{M}_\odot)$ and median absolute deviation of $0.38$ dex.  The median BH mass range for the ICRF3 sample agrees with other studies that have found that BH masses in radio-loud AGNs are typically more massive than those in radio-quiet AGNs \citep{1998MNRAS.297..817F,1999MNRAS.308..377M,2000MNRAS.317..249M}.  Similar studies of radio-loud quasars have found strong anti-correlations between radio loudness and BH mass, showing that a majority of radio-loud quasars have $M_{\rm{BH}}>10^8~\rm{M}_\odot$ \citep{2000ApJ...543L.111L,2001MNRAS.327.1111G}, while other studies have found that radio-loud quasars have a larger distribution of BH mass \citep{2002ApJ...564..120H}.  Our results appear to agree with the latter claim that BH mass can have a broad distribution across a wide range of redshifts regardless of estimator used, but we have made no attempt to correct for the effect of optical selection bias. A comprehensive study on the differences between radio-loud and radio-quiet objects with larger statistical samples that control for BH mass are needed to disentangle the apparent anti-correlation between BH mass and radio-loudness, but this is beyond the scope of this work.

\begin{figure}[htb!]
\includegraphics[width=\columnwidth]{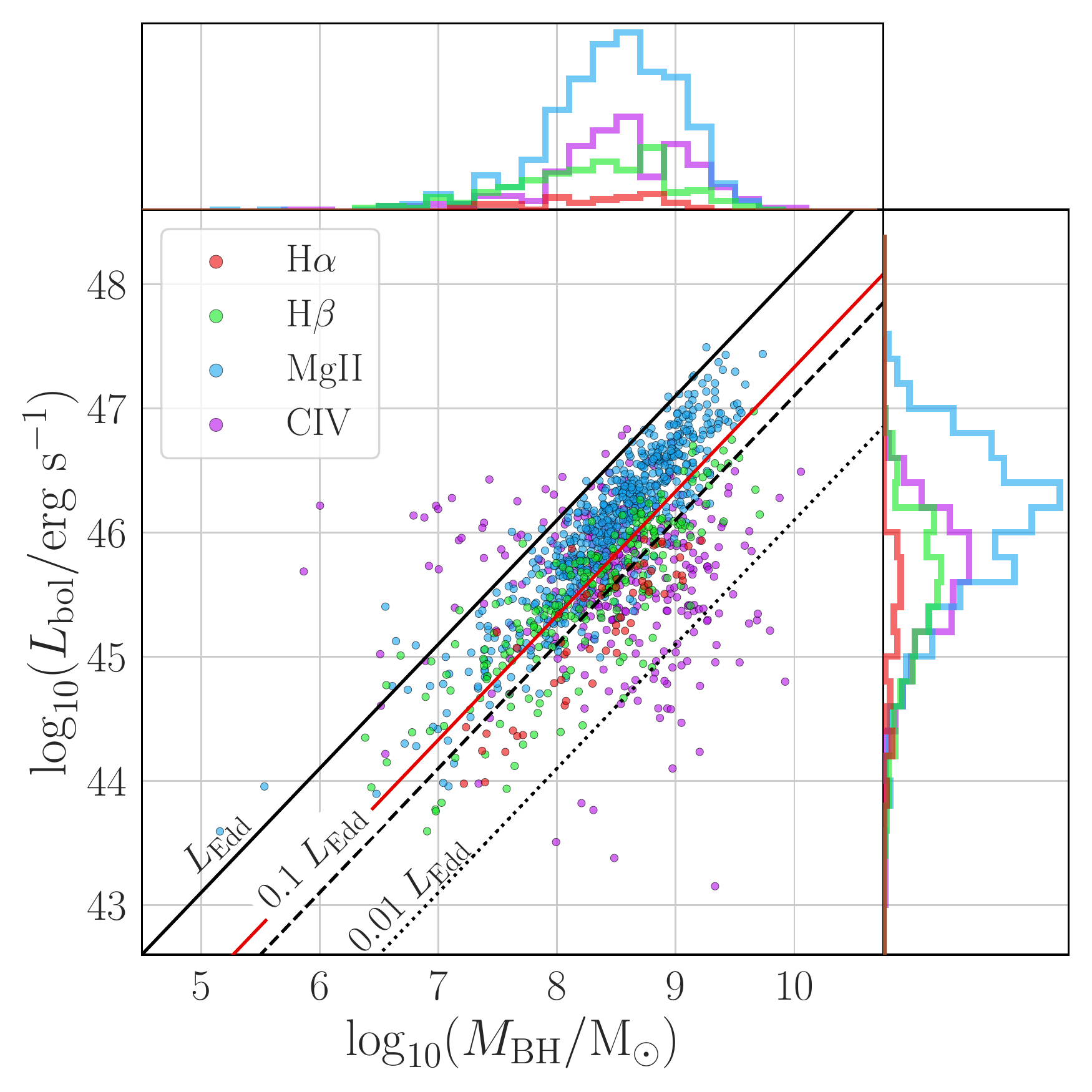}
\caption{Distributions of Eddington ratios by BH mass calibrator.  Marginalized distribution of $L_{\rm{bol}}$ and $M_{\rm{BH}}$ are shown in the right and top histograms, respectively.  The red solid line represents the median accretion rate of $\sim17\%$ for the entire sample.}
\label{fig:edd_ratios}
\end{figure}

\begin{figure*}[htb!]
\includegraphics[width=1.0\textwidth]{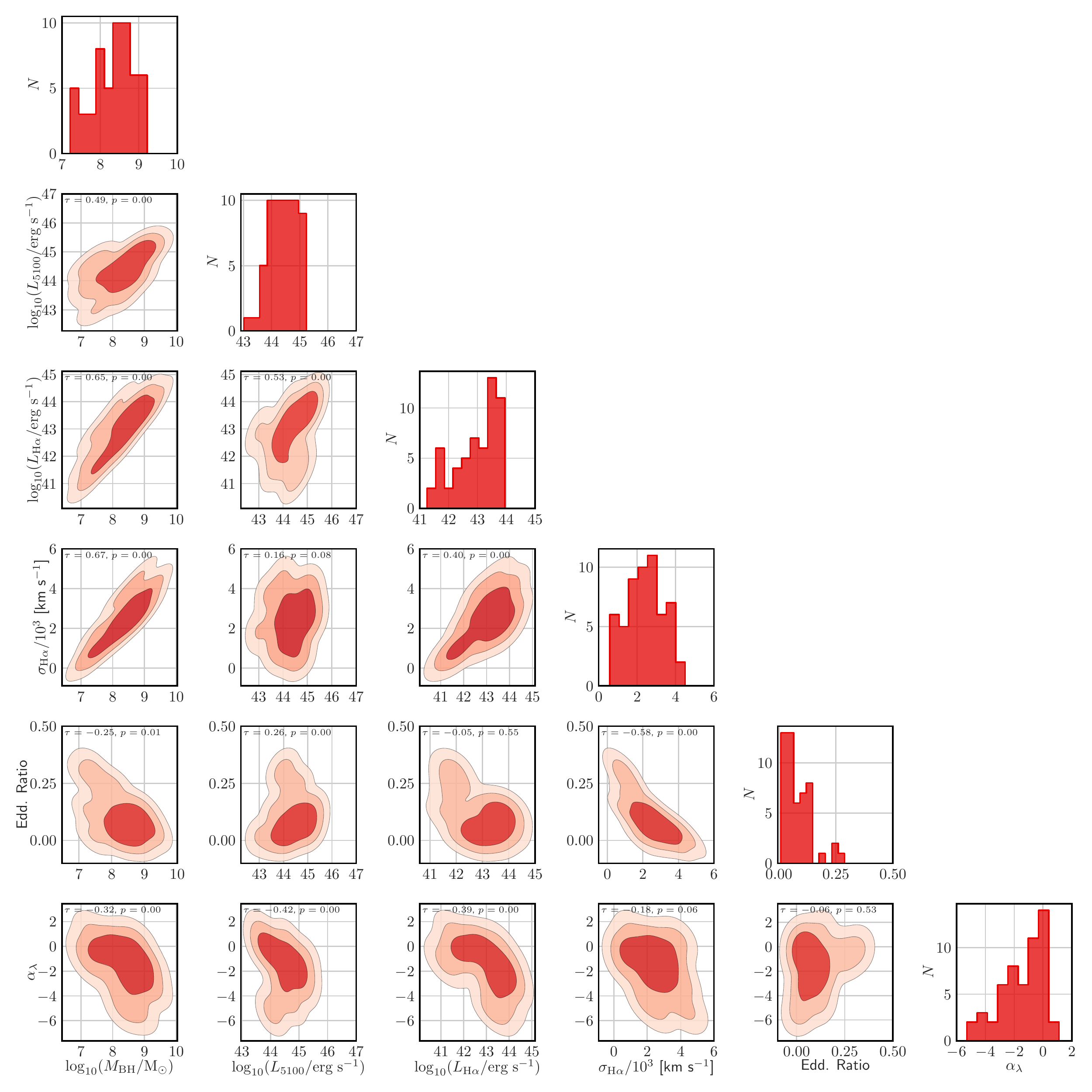}
\caption{Covariance matrix of quantities associated with $M_{\rm{BH}}$ as derived from the H$\alpha$ emission line.}
\label{fig:H_alpha_MBH_covar}
\end{figure*}

\begin{figure*}[htb!]
\includegraphics[width=1.0\textwidth]{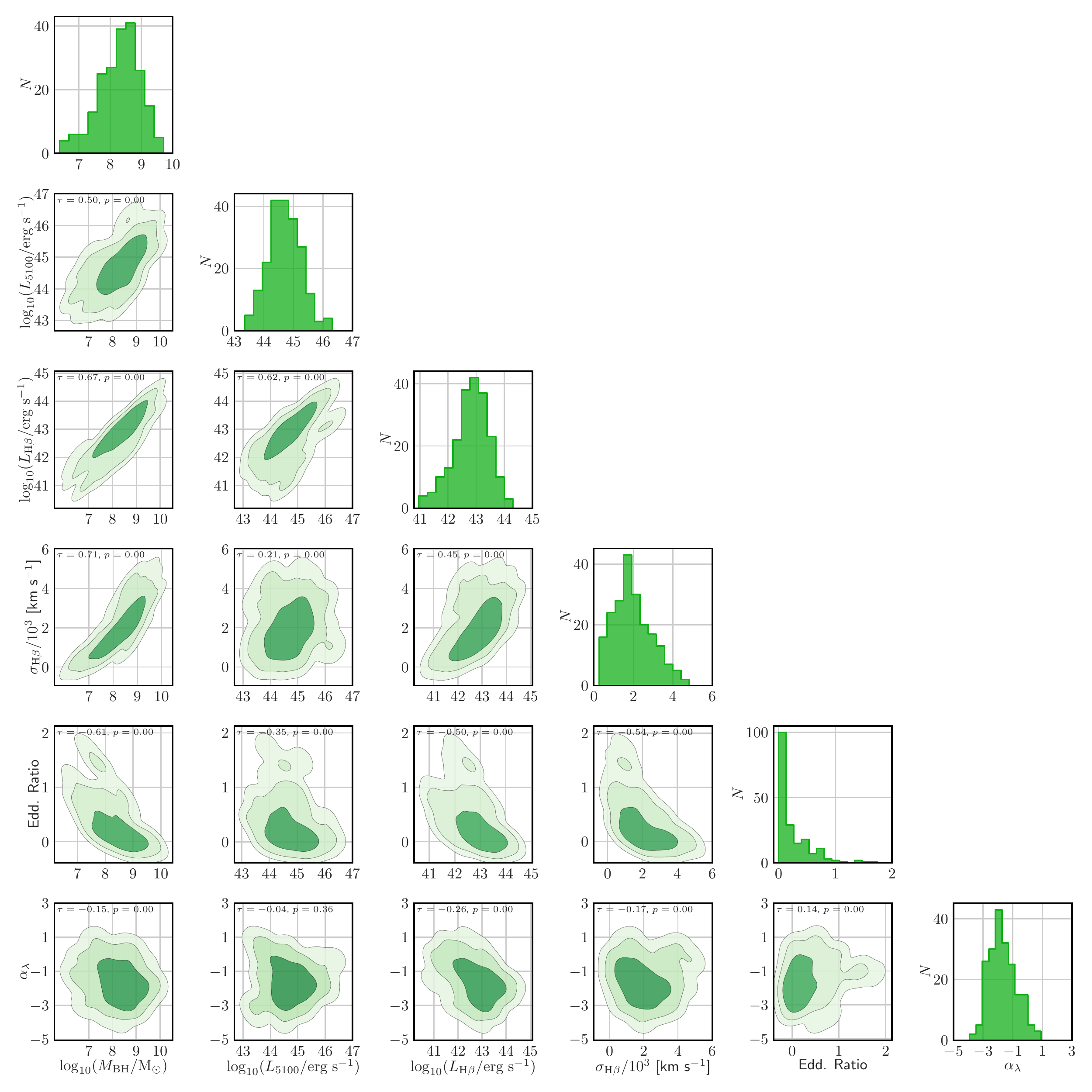}
\caption{Covariance matrix of quantities associated with $M_{\rm{BH}}$ as derived from the H$\beta$ emission line.}
\label{fig:H_beta_MBH_covar}
\end{figure*}

\begin{figure*}[htb!]
\includegraphics[width=1.0\textwidth]{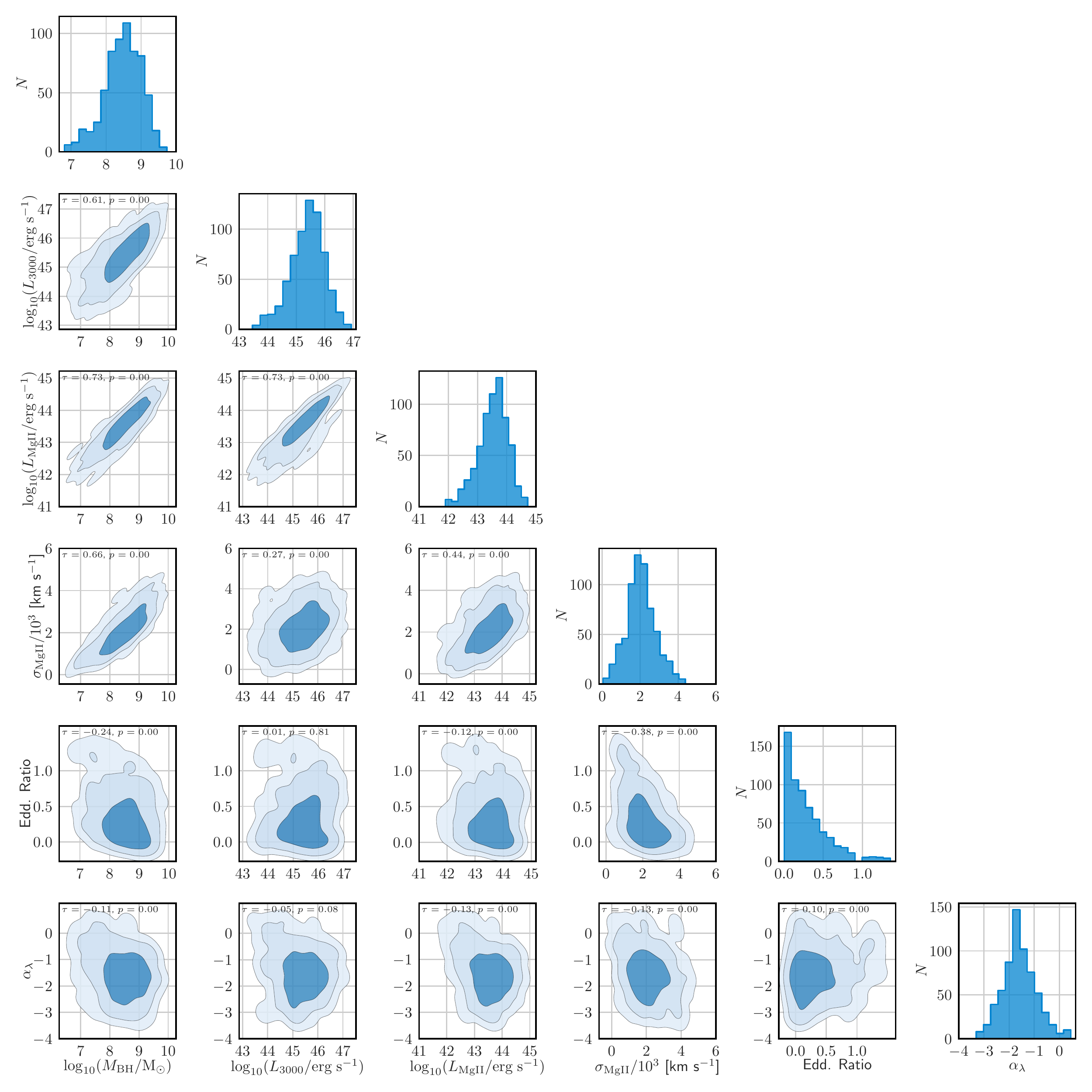}
\caption{Covariance matrix of quantities associated with $M_{\rm{BH}}$ as derived from the \ion{Mg}{2} emission line.}
\label{fig:MgII_MBH_covar}
\end{figure*}

\begin{figure*}[htb!]
\includegraphics[width=1.0\textwidth]{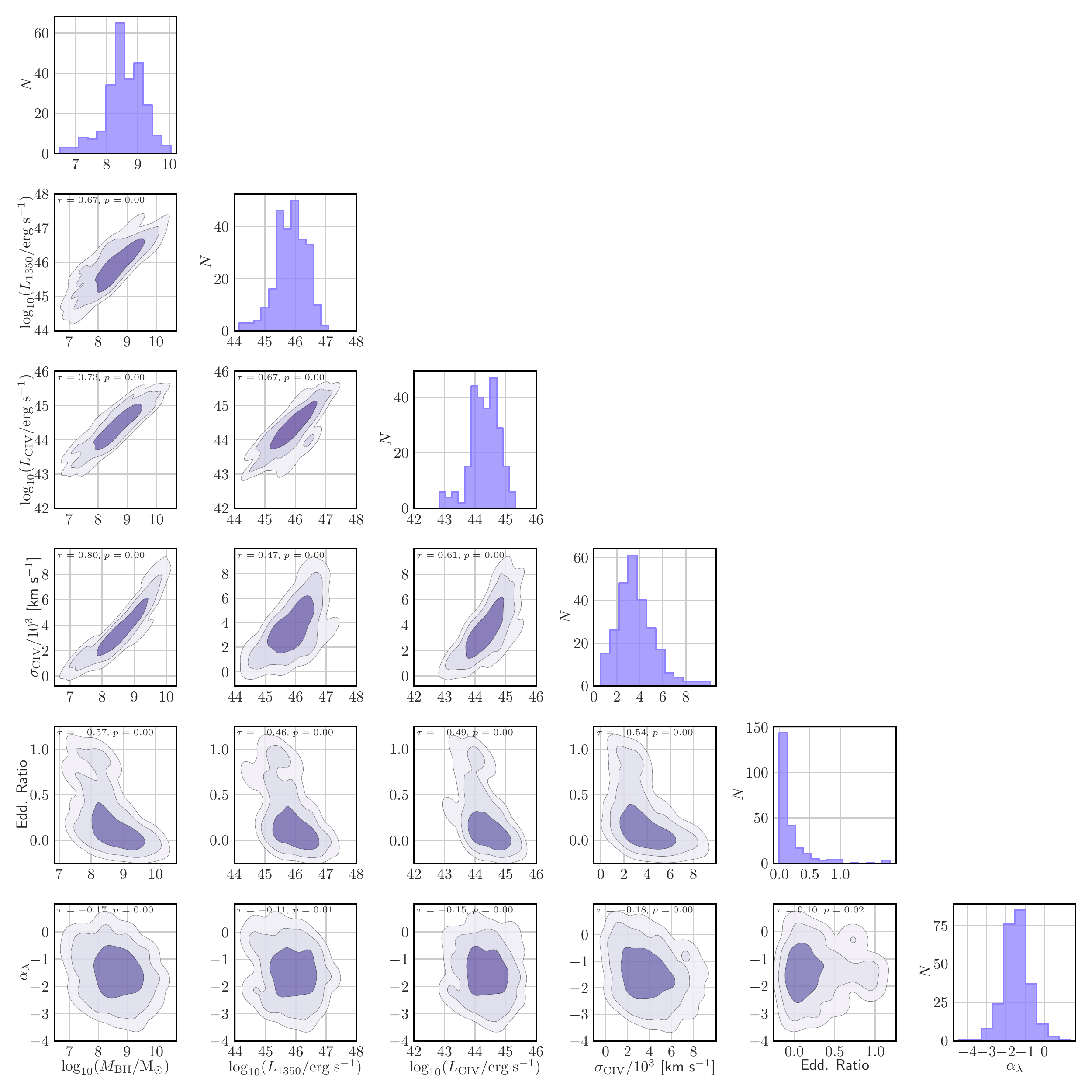}
\caption{Covariance matrix of quantities associated with $M_{\rm{BH}}$ as derived from the \ion{C}{4} emission line.}
\label{fig:CIV_MBH_covar}
\end{figure*}

In Figure~\ref{fig:edd_ratios} we plot bolometric luminosities as a function of BH mass to assess the distribution of Eddington radio for the ICRF3 sample.  Bolometric luminosities are estimated using the relations derived by \citet{2012MNRAS.422..478R,2012MNRAS.427.1800R}, who found that the monochromatic bolometric corrections between radio-loud and radio-quiet objects are consistent with 95\% confidence. Instead of using luminosities measured directly from the spectral continuum, we use emission line luminosities following \citet{2004A&A...424..793W}, who suggested that the presence of strong optical jets in radio-loud quasars can significantly boost apparent AGN luminosity as estimated from the optical continuum.  The luminosity of the broad line emission used to estimate BH mass may be a more direct estimate of the intrinsic AGN accretion luminosity than that measured from the continuum, the latter of which has contributions from many components (e.g, \ion{Fe}{2}, UV iron, Balmer pseudo-continuum, jet emission, and stellar emission).  Indeed, we observe slightly larger scatter in bolometric luminosities when using the optical continuum luminosities, and there is a tighter correlation observed between $M_{\rm{BH}}$ and line luminosities (also see Figures \ref{fig:H_alpha_MBH_covar} - \ref{fig:CIV_MBH_covar}) than with continuum luminosities. We note, however, that estimates of $M_{\rm{BH}}$ have a greater dependence on line width than continuum luminosity, and the line widths and line luminosities are strongly correlated.

We find a similar result to \citet{2013BASI...41...61S}, which is a general stratification of Eddington ratio with line calibrator, likely due to luminosity-dependent selection effects with increased redshift. The H$\alpha$ line calibrator typically provides the lowest Eddington ratios at $\sim0.1~L_{\rm{Edd}}$, and rises toward $L_{\rm{Edd}}$ in sequence with H$\beta$, \ion{Mg}{2}, and \ion{C}{4}.  The median accretion rate for the entire sample is $\sim 17\%$, however with considerable scatter.  The \ion{C}{4} line shows the greatest scatter, likely due to the fitting peculiarities mentioned in Section \ref{sec:blr_emission}.  We note that estimates of bolometric luminosity have considerable uncertainties, and that these first-order monochromatic bolometric corrections should be used with caution.  A complete modelling of the full spectral energy distribution for ICRF3 targets is necessary to acquire more accurate estimates of the bolometric luminosity and accretion rate, which we will perform in a future study.

In Figures \ref{fig:H_alpha_MBH_covar} - \ref{fig:CIV_MBH_covar} we plot the covariance matrices for all BH mass estimators and relevant parameters.  Among the strongest of these correlations with BH mass are continuum luminosity, line luminosity, line width, and Eddington ratio, with Kendall's $\tau\sim0.5$ across all estimators.  Other correlations, such as those with power-law slope $\alpha_\lambda$, are significantly weaker. 

\section{Catalog Structure}
\label{sec:catalog_structure}

We summarize the structure of the catalog.  There exist two version of the USNO VLBI Spectroscopic Catalog: (1) a shorter form that provides all parameters and derived quantities pertaining to BH masses, BPT diagrams, luminosities, Eddington ratios, and lines listed in boldface in Table \ref{tab:emline}, and (2) a longer form that includes all quantities included in the shorter table with additional parameters for all emission lines.  Individual column numbers, names, data types, and descriptions for the short table are given in Table \ref{tab:column_descrip}. 

There are a few caveats of note when using this catalog:

\begin{enumerate}

\item The cosmology adopted for this work ($H_0=70$ km s$^{-1}$ Mpc$^{-1}$, $\Omega_\mathrm{m}=0.3$) is used for reported continuum and emission line luminosities, and hence affects derived values of $M_{\rm{BH}}$, bolometric luminosities, and Eddington ratios.

\item Uncertainties for BH mass include uncertainties from width and luminosity estimators, as well as uncertainties derived from their respective calibrations.  Since all BH mass calibrations rely on reverberation-mapped and dynamical masses derived using H$\beta$, the minimum uncertainty on all reported BH masses is 0.44 dex.

\item Quantities such as bolometric luminosity and Eddington ratio have significant uncertainties that can be attributed to the uncertainties attributed to using monochromatic bolometric corrections as opposed to modeling the full spectral energy distribution, so should be used with caution.

\item Stellar kinematics, while reported for all objects whose ranges coincide with the \ion{Ca}{0}K+H, \ion{Mg}{1}b, and \ion{Ca}{0}T regions, should only be trusted for Sy~2 objects for which there is little to no AGN dilution of the continuum.  These objects can be queried using the \texttt{AGN\_TYPE} column.

\item Caution should be taken when using lines with best-fit parameters close to the hard limits imposed on the fits (Section~\ref{subsec: emission and absorption line modelling}).
\end{enumerate}

\newpage 

\newpage 

\newpage 

\section{Summary}
We have produced the first dedicated catalog of emission line properties of ICRF3 objects, such as emission line fluxes, stellar kinematics, AGN spectral types, and black hole masses. This catalog will aid in building a more complete physical understanding of the radio-loud AGNs that comprise the ICRF, aiding in the further development and refinement of the celestial reference frame. 


Development and refinement of the ICRF is a continuous and indefinite process, and outstanding issues exist with the effort to make the ICRF wavelength-independent, such as the presence of discrepancies between the optical (i.e., Gaia) and radio VLBI source positions. These issues are likely to be in large part astrophysical in nature, motivating efforts to physically characterize the objects that comprise the ICRF. Reliable spectroscopic redshifts, radio loudness, jet orientation angle, and perhaps even the relative accretion rate may prove to be important considerations in source selection and weighting for a multi-wavelength ICRF. This catalog provides an important reference for such work.

In producing this catalog, we have additionally derived black hole virial mass scaling relations using \ion{C}{4}, \ion{Mg}{2}, and H$\beta$ that are self-consistent, which will be useful to studies of radio-loud AGNs that span a large range in redshifts.

This catalog was produced using state-of-the-art Bayesian spectral fitting code that simultaneously fits all spectral parameters and performs posterior sampling using Markov chains. This code is, written in Python, has been updated since the last release \citep{2021MNRAS.500.2871S} to handle moderate-to-high redshift AGNs, and is a powerful tool for the AGN community.\footnote{\url{https://github.com/remingtonsexton/BADASS3}}

Finally, we note that this will be a living catalog, with updates and additions to occur as archival data from other facilities is incorporated and new data becomes available. Where available, data spanning multiple epochs will be analyzed and time domain behavior noted. These updates and additions will be announced in brief publications or research notes.

\acknowledgements

We thank Valeri Makarov and Jeffery Munn for their helpful and insightful comments in the review of this catalog and manuscript. 

This research was funded primarily by USNO to support its ongoing research into reference frame objects.

This research made use of Astropy,\footnote{\url{http://www.astropy.org}} a community-developed core Python package for Astronomy \citep{2013A&A...558A..33A, 2018AJ....156..123A}, as well as \textsc{topcat} \citep{2005ASPC..347...29T}.

Funding for the Sloan Digital Sky Survey IV has been provided by the Alfred P. Sloan Foundation, the U.S. Department of Energy Office of Science, and the Participating Institutions. SDSS acknowledges support and resources from the Center for High-Performance Computing at the University of Utah. The SDSS web site is www.sdss.org.

SDSS is managed by the Astrophysical Research Consortium for the Participating Institutions of the SDSS Collaboration including the Brazilian Participation Group, the Carnegie Institution for Science, Carnegie Mellon University, Center for Astrophysics | Harvard \& Smithsonian (CfA), the Chilean Participation Group, the French Participation Group, Instituto de Astrofísica de Canarias, The Johns Hopkins University, Kavli Institute for the Physics and Mathematics of the Universe (IPMU) / University of Tokyo, the Korean Participation Group, Lawrence Berkeley National Laboratory, Leibniz Institut für Astrophysik Potsdam (AIP), Max-Planck-Institut für Astronomie (MPIA Heidelberg), Max-Planck-Institut für Astrophysik (MPA Garching), Max-Planck-Institut für Extraterrestrische Physik (MPE), National Astronomical Observatories of China, New Mexico State University, New York University, University of Notre Dame, Observatório Nacional / MCTI, The Ohio State University, Pennsylvania State University, Shanghai Astronomical Observatory, United Kingdom Participation Group, Universidad Nacional Autónoma de México, University of Arizona, University of Colorado Boulder, University of Oxford, University of Portsmouth, University of Utah, University of Virginia, University of Washington, University of Wisconsin, Vanderbilt University, and Yale University.

\facilities{VLBA, SDSS}

\software{
Astropy \citep{2013A&A...558A..33A,2018AJ....156..123A}, 
\textsc{topcat} \citep{2005ASPC..347...29T}, \textsc{badass} \citep{2021MNRAS.500.2871S},
\textsc{aips}}

\clearpage
\bibliography{Sexton_icrf_cat}{}

\begin{thebibliography}{}
\expandafter\ifx\csname natexlab\endcsname\relax\def\natexlab#1{#1}\fi
\providecommand{\url}[1]{\href{#1}{#1}}
\providecommand{\dodoi}[1]{doi:~\href{http://doi.org/#1}{\nolinkurl{#1}}}
\providecommand{\doeprint}[1]{\href{http://ascl.net/#1}{\nolinkurl{http://ascl.net/#1}}}
\providecommand{\doarXiv}[1]{\href{https://arxiv.org/abs/#1}{\nolinkurl{https://arxiv.org/abs/#1}}}

\bibitem[{{Arias} {et~al.}(1995){Arias}, {Charlot}, {Feissel}, \&
  {Lestrade}}]{1995A&A...303..604A}
{Arias}, E.~F., {Charlot}, P., {Feissel}, M., \& {Lestrade}, J.~F. 1995, \aap,
  303, 604

\bibitem[{{Assafin} {et~al.}(2013){Assafin}, {Vieira-Martins}, {Andrei},
  {Camargo}, \& {da Silva Neto}}]{2013MNRAS.430.2797A}
{Assafin}, M., {Vieira-Martins}, R., {Andrei}, A.~H., {Camargo}, J.~I.~B., \&
  {da Silva Neto}, D.~N. 2013, \mnras, 430, 2797, \dodoi{10.1093/mnras/stt081}

\bibitem[{{Astropy Collaboration} {et~al.}(2013){Astropy Collaboration},
  {Robitaille}, {Tollerud}, {Greenfield}, {Droettboom}, {Bray}, {Aldcroft},
  {Davis}, {Ginsburg}, {Price-Whelan}, {Kerzendorf}, {Conley}, {Crighton},
  {Barbary}, {Muna}, {Ferguson}, {Grollier}, {Parikh}, {Nair}, {Unther},
  {Deil}, {Woillez}, {Conseil}, {Kramer}, {Turner}, {Singer}, {Fox}, {Weaver},
  {Zabalza}, {Edwards}, {Azalee Bostroem}, {Burke}, {Casey}, {Crawford},
  {Dencheva}, {Ely}, {Jenness}, {Labrie}, {Lim}, {Pierfederici}, {Pontzen},
  {Ptak}, {Refsdal}, {Servillat}, \& {Streicher}}]{2013A&A...558A..33A}
{Astropy Collaboration}, {Robitaille}, T.~P., {Tollerud}, E.~J., {et~al.} 2013,
  \aap, 558, A33, \dodoi{10.1051/0004-6361/201322068}

\bibitem[{{Astropy Collaboration} {et~al.}(2018){Astropy Collaboration},
  {Price-Whelan}, {Sip{\H{o}}cz}, {G{\"u}nther}, {Lim}, {Crawford}, {Conseil},
  {Shupe}, {Craig}, {Dencheva}, {Ginsburg}, {Vand erPlas}, {Bradley},
  {P{\'e}rez-Su{\'a}rez}, {de Val-Borro}, {Aldcroft}, {Cruz}, {Robitaille},
  {Tollerud}, {Ardelean}, {Babej}, {Bach}, {Bachetti}, {Bakanov}, {Bamford},
  {Barentsen}, {Barmby}, {Baumbach}, {Berry}, {Biscani}, {Boquien}, {Bostroem},
  {Bouma}, {Brammer}, {Bray}, {Breytenbach}, {Buddelmeijer}, {Burke},
  {Calderone}, {Cano Rodr{\'\i}guez}, {Cara}, {Cardoso}, {Cheedella}, {Copin},
  {Corrales}, {Crichton}, {D'Avella}, {Deil}, {Depagne}, {Dietrich}, {Donath},
  {Droettboom}, {Earl}, {Erben}, {Fabbro}, {Ferreira}, {Finethy}, {Fox},
  {Garrison}, {Gibbons}, {Goldstein}, {Gommers}, {Greco}, {Greenfield},
  {Groener}, {Grollier}, {Hagen}, {Hirst}, {Homeier}, {Horton}, {Hosseinzadeh},
  {Hu}, {Hunkeler}, {Ivezi{\'c}}, {Jain}, {Jenness}, {Kanarek}, {Kendrew},
  {Kern}, {Kerzendorf}, {Khvalko}, {King}, {Kirkby}, {Kulkarni}, {Kumar},
  {Lee}, {Lenz}, {Littlefair}, {Ma}, {Macleod}, {Mastropietro}, {McCully},
  {Montagnac}, {Morris}, {Mueller}, {Mumford}, {Muna}, {Murphy}, {Nelson},
  {Nguyen}, {Ninan}, {N{\"o}the}, {Ogaz}, {Oh}, {Parejko}, {Parley}, {Pascual},
  {Patil}, {Patil}, {Plunkett}, {Prochaska}, {Rastogi}, {Reddy Janga},
  {Sabater}, {Sakurikar}, {Seifert}, {Sherbert}, {Sherwood-Taylor}, {Shih},
  {Sick}, {Silbiger}, {Singanamalla}, {Singer}, {Sladen}, {Sooley},
  {Sornarajah}, {Streicher}, {Teuben}, {Thomas}, {Tremblay}, {Turner},
  {Terr{\'o}n}, {van Kerkwijk}, {de la Vega}, {Watkins}, {Weaver}, {Whitmore},
  {Woillez}, {Zabalza}, \& {Astropy Contributors}}]{2018AJ....156..123A}
{Astropy Collaboration}, {Price-Whelan}, A.~M., {Sip{\H{o}}cz}, B.~M., {et~al.}
  2018, \aj, 156, 123, \dodoi{10.3847/1538-3881/aabc4f}

\bibitem[{{Baldwin} {et~al.}(1981){Baldwin}, {Phillips}, \&
  {Terlevich}}]{1981PASP...93....5B}
{Baldwin}, J.~A., {Phillips}, M.~M., \& {Terlevich}, R. 1981, \pasp, 93, 5,
  \dodoi{10.1086/130766}

\bibitem[{{Bentz} \& {Manne-Nicholas}(2018)}]{2018ApJ...864..146B}
{Bentz}, M.~C., \& {Manne-Nicholas}, E. 2018, \apj, 864, 146,
  \dodoi{10.3847/1538-4357/aad808}

\bibitem[{{Bentz} {et~al.}(2013){Bentz}, {Denney}, {Grier}, {Barth},
  {Peterson}, {Vestergaard}, {Bennert}, {Canalizo}, {De Rosa}, {Filippenko},
  {Gates}, {Greene}, {Li}, {Malkan}, {Pogge}, {Stern}, {Treu}, \&
  {Woo}}]{2013ApJ...767..149B}
{Bentz}, M.~C., {Denney}, K.~D., {Grier}, C.~J., {et~al.} 2013, \apj, 767, 149,
  \dodoi{10.1088/0004-637X/767/2/149}

\bibitem[{{Berton} {et~al.}(2020){Berton}, {J{\"a}rvel{\"a}}, {Crepaldi},
  {L{\"a}hteenm{\"a}ki}, {Tornikoski}, {Congiu}, {Kharb}, {Terreran}, \&
  {Vietri}}]{2020A&A...636A..64B}
{Berton}, M., {J{\"a}rvel{\"a}}, E., {Crepaldi}, L., {et~al.} 2020, \aap, 636,
  A64, \dodoi{10.1051/0004-6361/202037793}

\bibitem[{{Blandford} {et~al.}(2019){Blandford}, {Meier}, \&
  {Readhead}}]{2019ARA&A..57..467B}
{Blandford}, R., {Meier}, D., \& {Readhead}, A. 2019, \araa, 57, 467,
  \dodoi{10.1146/annurev-astro-081817-051948}

\bibitem[{{Blanton} {et~al.}(2017){Blanton}, {Bershady}, {Abolfathi},
  {Albareti}, {Allende Prieto}, {Almeida}, {Alonso-Garc{\'\i}a}, {Anders},
  {Anderson}, {Andrews}, {Aquino-Ort{\'\i}z}, {Arag{\'o}n-Salamanca},
  {Argudo-Fern{\'a}ndez}, {Armengaud}, {Aubourg}, {Avila-Reese}, {Badenes},
  {Bailey}, {Barger}, {Barrera-Ballesteros}, {Bartosz}, {Bates}, {Baumgarten},
  {Bautista}, {Beaton}, {Beers}, {Belfiore}, {Bender}, {Berlind}, {Bernardi},
  {Beutler}, {Bird}, {Bizyaev}, {Blanc}, {Blomqvist}, {Bolton}, {Boquien},
  {Borissova}, {van den Bosch}, {Bovy}, {Brandt}, {Brinkmann}, {Brownstein},
  {Bundy}, {Burgasser}, {Burtin}, {Busca}, {Cappellari}, {Delgado Carigi},
  {Carlberg}, {Carnero Rosell}, {Carrera}, {Chanover}, {Cherinka}, {Cheung},
  {G{\'o}mez Maqueo Chew}, {Chiappini}, {Choi}, {Chojnowski}, {Chuang},
  {Chung}, {Cirolini}, {Clerc}, {Cohen}, {Comparat}, {da Costa}, {Cousinou},
  {Covey}, {Crane}, {Croft}, {Cruz-Gonzalez}, {Garrido Cuadra}, {Cunha},
  {Damke}, {Darling}, {Davies}, {Dawson}, {de la Macorra}, {Dell'Agli}, {De
  Lee}, {Delubac}, {Di Mille}, {Diamond-Stanic}, {Cano-D{\'\i}az}, {Donor},
  {Downes}, {Drory}, {du Mas des Bourboux}, {Duckworth}, {Dwelly}, {Dyer},
  {Ebelke}, {Eigenbrot}, {Eisenstein}, {Emsellem}, {Eracleous}, {Escoffier},
  {Evans}, {Fan}, {Fern{\'a}ndez-Alvar}, {Fernandez-Trincado}, {Feuillet},
  {Finoguenov}, {Fleming}, {Font-Ribera}, {Fredrickson}, {Freischlad},
  {Frinchaboy}, {Fuentes}, {Galbany}, {Garcia-Dias},
  {Garc{\'\i}a-Hern{\'a}ndez}, {Gaulme}, {Geisler}, {Gelfand},
  {Gil-Mar{\'\i}n}, {Gillespie}, {Goddard}, {Gonzalez-Perez}, {Grabowski},
  {Green}, {Grier}, {Gunn}, {Guo}, {Guy}, {Hagen}, {Hahn}, {Hall}, {Harding},
  {Hasselquist}, {Hawley}, {Hearty}, {Gonzalez Hern{\'a}ndez}, {Ho}, {Hogg},
  {Holley-Bockelmann}, {Holtzman}, {Holzer}, {Huehnerhoff}, {Hutchinson},
  {Hwang}, {Ibarra-Medel}, {da Silva Ilha}, {Ivans}, {Ivory}, {Jackson},
  {Jensen}, {Johnson}, {Jones}, {J{\"o}nsson}, {Jullo}, {Kamble}, {Kinemuchi},
  {Kirkby}, {Kitaura}, {Klaene}, {Knapp}, {Kneib}, {Kollmeier}, {Lacerna},
  {Lane}, {Lang}, {Law}, {Lazarz}, {Lee}, {Le Goff}, {Liang}, {Li}, {Li},
  {Lian}, {Lima}, {Lin}, {Lin}, {Bertran de Lis}, {Liu}, {de Icaza Lizaola},
  {Long}, {Lucatello}, {Lundgren}, {MacDonald}, {Deconto Machado}, {MacLeod},
  {Mahadevan}, {Geimba Maia}, {Maiolino}, {Majewski}, {Malanushenko},
  {Malanushenko}, {Manchado}, {Mao}, {Maraston}, {Marques-Chaves}, {Masseron},
  {Masters}, {McBride}, {McDermid}, {McGrath}, {McGreer}, {Medina Pe{\~n}a},
  {Melendez}, {Merloni}, {Merrifield}, {Meszaros}, {Meza}, {Minchev},
  {Minniti}, {Miyaji}, {More}, {Mulchaey}, {M{\"u}ller-S{\'a}nchez}, {Muna},
  {Munoz}, {Myers}, {Nair}, {Nandra}, {Correa do Nascimento}, {Negrete},
  {Ness}, {Newman}, {Nichol}, {Nidever}, {Nitschelm}, {Ntelis}, {O'Connell},
  {Oelkers}, {Oravetz}, {Oravetz}, {Pace}, {Padilla}, {Palanque-Delabrouille},
  {Alonso Palicio}, {Pan}, {Parejko}, {Parikh}, {P{\^a}ris}, {Park}, {Patten},
  {Peirani}, {Pellejero-Ibanez}, {Penny}, {Percival}, {Perez-Fournon},
  {Petitjean}, {Pieri}, {Pinsonneault}, {Pisani}, {Poleski}, {Prada},
  {Prakash}, {Queiroz}, {Raddick}, {Raichoor}, {Barboza Rembold}, {Richstein},
  {Riffel}, {Riffel}, {Rix}, {Robin}, {Rockosi}, {Rodr{\'\i}guez-Torres},
  {Roman-Lopes}, {Rom{\'a}n-Z{\'u}{\~n}iga}, {Rosado}, {Ross}, {Rossi}, {Ruan},
  {Ruggeri}, {Rykoff}, {Salazar-Albornoz}, {Salvato}, {S{\'a}nchez}, {Aguado},
  {S{\'a}nchez-Gallego}, {Santana}, {Santiago}, {Sayres}, {Schiavon}, {da Silva
  Schimoia}, {Schlafly}, {Schlegel}, {Schneider}, {Schultheis}, {Schuster},
  {Schwope}, {Seo}, {Shao}, {Shen}, {Shetrone}, {Shull}, {Simon}, {Skinner},
  {Skrutskie}, {Slosar}, {Smith}, {Sobeck}, {Sobreira}, {Somers}, {Souto},
  {Stark}, {Stassun}, {Stauffer}, {Steinmetz}, {Storchi-Bergmann},
  {Streblyanska}, {Stringfellow}, {Su{\'a}rez}, {Sun}, {Suzuki}, {Szigeti},
  {Taghizadeh-Popp}, {Tang}, {Tao}, {Tayar}, {Tembe}, {Teske}, {Thakar},
  {Thomas}, {Thompson}, {Tinker}, {Tissera}, {Tojeiro}, {Hernandez Toledo}, {de
  la Torre}, {Tremonti}, {Troup}, {Valenzuela}, {Martinez Valpuesta},
  {Vargas-Gonz{\'a}lez}, {Vargas-Maga{\~n}a}, {Vazquez}, {Villanova}, {Vivek},
  {Vogt}, {Wake}, {Walterbos}, {Wang}, {Weaver}, {Weijmans}, {Weinberg},
  {Westfall}, {Whelan}, {Wild}, {Wilson}, {Wood-Vasey}, {Wylezalek}, {Xiao},
  {Yan}, {Yang}, {Ybarra}, {Y{\`e}che}, {Zakamska}, {Zamora}, {Zarrouk},
  {Zasowski}, {Zhang}, {Zhao}, {Zheng}, {Zheng}, {Zhou}, {Zhou}, {Zhu},
  {Zoccali}, \& {Zou}}]{2017AJ....154...28B}
{Blanton}, M.~R., {Bershady}, M.~A., {Abolfathi}, B., {et~al.} 2017, \aj, 154,
  28, \dodoi{10.3847/1538-3881/aa7567}

\bibitem[{{Boroson} \& {Green}(1992)}]{1992ApJS...80..109B}
{Boroson}, T.~A., \& {Green}, R.~F. 1992, \apjs, 80, 109,
  \dodoi{10.1086/191661}

\bibitem[{{Brotherton}(1996)}]{1996ApJS..102....1B}
{Brotherton}, M.~S. 1996, \apjs, 102, 1, \dodoi{10.1086/192249}

\bibitem[{{Calderone} {et~al.}(2017){Calderone}, {Nicastro}, {Ghisellini},
  {Dotti}, {Sbarrato}, {Shankar}, \& {Colpi}}]{2017MNRAS.472.4051C}
{Calderone}, G., {Nicastro}, L., {Ghisellini}, G., {et~al.} 2017, \mnras, 472,
  4051, \dodoi{10.1093/mnras/stx2239}

\bibitem[{{Canalizo} {et~al.}(2012){Canalizo}, {Wold}, {Hiner}, {Lazarova},
  {Lacy}, \& {Aylor}}]{2012ApJ...760...38C}
{Canalizo}, G., {Wold}, M., {Hiner}, K.~D., {et~al.} 2012, \apj, 760, 38,
  \dodoi{10.1088/0004-637X/760/1/38}

\bibitem[{{Cappellari}(2017)}]{2017MNRAS.466..798C}
{Cappellari}, M. 2017, \mnras, 466, 798, \dodoi{10.1093/mnras/stw3020}

\bibitem[{{Cardelli} {et~al.}(1989){Cardelli}, {Clayton}, \&
  {Mathis}}]{1989ApJ...345..245C}
{Cardelli}, J.~A., {Clayton}, G.~C., \& {Mathis}, J.~S. 1989, \apj, 345, 245,
  \dodoi{10.1086/167900}

\bibitem[{{Charlot} {et~al.}(2020){Charlot}, {Jacobs}, {Gordon}, {Lambert}, {de
  Witt}, {B{\"o}hm}, {Fey}, {Heinkelmann}, {Skurikhina}, {Titov}, {Arias},
  {Bolotin}, {Bourda}, {Ma}, {Malkin}, {Nothnagel}, {Mayer}, {MacMillan},
  {Nilsson}, \& {Gaume}}]{2020A&A...644A.159C}
{Charlot}, P., {Jacobs}, C.~S., {Gordon}, D., {et~al.} 2020, \aap, 644, A159,
  \dodoi{10.1051/0004-6361/202038368}

\bibitem[{{da Silva Neto} {et~al.}(2002){da Silva Neto}, {Andrei}, {Vieira
  Martins}, \& {Assafin}}]{2002AJ....124..612D}
{da Silva Neto}, D.~N., {Andrei}, A.~H., {Vieira Martins}, R., \& {Assafin}, M.
  2002, \aj, 124, 612, \dodoi{10.1086/341163}

\bibitem[{{Dawson} {et~al.}(2013){Dawson}, {Schlegel}, {Ahn}, {Anderson},
  {Aubourg}, {Bailey}, {Barkhouser}, {Bautista}, {Beifiori}, {Berlind},
  {Bhardwaj}, {Bizyaev}, {Blake}, {Blanton}, {Blomqvist}, {Bolton}, {Borde},
  {Bovy}, {Brandt}, {Brewington}, {Brinkmann}, {Brown}, {Brownstein}, {Bundy},
  {Busca}, {Carithers}, {Carnero}, {Carr}, {Chen}, {Comparat}, {Connolly},
  {Cope}, {Croft}, {Cuesta}, {da Costa}, {Davenport}, {Delubac}, {de Putter},
  {Dhital}, {Ealet}, {Ebelke}, {Eisenstein}, {Escoffier}, {Fan}, {Filiz Ak},
  {Finley}, {Font-Ribera}, {G{\'e}nova-Santos}, {Gunn}, {Guo}, {Haggard},
  {Hall}, {Hamilton}, {Harris}, {Harris}, {Ho}, {Hogg}, {Holder}, {Honscheid},
  {Huehnerhoff}, {Jordan}, {Jordan}, {Kauffmann}, {Kazin}, {Kirkby}, {Klaene},
  {Kneib}, {Le Goff}, {Lee}, {Long}, {Loomis}, {Lundgren}, {Lupton}, {Maia},
  {Makler}, {Malanushenko}, {Malanushenko}, {Mandelbaum}, {Manera}, {Maraston},
  {Margala}, {Masters}, {McBride}, {McDonald}, {McGreer}, {McMahon}, {Mena},
  {Miralda-Escud{\'e}}, {Montero-Dorta}, {Montesano}, {Muna}, {Myers},
  {Naugle}, {Nichol}, {Noterdaeme}, {Nuza}, {Olmstead}, {Oravetz}, {Oravetz},
  {Owen}, {Padmanabhan}, {Palanque-Delabrouille}, {Pan}, {Parejko},
  {P{\^a}ris}, {Percival}, {P{\'e}rez-Fournon}, {P{\'e}rez-R{\`a}fols},
  {Petitjean}, {Pfaffenberger}, {Pforr}, {Pieri}, {Prada}, {Price-Whelan},
  {Raddick}, {Rebolo}, {Rich}, {Richards}, {Rockosi}, {Roe}, {Ross}, {Ross},
  {Rossi}, {Rubi{\~n}o-Martin}, {Samushia}, {S{\'a}nchez}, {Sayres}, {Schmidt},
  {Schneider}, {Sc{\'o}ccola}, {Seo}, {Shelden}, {Sheldon}, {Shen}, {Shu},
  {Slosar}, {Smee}, {Snedden}, {Stauffer}, {Steele}, {Strauss}, {Streblyanska},
  {Suzuki}, {Swanson}, {Tal}, {Tanaka}, {Thomas}, {Tinker}, {Tojeiro},
  {Tremonti}, {Vargas Maga{\~n}a}, {Verde}, {Viel}, {Wake}, {Watson}, {Weaver},
  {Weinberg}, {Weiner}, {West}, {White}, {Wood-Vasey}, {Yeche}, {Zehavi},
  {Zhao}, \& {Zheng}}]{2013AJ....145...10D}
{Dawson}, K.~S., {Schlegel}, D.~J., {Ahn}, C.~P., {et~al.} 2013, \aj, 145, 10,
  \dodoi{10.1088/0004-6256/145/1/10}

\bibitem[{{Dawson} {et~al.}(2016){Dawson}, {Kneib}, {Percival}, {Alam},
  {Albareti}, {Anderson}, {Armengaud}, {Aubourg}, {Bailey}, {Bautista},
  {Berlind}, {Bershady}, {Beutler}, {Bizyaev}, {Blanton}, {Blomqvist},
  {Bolton}, {Bovy}, {Brandt}, {Brinkmann}, {Brownstein}, {Burtin}, {Busca},
  {Cai}, {Chuang}, {Clerc}, {Comparat}, {Cope}, {Croft}, {Cruz-Gonzalez}, {da
  Costa}, {Cousinou}, {Darling}, {de la Macorra}, {de la Torre}, {Delubac}, {du
  Mas des Bourboux}, {Dwelly}, {Ealet}, {Eisenstein}, {Eracleous}, {Escoffier},
  {Fan}, {Finoguenov}, {Font-Ribera}, {Frinchaboy}, {Gaulme}, {Georgakakis},
  {Green}, {Guo}, {Guy}, {Ho}, {Holder}, {Huehnerhoff}, {Hutchinson}, {Jing},
  {Jullo}, {Kamble}, {Kinemuchi}, {Kirkby}, {Kitaura}, {Klaene}, {Laher},
  {Lang}, {Laurent}, {Le Goff}, {Li}, {Liang}, {Lima}, {Lin}, {Lin}, {Lin},
  {Long}, {Lundgren}, {MacDonald}, {Geimba Maia}, {Malanushenko},
  {Malanushenko}, {Mariappan}, {McBride}, {McGreer}, {M{\'e}nard}, {Merloni},
  {Meza}, {Montero-Dorta}, {Muna}, {Myers}, {Nandra}, {Naugle}, {Newman},
  {Noterdaeme}, {Nugent}, {Ogando}, {Olmstead}, {Oravetz}, {Oravetz},
  {Padmanabhan}, {Palanque-Delabrouille}, {Pan}, {Parejko}, {P{\^a}ris},
  {Peacock}, {Petitjean}, {Pieri}, {Pisani}, {Prada}, {Prakash}, {Raichoor},
  {Reid}, {Rich}, {Ridl}, {Rodriguez-Torres}, {Carnero Rosell}, {Ross},
  {Rossi}, {Ruan}, {Salvato}, {Sayres}, {Schneider}, {Schlegel}, {Seljak},
  {Seo}, {Sesar}, {Shandera}, {Shu}, {Slosar}, {Sobreira}, {Streblyanska},
  {Suzuki}, {Taylor}, {Tao}, {Tinker}, {Tojeiro}, {Vargas-Maga{\~n}a}, {Wang},
  {Weaver}, {Weinberg}, {White}, {Wood-Vasey}, {Yeche}, {Zhai}, {Zhao}, {Zhao},
  {Zheng}, {Ben Zhu}, \& {Zou}}]{2016AJ....151...44D}
{Dawson}, K.~S., {Kneib}, J.-P., {Percival}, W.~J., {et~al.} 2016, \aj, 151,
  44, \dodoi{10.3847/0004-6256/151/2/44}

\bibitem[{{Dietrich} {et~al.}(2002){Dietrich}, {Appenzeller}, {Vestergaard}, \&
  {Wagner}}]{2002ApJ...564..581D}
{Dietrich}, M., {Appenzeller}, I., {Vestergaard}, M., \& {Wagner}, S.~J. 2002,
  \apj, 564, 581, \dodoi{10.1086/324337}

\bibitem[{{Dorland} {et~al.}(2020){Dorland}, {Secrest}, {Johnson}, {Fischer},
  {Zacharias}, {Souchay}, {Lambert}, {Barache}, \&
  {Taris}}]{2020jsrs.conf..165D}
{Dorland}, B., {Secrest}, N., {Johnson}, M., {et~al.} 2020, in Astrometry,
  Earth Rotation, and Reference Systems in the GAIA era, ed. C.~{Bizouard},
  165--171.
\newblock \doarXiv{2009.02169}

\bibitem[{{Eisenstein} {et~al.}(2011){Eisenstein}, {Weinberg}, {Agol},
  {Aihara}, {Allende Prieto}, {Anderson}, {Arns}, {Aubourg}, {Bailey},
  {Balbinot}, {Barkhouser}, {Beers}, {Berlind}, {Bickerton}, {Bizyaev},
  {Blanton}, {Bochanski}, {Bolton}, {Bosman}, {Bovy}, {Brandt}, {Breslauer},
  {Brewington}, {Brinkmann}, {Brown}, {Brownstein}, {Burger}, {Busca},
  {Campbell}, {Cargile}, {Carithers}, {Carlberg}, {Carr}, {Chang}, {Chen},
  {Chiappini}, {Comparat}, {Connolly}, {Cortes}, {Croft}, {Cunha}, {da Costa},
  {Davenport}, {Dawson}, {De Lee}, {Porto de Mello}, {de Simoni}, {Dean},
  {Dhital}, {Ealet}, {Ebelke}, {Edmondson}, {Eiting}, {Escoffier}, {Esposito},
  {Evans}, {Fan}, {Femen{\'\i}a Castell{\'a}}, {Dutra Ferreira}, {Fitzgerald},
  {Fleming}, {Font-Ribera}, {Ford}, {Frinchaboy}, {Garc{\'\i}a P{\'e}rez},
  {Gaudi}, {Ge}, {Ghezzi}, {Gillespie}, {Gilmore}, {Girardi}, {Gott}, {Gould},
  {Grebel}, {Gunn}, {Hamilton}, {Harding}, {Harris}, {Hawley}, {Hearty},
  {Hennawi}, {Gonz{\'a}lez Hern{\'a}ndez}, {Ho}, {Hogg}, {Holtzman},
  {Honscheid}, {Inada}, {Ivans}, {Jiang}, {Jiang}, {Johnson}, {Jordan},
  {Jordan}, {Kauffmann}, {Kazin}, {Kirkby}, {Klaene}, {Knapp}, {Kneib},
  {Kochanek}, {Koesterke}, {Kollmeier}, {Kron}, {Lampeitl}, {Lang}, {Lawler},
  {Le Goff}, {Lee}, {Lee}, {Leisenring}, {Lin}, {Liu}, {Long}, {Loomis},
  {Lucatello}, {Lundgren}, {Lupton}, {Ma}, {Ma}, {MacDonald}, {Mack},
  {Mahadevan}, {Maia}, {Majewski}, {Makler}, {Malanushenko}, {Malanushenko},
  {Mandelbaum}, {Maraston}, {Margala}, {Maseman}, {Masters}, {McBride},
  {McDonald}, {McGreer}, {McMahon}, {Mena Requejo}, {M{\'e}nard},
  {Miralda-Escud{\'e}}, {Morrison}, {Mullally}, {Muna}, {Murayama}, {Myers},
  {Naugle}, {Neto}, {Nguyen}, {Nichol}, {Nidever}, {O'Connell}, {Ogando},
  {Olmstead}, {Oravetz}, {Padmanabhan}, {Paegert}, {Palanque-Delabrouille},
  {Pan}, {Pandey}, {Parejko}, {P{\^a}ris}, {Pellegrini}, {Pepper}, {Percival},
  {Petitjean}, {Pfaffenberger}, {Pforr}, {Phleps}, {Pichon}, {Pieri}, {Prada},
  {Price-Whelan}, {Raddick}, {Ramos}, {Reid}, {Reyle}, {Rich}, {Richards},
  {Rieke}, {Rieke}, {Rix}, {Robin}, {Rocha-Pinto}, {Rockosi}, {Roe},
  {Rollinde}, {Ross}, {Ross}, {Rossetto}, {S{\'a}nchez}, {Santiago}, {Sayres},
  {Schiavon}, {Schlegel}, {Schlesinger}, {Schmidt}, {Schneider}, {Sellgren},
  {Shelden}, {Sheldon}, {Shetrone}, {Shu}, {Silverman}, {Simmerer}, {Simmons},
  {Sivarani}, {Skrutskie}, {Slosar}, {Smee}, {Smith}, {Snedden}, {Stassun},
  {Steele}, {Steinmetz}, {Stockett}, {Stollberg}, {Strauss}, {Szalay},
  {Tanaka}, {Thakar}, {Thomas}, {Tinker}, {Tofflemire}, {Tojeiro}, {Tremonti},
  {Vargas Maga{\~n}a}, {Verde}, {Vogt}, {Wake}, {Wan}, {Wang}, {Weaver},
  {White}, {White}, {Wilson}, {Wisniewski}, {Wood-Vasey}, {Yanny}, {Yasuda},
  {Y{\`e}che}, {York}, {Young}, {Zasowski}, {Zehavi}, \&
  {Zhao}}]{2011AJ....142...72E}
{Eisenstein}, D.~J., {Weinberg}, D.~H., {Agol}, E., {et~al.} 2011, \aj, 142,
  72, \dodoi{10.1088/0004-6256/142/3/72}

\bibitem[{{Fernandez} {et~al.}(2022){Fernandez}, {Secrest}, {Johnson},
  {Schmitt}, {Fischer}, {Cigan}, \& {Dorland}}]{2022arXiv220105152F}
{Fernandez}, L.~C., {Secrest}, N.~J., {Johnson}, M.~C., {et~al.} 2022, arXiv
  e-prints, arXiv:2201.05152.
\newblock \doarXiv{2201.05152}

\bibitem[{{Fey} {et~al.}(2015){Fey}, {Gordon}, {Jacobs}, {Ma}, {Gaume},
  {Arias}, {Bianco}, {Boboltz}, {B{\"o}ckmann}, {Bolotin}, {Charlot},
  {Collioud}, {Engelhardt}, {Gipson}, {Gontier}, {Heinkelmann}, {Kurdubov},
  {Lambert}, {Lytvyn}, {MacMillan}, {Malkin}, {Nothnagel}, {Ojha},
  {Skurikhina}, {Sokolova}, {Souchay}, {Sovers}, {Tesmer}, {Titov}, {Wang}, \&
  {Zharov}}]{2015AJ....150...58F}
{Fey}, A.~L., {Gordon}, D., {Jacobs}, C.~S., {et~al.} 2015, \aj, 150, 58,
  \dodoi{10.1088/0004-6256/150/2/58}

\bibitem[{{Fischer} {et~al.}(2021){Fischer}, {Secrest}, {Johnson}, {Dorland},
  {Cigan}, {Fernandez}, {Hunt}, {Koss}, {Schmitt}, \&
  {Zacharias}}]{2021ApJ...906...88F}
{Fischer}, T.~C., {Secrest}, N.~J., {Johnson}, M.~C., {et~al.} 2021, \apj, 906,
  88, \dodoi{10.3847/1538-4357/abca3c}

\bibitem[{{Foreman-Mackey} {et~al.}(2019){Foreman-Mackey}, {Farr}, {Sinha},
  {Archibald}, {Hogg}, {Sanders}, {Zuntz}, {Williams}, {Nelson}, {de
  Val-Borro}, {Erhardt}, {Pashchenko}, \& {Pla}}]{2019JOSS....4.1864F}
{Foreman-Mackey}, D., {Farr}, W., {Sinha}, M., {et~al.} 2019, The Journal of
  Open Source Software, 4, 1864, \dodoi{10.21105/joss.01864}

\bibitem[{{Franceschini} {et~al.}(1998){Franceschini}, {Vercellone}, \&
  {Fabian}}]{1998MNRAS.297..817F}
{Franceschini}, A., {Vercellone}, S., \& {Fabian}, A.~C. 1998, \mnras, 297,
  817, \dodoi{10.1046/j.1365-8711.1998.01534.x}

\bibitem[{{Fricke} {et~al.}(1988){Fricke}, {Schwan}, {Lederle}, {Bastian},
  {Bien}, {Burkhardt}, {Du Mont}, {Hering}, {J{\"a}hrling}, {Jahrei{\ss}},
  {R{\"o}ser}, {Schwerdtfeger}, \& {Walter}}]{1988VeARI..32....1F}
{Fricke}, W., {Schwan}, H., {Lederle}, T., {et~al.} 1988, Veroeffentlichungen
  des Astronomischen Rechen-Instituts Heidelberg, 32, 1

\bibitem[{{Grandi}(1982)}]{1982ApJ...255...25G}
{Grandi}, S.~A. 1982, \apj, 255, 25, \dodoi{10.1086/159799}

\bibitem[{{Greene} \& {Ho}(2005)}]{2005ApJ...630..122G}
{Greene}, J.~E., \& {Ho}, L.~C. 2005, \apj, 630, 122, \dodoi{10.1086/431897}

\bibitem[{{Greene} \& {Ho}(2006)}]{2006ApJ...641..117G}
---. 2006, \apj, 641, 117, \dodoi{10.1086/500353}

\bibitem[{{Gu} {et~al.}(2001){Gu}, {Cao}, \& {Jiang}}]{2001MNRAS.327.1111G}
{Gu}, M., {Cao}, X., \& {Jiang}, D.~R. 2001, \mnras, 327, 1111,
  \dodoi{10.1046/j.1365-8711.2001.04795.x}

\bibitem[{{Heckman}(1980)}]{1980A&A....87..152H}
{Heckman}, T.~M. 1980, \aap, 500, 187

\bibitem[{{Ho}(2002)}]{2002ApJ...564..120H}
{Ho}, L.~C. 2002, \apj, 564, 120, \dodoi{10.1086/324399}

\bibitem[{{Kauffmann} {et~al.}(2003){Kauffmann}, {Heckman}, {Tremonti},
  {Brinchmann}, {Charlot}, {White}, {Ridgway}, {Brinkmann}, {Fukugita}, {Hall},
  {Ivezi{\'c}}, {Richards}, \& {Schneider}}]{2003MNRAS.346.1055K}
{Kauffmann}, G., {Heckman}, T.~M., {Tremonti}, C., {et~al.} 2003, \mnras, 346,
  1055, \dodoi{10.1111/j.1365-2966.2003.07154.x}

\bibitem[{{Kewley} {et~al.}(2001){Kewley}, {Dopita}, {Sutherland}, {Heisler},
  \& {Trevena}}]{2001ApJ...556..121K}
{Kewley}, L.~J., {Dopita}, M.~A., {Sutherland}, R.~S., {Heisler}, C.~A., \&
  {Trevena}, J. 2001, \apj, 556, 121, \dodoi{10.1086/321545}

\bibitem[{{Kewley} {et~al.}(2006){Kewley}, {Groves}, {Kauffmann}, \&
  {Heckman}}]{2006MNRAS.372..961K}
{Kewley}, L.~J., {Groves}, B., {Kauffmann}, G., \& {Heckman}, T. 2006, \mnras,
  372, 961, \dodoi{10.1111/j.1365-2966.2006.10859.x}

\bibitem[{{Kovalev} {et~al.}(2017){Kovalev}, {Petrov}, \&
  {Plavin}}]{2017A&A...598L...1K}
{Kovalev}, Y.~Y., {Petrov}, L., \& {Plavin}, A.~V. 2017, \aap, 598, L1,
  \dodoi{10.1051/0004-6361/201630031}

\bibitem[{{Laor}(2000)}]{2000ApJ...543L.111L}
{Laor}, A. 2000, \apjl, 543, L111, \dodoi{10.1086/317280}

\bibitem[{{Lauer} {et~al.}(2007){Lauer}, {Tremaine}, {Richstone}, \&
  {Faber}}]{2007ApJ...670..249L}
{Lauer}, T.~R., {Tremaine}, S., {Richstone}, D., \& {Faber}, S.~M. 2007, \apj,
  670, 249, \dodoi{10.1086/522083}

\bibitem[{{Le} {et~al.}(2020){Le}, {Woo}, \& {Xue}}]{2020ApJ...901...35L}
{Le}, H. A.~N., {Woo}, J.-H., \& {Xue}, Y. 2020, \apj, 901, 35,
  \dodoi{10.3847/1538-4357/abada0}

\bibitem[{{Leighly}(1999)}]{1999ApJS..125..297L}
{Leighly}, K.~M. 1999, \apjs, 125, 297, \dodoi{10.1086/313277}

\bibitem[{{Ma} {et~al.}(1998){Ma}, {Arias}, {Eubanks}, {Fey}, {Gontier},
  {Jacobs}, {Sovers}, {Archinal}, \& {Charlot}}]{1998AJ....116..516M}
{Ma}, C., {Arias}, E.~F., {Eubanks}, T.~M., {et~al.} 1998, \aj, 116, 516,
  \dodoi{10.1086/300408}

\bibitem[{{Makarov} {et~al.}(2019){Makarov}, {Berghea}, {Frouard}, {Fey}, \&
  {Schmitt}}]{2019ApJ...873..132M}
{Makarov}, V.~V., {Berghea}, C.~T., {Frouard}, J., {Fey}, A., \& {Schmitt},
  H.~R. 2019, \apj, 873, 132, \dodoi{10.3847/1538-4357/aafa1c}

\bibitem[{{Malkin}(2018)}]{2018ApJS..239...20M}
{Malkin}, Z. 2018, \apjs, 239, 20, \dodoi{10.3847/1538-4365/aae777}

\bibitem[{{McLure} \& {Dunlop}(2000)}]{2000MNRAS.317..249M}
{McLure}, R.~J., \& {Dunlop}, J.~S. 2000, \mnras, 317, 249,
  \dodoi{10.1046/j.1365-8711.2000.03509.x}

\bibitem[{{McLure} {et~al.}(1999){McLure}, {Kukula}, {Dunlop}, {Baum}, {O'Dea},
  \& {Hughes}}]{1999MNRAS.308..377M}
{McLure}, R.~J., {Kukula}, M.~J., {Dunlop}, J.~S., {et~al.} 1999, \mnras, 308,
  377, \dodoi{10.1046/j.1365-8711.1999.02676.x}

\bibitem[{{Mignard} {et~al.}(2016){Mignard}, {Klioner}, {Lindegren}, {Bastian},
  {Bombrun}, {Hern{\'a}ndez}, {Hobbs}, {Lammers}, {Michalik}, {Ramos-Lerate},
  {Biermann}, {Butkevich}, {Comoretto}, {Joliet}, {Holl}, {Hutton}, {Parsons},
  {Steidelm{\"u}ller}, {Andrei}, {Bourda}, \& {Charlot}}]{2016A&A...595A...5M}
{Mignard}, F., {Klioner}, S., {Lindegren}, L., {et~al.} 2016, \aap, 595, A5,
  \dodoi{10.1051/0004-6361/201629534}

\bibitem[{{Napier} {et~al.}(1994){Napier}, {Bagri}, {Clark}, {Rogers},
  {Romney}, {Thompson}, \& {Walker}}]{1994IEEEP..82..658N}
{Napier}, P.~J., {Bagri}, D.~S., {Clark}, B.~G., {et~al.} 1994, IEEE
  Proceedings, 82, 658, \dodoi{10.1109/5.284733}

\bibitem[{{Nelson} \& {Whittle}(1996)}]{1996ApJ...465...96N}
{Nelson}, C.~H., \& {Whittle}, M. 1996, \apj, 465, 96, \dodoi{10.1086/177405}

\bibitem[{{Orosz} \& {Frey}(2013)}]{2013A&A...553A..13O}
{Orosz}, G., \& {Frey}, S. 2013, \aap, 553, A13,
  \dodoi{10.1051/0004-6361/201321279}

\bibitem[{{Park} {et~al.}(2017){Park}, {Barth}, {Woo}, {Malkan}, {Treu},
  {Bennert}, {Assef}, \& {Pancoast}}]{2017ApJ...839...93P}
{Park}, D., {Barth}, A.~J., {Woo}, J.-H., {et~al.} 2017, \apj, 839, 93,
  \dodoi{10.3847/1538-4357/aa6a53}

\bibitem[{{Peterson}(2004)}]{2004IAUS..222...15P}
{Peterson}, B.~M. 2004, in The Interplay Among Black Holes, Stars and ISM in
  Galactic Nuclei, ed. T.~{Storchi-Bergmann}, L.~C. {Ho}, \& H.~R. {Schmitt},
  Vol. 222, 15--20, \dodoi{10.1017/S1743921304001358}

\bibitem[{{Petrov} \& {Kovalev}(2017)}]{2017MNRAS.467L..71P}
{Petrov}, L., \& {Kovalev}, Y.~Y. 2017, \mnras, 467, L71,
  \dodoi{10.1093/mnrasl/slx001}

\bibitem[{{Petrov} {et~al.}(2019){Petrov}, {Kovalev}, \&
  {Plavin}}]{2019MNRAS.482.3023P}
{Petrov}, L., {Kovalev}, Y.~Y., \& {Plavin}, A.~V. 2019, \mnras, 482, 3023,
  \dodoi{10.1093/mnras/sty2807}

\bibitem[{{Plavin} {et~al.}(2019){Plavin}, {Kovalev}, {Pushkarev}, \&
  {Lobanov}}]{2019MNRAS.485.1822P}
{Plavin}, A.~V., {Kovalev}, Y.~Y., {Pushkarev}, A.~B., \& {Lobanov}, A.~P.
  2019, \mnras, 485, 1822, \dodoi{10.1093/mnras/stz504}

\bibitem[{{Runnoe} {et~al.}(2012{\natexlab{a}}){Runnoe}, {Brotherton}, \&
  {Shang}}]{2012MNRAS.422..478R}
{Runnoe}, J.~C., {Brotherton}, M.~S., \& {Shang}, Z. 2012{\natexlab{a}},
  \mnras, 422, 478, \dodoi{10.1111/j.1365-2966.2012.20620.x}

\bibitem[{{Runnoe} {et~al.}(2012{\natexlab{b}}){Runnoe}, {Brotherton}, \&
  {Shang}}]{2012MNRAS.427.1800R}
---. 2012{\natexlab{b}}, \mnras, 427, 1800,
  \dodoi{10.1111/j.1365-2966.2012.21878.x}

\bibitem[{{Schlafly} \& {Finkbeiner}(2011)}]{2011ApJ...737..103S}
{Schlafly}, E.~F., \& {Finkbeiner}, D.~P. 2011, \apj, 737, 103,
  \dodoi{10.1088/0004-637X/737/2/103}

\bibitem[{{Sexton} {et~al.}(2019){Sexton}, {Canalizo}, {Hiner}, {Komossa},
  {Woo}, {Treister}, \& {Hiner Dimassimo}}]{2019ApJ...878..101S}
{Sexton}, R.~O., {Canalizo}, G., {Hiner}, K.~D., {et~al.} 2019, \apj, 878, 101,
  \dodoi{10.3847/1538-4357/ab21d5}

\bibitem[{{Sexton} {et~al.}(2021){Sexton}, {Matzko}, {Darden}, {Canalizo}, \&
  {Gorjian}}]{2021MNRAS.500.2871S}
{Sexton}, R.~O., {Matzko}, W., {Darden}, N., {Canalizo}, G., \& {Gorjian}, V.
  2021, \mnras, 500, 2871, \dodoi{10.1093/mnras/staa3278}

\bibitem[{{Shen}(2013)}]{2013BASI...41...61S}
{Shen}, Y. 2013, Bulletin of the Astronomical Society of India, 41, 61.
\newblock \doarXiv{1302.2643}

\bibitem[{{Shen} {et~al.}(2008){Shen}, {Greene}, {Strauss}, {Richards}, \&
  {Schneider}}]{2008ApJ...680..169S}
{Shen}, Y., {Greene}, J.~E., {Strauss}, M.~A., {Richards}, G.~T., \&
  {Schneider}, D.~P. 2008, \apj, 680, 169, \dodoi{10.1086/587475}

\bibitem[{{Shen} {et~al.}(2011){Shen}, {Richards}, {Strauss}, {Hall},
  {Schneider}, {Snedden}, {Bizyaev}, {Brewington}, {Malanushenko},
  {Malanushenko}, {Oravetz}, {Pan}, \& {Simmons}}]{2011ApJS..194...45S}
{Shen}, Y., {Richards}, G.~T., {Strauss}, M.~A., {et~al.} 2011, \apjs, 194, 45,
  \dodoi{10.1088/0067-0049/194/2/45}

\bibitem[{{Sulentic}(1989)}]{1989ApJ...343...54S}
{Sulentic}, J.~W. 1989, \apj, 343, 54, \dodoi{10.1086/167684}

\bibitem[{{Taylor}(2005)}]{2005ASPC..347...29T}
{Taylor}, M.~B. 2005, Astronomical Society of the Pacific Conference Series,
  Vol. 347, {TOPCAT \&amp; STIL: Starlink Table/VOTable Processing Software}
  (Astronomical Society of the Pacific), 29

\bibitem[{{Thomas} {et~al.}(2013){Thomas}, {Steele}, {Maraston}, {Johansson},
  {Beifiori}, {Pforr}, {Str{\"o}mb{\"a}ck}, {Tremonti}, {Wake}, {Bizyaev},
  {Bolton}, {Brewington}, {Brownstein}, {Comparat}, {Kneib}, {Malanushenko},
  {Malanushenko}, {Oravetz}, {Pan}, {Parejko}, {Schneider}, {Shelden},
  {Simmons}, {Snedden}, {Tanaka}, {Weaver}, \& {Yan}}]{2013MNRAS.431.1383T}
{Thomas}, D., {Steele}, O., {Maraston}, C., {et~al.} 2013, \mnras, 431, 1383,
  \dodoi{10.1093/mnras/stt261}

\bibitem[{{Titov} {et~al.}(2011){Titov}, {Lambert}, \&
  {Gontier}}]{2011A&A...529A..91T}
{Titov}, O., {Lambert}, S.~B., \& {Gontier}, A.~M. 2011, \aap, 529, A91,
  \dodoi{10.1051/0004-6361/201015718}

\bibitem[{{Titov} \& {Malkin}(2009)}]{2009A&A...506.1477T}
{Titov}, O., \& {Malkin}, Z. 2009, \aap, 506, 1477,
  \dodoi{10.1051/0004-6361/200912369}

\bibitem[{{Titov} {et~al.}(2017){Titov}, {Pursimo}, {Johnston}, {Stanford},
  {Hunstead}, {Jauncey}, \& {Zenere}}]{2017AJ....153..157T}
{Titov}, O., {Pursimo}, T., {Johnston}, H.~M., {et~al.} 2017, \aj, 153, 157,
  \dodoi{10.3847/1538-3881/aa61fd}

\bibitem[{{Tonry} \& {Davis}(1979)}]{1979AJ.....84.1511T}
{Tonry}, J., \& {Davis}, M. 1979, \aj, 84, 1511, \dodoi{10.1086/112569}

\bibitem[{{Ulrich} {et~al.}(1997){Ulrich}, {Maraschi}, \&
  {Urry}}]{1997ARA&A..35..445U}
{Ulrich}, M.-H., {Maraschi}, L., \& {Urry}, C.~M. 1997, \araa, 35, 445,
  \dodoi{10.1146/annurev.astro.35.1.445}

\bibitem[{{Valdes} {et~al.}(2004){Valdes}, {Gupta}, {Rose}, {Singh}, \&
  {Bell}}]{2004ApJS..152..251V}
{Valdes}, F., {Gupta}, R., {Rose}, J.~A., {Singh}, H.~P., \& {Bell}, D.~J.
  2004, \apjs, 152, 251, \dodoi{10.1086/386343}

\bibitem[{{Vazdekis} {et~al.}(2016){Vazdekis}, {Koleva}, {Ricciardelli},
  {R{\"o}ck}, \& {Falc{\'o}n-Barroso}}]{2016MNRAS.463.3409V}
{Vazdekis}, A., {Koleva}, M., {Ricciardelli}, E., {R{\"o}ck}, B., \&
  {Falc{\'o}n-Barroso}, J. 2016, \mnras, 463, 3409,
  \dodoi{10.1093/mnras/stw2231}

\bibitem[{{V{\'e}ron-Cetty} {et~al.}(2004){V{\'e}ron-Cetty}, {Joly}, \&
  {V{\'e}ron}}]{2004A&A...417..515V}
{V{\'e}ron-Cetty}, M.~P., {Joly}, M., \& {V{\'e}ron}, P. 2004, \aap, 417, 515,
  \dodoi{10.1051/0004-6361:20035714}

\bibitem[{{V{\'e}ron-Cetty} {et~al.}(2001){V{\'e}ron-Cetty}, {V{\'e}ron}, \&
  {Gon{\c{c}}alves}}]{2001A&A...372..730V}
{V{\'e}ron-Cetty}, M.~P., {V{\'e}ron}, P., \& {Gon{\c{c}}alves}, A.~C. 2001,
  \aap, 372, 730, \dodoi{10.1051/0004-6361:20010489}

\bibitem[{{Vestergaard} \& {Wilkes}(2001)}]{2001ApJS..134....1V}
{Vestergaard}, M., \& {Wilkes}, B.~J. 2001, \apjs, 134, 1,
  \dodoi{10.1086/320357}

\bibitem[{{Wills} {et~al.}(1985){Wills}, {Netzer}, \&
  {Wills}}]{1985ApJ...288...94W}
{Wills}, B.~J., {Netzer}, H., \& {Wills}, D. 1985, \apj, 288, 94,
  \dodoi{10.1086/162767}

\bibitem[{{Woo} {et~al.}(2018){Woo}, {Le}, {Karouzos}, {Park}, {Park},
  {Malkan}, {Treu}, \& {Bennert}}]{2018ApJ...859..138W}
{Woo}, J.-H., {Le}, H. A.~N., {Karouzos}, M., {et~al.} 2018, \apj, 859, 138,
  \dodoi{10.3847/1538-4357/aabf3e}

\bibitem[{{Woo} \& {Park}(2011)}]{2011nlsg.confE..39W}
{Woo}, J.~H., \& {Park}, D. 2011, in Narrow-Line Seyfert 1 Galaxies and their
  Place in the Universe, ed. L.~{Foschini}, M.~{Colpi}, L.~{Gallo}, D.~{Grupe},
  S.~{Komossa}, K.~{Leighly}, \& S.~{Mathur}, 39

\bibitem[{{Woo} {et~al.}(2015){Woo}, {Yoon}, {Park}, {Park}, \&
  {Kim}}]{2015ApJ...801...38W}
{Woo}, J.-H., {Yoon}, Y., {Park}, S., {Park}, D., \& {Kim}, S.~C. 2015, \apj,
  801, 38, \dodoi{10.1088/0004-637X/801/1/38}

\bibitem[{{Wu} {et~al.}(2004){Wu}, {Wang}, {Kong}, {Liu}, \&
  {Han}}]{2004A&A...424..793W}
{Wu}, X.~B., {Wang}, R., {Kong}, M.~Z., {Liu}, F.~K., \& {Han}, J.~L. 2004,
  \aap, 424, 793, \dodoi{10.1051/0004-6361:20035845}

\bibitem[{{Yanny} {et~al.}(2009){Yanny}, {Rockosi}, {Newberg}, {Knapp},
  {Adelman-McCarthy}, {Alcorn}, {Allam}, {Allende Prieto}, {An}, {Anderson},
  {Anderson}, {Bailer-Jones}, {Bastian}, {Beers}, {Bell}, {Belokurov},
  {Bizyaev}, {Blythe}, {Bochanski}, {Boroski}, {Brinchmann}, {Brinkmann},
  {Brewington}, {Carey}, {Cudworth}, {Evans}, {Evans}, {Gates}, {G{\"a}nsicke},
  {Gillespie}, {Gilmore}, {Nebot Gomez-Moran}, {Grebel}, {Greenwell}, {Gunn},
  {Jordan}, {Jordan}, {Harding}, {Harris}, {Hendry}, {Holder}, {Ivans},
  {Ivezi{\v{c}}}, {Jester}, {Johnson}, {Kent}, {Kleinman}, {Kniazev},
  {Krzesinski}, {Kron}, {Kuropatkin}, {Lebedeva}, {Lee}, {French Leger},
  {L{\'e}pine}, {Levine}, {Lin}, {Long}, {Loomis}, {Lupton}, {Malanushenko},
  {Malanushenko}, {Margon}, {Martinez-Delgado}, {McGehee}, {Monet}, {Morrison},
  {Munn}, {Neilsen}, {Nitta}, {Norris}, {Oravetz}, {Owen}, {Padmanabhan},
  {Pan}, {Peterson}, {Pier}, {Platson}, {Re Fiorentin}, {Richards}, {Rix},
  {Schlegel}, {Schneider}, {Schreiber}, {Schwope}, {Sibley}, {Simmons},
  {Snedden}, {Allyn Smith}, {Stark}, {Stauffer}, {Steinmetz}, {Stoughton},
  {SubbaRao}, {Szalay}, {Szkody}, {Thakar}, {Sivarani}, {Tucker}, {Uomoto},
  {Vanden Berk}, {Vidrih}, {Wadadekar}, {Watters}, {Wilhelm}, {Wyse}, {Yarger},
  \& {Zucker}}]{2009AJ....137.4377Y}
{Yanny}, B., {Rockosi}, C., {Newberg}, H.~J., {et~al.} 2009, \aj, 137, 4377,
  \dodoi{10.1088/0004-6256/137/5/4377}

\bibitem[{{York} {et~al.}(2000){York}, {Adelman}, {Anderson}, {Anderson},
  {Annis}, {Bahcall}, {Bakken}, {Barkhouser}, {Bastian}, {Berman}, {Boroski},
  {Bracker}, {Briegel}, {Briggs}, {Brinkmann}, {Brunner}, {Burles}, {Carey},
  {Carr}, {Castander}, {Chen}, {Colestock}, {Connolly}, {Crocker}, {Csabai},
  {Czarapata}, {Davis}, {Doi}, {Dombeck}, {Eisenstein}, {Ellman}, {Elms},
  {Evans}, {Fan}, {Federwitz}, {Fiscelli}, {Friedman}, {Frieman}, {Fukugita},
  {Gillespie}, {Gunn}, {Gurbani}, {de Haas}, {Haldeman}, {Harris}, {Hayes},
  {Heckman}, {Hennessy}, {Hindsley}, {Holm}, {Holmgren}, {Huang}, {Hull},
  {Husby}, {Ichikawa}, {Ichikawa}, {Ivezi{\'c}}, {Kent}, {Kim}, {Kinney},
  {Klaene}, {Kleinman}, {Kleinman}, {Knapp}, {Korienek}, {Kron}, {Kunszt},
  {Lamb}, {Lee}, {Leger}, {Limmongkol}, {Lindenmeyer}, {Long}, {Loomis},
  {Loveday}, {Lucinio}, {Lupton}, {MacKinnon}, {Mannery}, {Mantsch}, {Margon},
  {McGehee}, {McKay}, {Meiksin}, {Merelli}, {Monet}, {Munn}, {Narayanan},
  {Nash}, {Neilsen}, {Neswold}, {Newberg}, {Nichol}, {Nicinski}, {Nonino},
  {Okada}, {Okamura}, {Ostriker}, {Owen}, {Pauls}, {Peoples}, {Peterson},
  {Petravick}, {Pier}, {Pope}, {Pordes}, {Prosapio}, {Rechenmacher}, {Quinn},
  {Richards}, {Richmond}, {Rivetta}, {Rockosi}, {Ruthmansdorfer}, {Sandford},
  {Schlegel}, {Schneider}, {Sekiguchi}, {Sergey}, {Shimasaku}, {Siegmund},
  {Smee}, {Smith}, {Snedden}, {Stone}, {Stoughton}, {Strauss}, {Stubbs},
  {SubbaRao}, {Szalay}, {Szapudi}, {Szokoly}, {Thakar}, {Tremonti}, {Tucker},
  {Uomoto}, {Vanden Berk}, {Vogeley}, {Waddell}, {Wang}, {Watanabe},
  {Weinberg}, {Yanny}, {Yasuda}, \& {SDSS Collaboration}}]{2000AJ....120.1579Y}
{York}, D.~G., {Adelman}, J., {Anderson}, John~E., J., {et~al.} 2000, \aj, 120,
  1579, \dodoi{10.1086/301513}

\bibitem[{{Zacharias} {et~al.}(2020){Zacharias}, {Finch}, {Dorland}, {Secrest},
  \& {Johnson}}]{2020jsrs.conf..179Z}
{Zacharias}, N., {Finch}, C., {Dorland}, B., {Secrest}, N., \& {Johnson}, M.
  2020, in Astrometry, Earth Rotation, and Reference Systems in the GAIA era,
  ed. C.~{Bizouard}, 179--182

\bibitem[{{Zacharias} \& {Zacharias}(2014)}]{2014AJ....147...95Z}
{Zacharias}, N., \& {Zacharias}, M.~I. 2014, \aj, 147, 95,
  \dodoi{10.1088/0004-6256/147/5/95}

\end{thebibliography}
\bibliographystyle{aasjournal}

\appendix

\section{Emission Lines}

Table \ref{tab:emline} shows the full list of emission lines fit for all objects in the full USNO VLBI Spectroscopic Catalog.  

\startlongtable
\begin{deluxetable}{clll}
\label{tab:emline}
\tablecaption{Fitted Emission Lines in the USNO VLBI Spectroscopic Catalog}
\tablehead{
\colhead{Region} &  \colhead{Ion} & \colhead{$\lambda$ [$\textrm{\AA}$]} & \colhead{$N_{95\%}$} }
\colnumbers
\startdata
\multirow{5}{*}{1}  & [\ion{S}{3}]                  & 9068.600      & 3\\
                    & [\ion{Fe}{2}]                 & 8891.910      & 3\\
                    & [\ion{Fe}{2}]                 & 8616.950      & 0\\
                    & \ion{O}{1}                    & 8446.359      & 2\\
                    & \ion{He}{2}                   & 8236.790      & 3\\
\tableline
\multirow{6}{*}{2}  & [\ion{Fe}{11}]                & 7891.800      & 1\\
                    & [\ion{Ni}{3}]                 & 7889.900      & 0\\
                    & [\ion{O}{2}]                  & 7330.730      & 2\\
                    & [\ion{O}{2}]                  & 7319.990      & 3\\
                    & [\ion{Ar}{3}]                 & 7135.790      & 14\\
                    & \ion{He}{1}                   & 7065.196      & 4\\
\tableline
\multirow{9}{*}{3}  & \bf{[\ion{S}{2}]}             & \bf{6732.668} & \bf{76}\\
                    & [\ion{S}{2}]                  & 6718.294      & 86\\
                    & \bf{[\ion{N}{2}]}             & \bf{6585.278} & \bf{44}\\
                    & \bf{Na. H$\alpha$}            & \bf{6564.632} & \bf{58}\\
                    & \bf{Br. H$\alpha$}            & \bf{6564.632} & \bf{82}\\
                    & [\ion{N}{2}]                  & 6549.859      & 37\\
                    & [\ion{Fe}{10}]                & 6374.510      & 1\\
                    & [\ion{O}{1}]                  & 6365.535      & 2\\
                    & \bf{[\ion{O}{1}]}             & \bf{6302.046} & \bf{89}\\
\tableline
\multirow{5}{*}{4}  & [\ion{Fe}{7}]                 & 6087.000      & 15\\
                    & \ion{He}{1}                   & 5875.624      & 60\\
                    & [\ion{Fe}{7}]                 & 5720.700      & 11\\
                    & [\ion{Fe}{6}]                 & 5677.000      & 1\\
                    & [\ion{Fe}{6}]                 & 5637.600      & 1\\
\tableline
\multirow{5}{*}{5}  & \bf{[\ion{O}{3}]}             & \bf{5008.240} & \bf{308}\\
                    & [\ion{O}{3}]                  & 4960.295      & 243\\
                    & \bf{Na. H$\beta$}             & \bf{4862.691} & \bf{196}\\
                    & \bf{Br. H$\beta$}             & \bf{4862.691} & \bf{206}\\
                    & \ion{He}{2}                   & 4687.021      & 42\\
                    & \ion{He}{1}                   & 4471.479      & 1\\
\tableline
\multirow{8}{*}{6}  & [\ion{O}{3}]                  & 4364.436      & 29\\
                    & Na. H$\gamma$                 & 4341.691      & 73\\
                    & Br. H$\gamma$                 & 4341.691      & 83\\
                    & Na. H$\delta$                 & 4102.900      & 99\\
                    & Br. H$\delta$                 & 4102.900      & 118\\
                    & [\ion{Ne}{3}]                 & 3968.593      & 202\\
                    & \ion{He}{1}                   & 3888.647      & 97\\
                    & [\ion{Ne}{3}]                 & 3869.857      & 256\\
                    & [\ion{O}{2}]\tablenotemark{a} & 3729.875      & 294\\
                    & [\ion{O}{2}]\tablenotemark{a} & 3727.092      & 294\\
\tableline
\multirow{6}{*}{7}  & [\ion{Ne}{5}]                 & 3426.863      & 149\\
                    & [\ion{Ne}{5}]                 & 3346.783      & 54\\
                    & \ion{He}{2}                   & 3203.100      & 14\\
                    & \ion{O}{3}                    & 3132.794      & 50\\
                    & \bf{\ion{Mg}{2}]}             & \bf{2799.117} & \bf{675}\\
                    & \ion{C}{2}]                   & 2326.000      & 11\\
\tableline
\multirow{10}{*}{8} & \ion{C}{3}]                   & 1908.734      & 486\\
                    & \ion{Al}{3}                   & 1857.400      & 113\\
                    & \ion{O}{3}]                   & 1665.850      & 66\\
                    & \ion{He}{2}                   & 1640.400      & 55\\
                    & \bf{\ion{C}{4}}               & \bf{1549.480} & \bf{364}\\
                    & \ion{C}{2}                    & 1335.310      & 63\\
                    & \ion{O}{1}                    & 1305.530      & 96\\
                    & \ion{N}{5}                    & 1240.810      & 161\\
                    & Ly$\alpha$                    & 1215.240      & 164\\
                    & \ion{O}{6}                    & 1033.820      & 11\\
\enddata
\tablecomments{Columns 1: Region number.  Column 2: Emission line. Column 3: Reference rest wavelength used during fit. Column 4: Number of detections at the 95\% confidence level.  Emission lines in boldface font are included in the short catalog.}
\tablenotetext{a}{blended emission lines share identical probabilities.}
\end{deluxetable}

\newpage

\section{Catalog Structure}

\begin{longrotatetable}
\centerwidetable
\begin{deluxetable}{clll}
\label{tab:column_descrip}
\tablecaption{USNO VLBI Spectroscopic ``Short'' Catalog Column Descriptions}
\tablehead{
\colhead{Number} & \colhead{Format} & \colhead{Column Name} & \colhead{Description}
}
\colnumbers
\startdata
    \multicolumn{4}{c}{ \bf{ICRF3/SDSS metadata}}\\
    \hline
1 & IERS                                 & STRING & IERS designation \\
2 & ICRF                                 & STRING & ICRF3 designation \\
3 & DEFINING                             & BOOL & ICRF3 defining source? \\
4 & ICRF\_RA                             & STRING & ICRF3 right ascension [HH:SS:SS.S] \\
5 & ICRF\_DEC                            & STRING & ICRF3 declination [DD:MM:SS.S] \\
6 & SDSS\_RA                             & FLOAT64 & SDSS right ascension [degrees]\\
7 & SDSS\_DEC                            & FLOAT64 & SDSS declination [degrees]\\
8 & Z\_SDSS                              & FLOAT64 & SDSS redshift used for fit \\
9 & PLATE                                & INT64 & SDSS plate number \\
10 & MJD                                  & INT64 & SDSS observation MJD \\
11 & FIBERID                              & INT64 & SDSS object fiber ID \\
12 & RUN2D                                & STRING & SDSS 2D reduction run number \\
13 & AGN\_TYPE                            & STRING & AGN Type (``QSO'', ``Sy 1'', ``Sy 2'', or ``featureless'') \\
14 & CLASS                                & STRING & SDSS SpecObj class \\
15 & SUBCLASS                             & STRING & SDSS SpecObj subclass \\
16 & SN\_MEDIAN\_R                        & FLOAT64 & SDSS median signal-to-noise \\
17 & WAVE\_MIN                            & FLOAT64 & minimum rest-frame spectral wavelength [$\rm{\AA}$]\\
18 & WAVE\_MAX                            & FLOAT64 & maximum rest-frame spectral wavelength [$\rm{\AA}$]\\
19 & SURVEY                               & STRING & SDSS observation survey \\
20 & INSTRUMENT                           & STRING & SDSS observation instrument \\
21 & PROGRAM\_NAME                        & STRING & SDSS program name \\
    \hline
    \multicolumn{4}{c}{ \bf{BH mass estimates}}\\
    \hline
22 & LOG\_MBH\_H\_ALPHA\_DISP             & FLOAT64 & Virial BH mass based on H$\alpha$ dispersion [$\log_{10}(M_{\rm{BH}}/M_\odot)$] \\
23 & LOG\_MBH\_H\_ALPHA\_DISP\_ERR        & FLOAT64 & Uncertainty virial in BH mass based on H$\alpha$ dispersion \\
24 & LOG\_MBH\_H\_ALPHA\_FWHM             & FLOAT64 & Virial BH mass based on H$\alpha$ FWHM [$\log_{10}(M_{\rm{BH}}/M_\odot)$] \\
25 & LOG\_MBH\_H\_ALPHA\_FWHM\_ERR        & FLOAT64 & Uncertainty virial in BH mass based on H$\alpha$ FWHM \\
26 & MBH\_H\_ALPHA\_FLAG                  & INT64   & Quality flag for H$\alpha$ BH mass \\
27 & LOG\_MBH\_H\_BETA\_DISP              & FLOAT64 & Virial BH mass based on H$\beta$ dispersion [$\log_{10}(M_{\rm{BH}}/M_\odot)$] \\
28 & LOG\_MBH\_H\_BETA\_DISP\_ERR         & FLOAT64 & Uncertainty in virial BH mass based on H$\beta$ dispersion\\
29 & LOG\_MBH\_H\_BETA\_FWHM              & FLOAT64 & Virial BH mass based on H$\beta$ FWHM [$\log_{10}(M_{\rm{BH}}/M_\odot)$] \\
30 & LOG\_MBH\_H\_BETA\_FWHM\_ERR         & FLOAT64 & Uncertainty in virial BH mass based on H$\beta$ FWHM \\
31 & MBH\_H\_BETA\_FLAG                   & INT64   & Quality flag for H$\beta$ BH mass \\
32 & LOG\_MBH\_MGII\_DISP                 & FLOAT64 & Virial BH mass based on \ion{Mg}{2} dispersion [$\log_{10}(M_{\rm{BH}}/M_\odot)$] \\
33 & LOG\_MBH\_MGII\_DISP\_ERR            & FLOAT64 & Uncertainty in virial BH mass based on \ion{Mg}{2} dispersion \\
34 & LOG\_MBH\_MGII\_FWHM                 & FLOAT64 & Virial BH mass based on \ion{Mg}{2} FWHM [$\log_{10}(M_{\rm{BH}}/M_\odot)$] \\
35 & LOG\_MBH\_MGII\_FWHM\_ERR            & FLOAT64 & Uncertainty in virial BH mass based on \ion{Mg}{2} FWHM \\
36 & MBH\_MGII\_FLAG                      & INT64   & Quality flag for \ion{Mg}{2} BH mass \\
37 & LOG\_MBH\_CIV\_DISP                  & FLOAT64 & Virial BH mass based on \ion{C}{4} dispersion [$\log_{10}(M_{\rm{BH}}/M_\odot)$] \\
38 & LOG\_MBH\_CIV\_DISP\_ERR             & FLOAT64 & Uncertainty in virial BH mass based on \ion{C}{4} dispersion \\
39 & LOG\_MBH\_CIV\_FWHM                  & FLOAT64 & Virial BH mass based on \ion{C}{4} FWHM [$\log_{10}(M_{\rm{BH}}/M_\odot)$] \\
40 & LOG\_MBH\_CIV\_FWHM\_ERR             & FLOAT64 & Uncertainty in virial BH mass based on \ion{C}{4} FWHM \\
41 & MBH\_CIV\_FLAG                       & INT64   & Quality flag for \ion{C}{4} BH mass \\
    \hline
    \multicolumn{4}{c}{ \bf{Continuum measurements}}\\
    \hline
42 & ALPHA                                 & FLOAT64 & Power-law index (slope) of the AGN continuum $\alpha_\lambda$ \\
43 & ALPHA\_ERR                            & FLOAT64 & Uncertainty in $\alpha_\lambda$  \\
44 & AGN\_FRAC\_4000                       & FLOAT64 & AGN fraction of the continuum at 4000 $\rm{\AA}$ \\
45 & AGN\_FRAC\_4000\_ERR                  & FLOAT64 & Uncertainty in the AGN fraction at 4000 $\rm{\AA}$ \\
46 & AGN\_FRAC\_7000                       & FLOAT64 & AGN fraction of the continuum at 7000 $\rm{\AA}$ \\
47 & AGN\_FRAC\_7000\_ERR                  & FLOAT64 & Uncertainty in the AGN fraction at 7000 $\rm{\AA}$ \\
48 & LOG\_L\_CONT\_AGN\_5100               & FLOAT64 & Monochromatic AGN luminosity at 5100 $\rm{\AA}$ [$\log_{10}(L/\rm{erg\;s}^{-1})$] \\
49 & LOG\_L\_CONT\_AGN\_5100\_ERR          & FLOAT64 & Uncertainty in the monochromatic AGN luminosity at 5100 $\rm{\AA}$ \\
50 & LOG\_L\_CONT\_AGN\_3000               & FLOAT64 & Monochromatic AGN luminosity at 3000 $\rm{\AA}$ [$\log_{10}(L/\rm{erg\;s}^{-1})$] \\
51 & LOG\_L\_CONT\_AGN\_3000\_ERR          & FLOAT64 & Uncertainty in the monochromatic AGN luminosity at 3000 $\rm{\AA}$ \\
52 & LOG\_L\_CONT\_AGN\_1350               & FLOAT64 & Monochromatic AGN luminosity at 1350 $\rm{\AA}$ [$\log_{10}(L/\rm{erg\;s}^{-1})$] \\
53 & LOG\_L\_CONT\_AGN\_1350\_ERR          & FLOAT64 & Uncertainty in the monochromatic AGN luminosity at 1350 $\rm{\AA}$ \\
    \hline
    \multicolumn{4}{c}{ \bf{Eddington luminosities \& ratios}}\\
    \hline
54 & LOG\_L\_BOL\_H\_ALPHA                 & FLOAT64 & Bolometric AGN luminosity at 5100 $\rm{\AA}$ for H$\alpha$ [$\log_{10}(L/\rm{erg\;s}^{-1})$] \\
55 & LOG\_L\_BOL\_H\_ALPHA\_ERR            & FLOAT64 & Uncertainty in the bolometric AGN luminosity at 5100 $\rm{\AA}$ for H$\alpha$ \\
56 & LOG\_L\_BOL\_H\_BETA                  & FLOAT64 & Bolometric AGN luminosity at 5100 $\rm{\AA}$ for H$\beta$ [$\log_{10}(L/\rm{erg\;s}^{-1})$] \\
57 & LOG\_L\_BOL\_H\_BETA\_ERR             & FLOAT64 & Uncertainty in the bolometric AGN luminosity at 5100 $\rm{\AA}$ for H$\beta$\\
58 & LOG\_L\_BOL\_MGII                     & FLOAT64 & Bolometric AGN luminosity at 3000 $\rm{\AA}$ for \ion{Mg}{2} [$\log_{10}(L/\rm{erg\;s}^{-1})$] \\
59 & LOG\_L\_BOL\_MGII\_ERR                & FLOAT64 & Uncertainty in the bolometric AGN luminosity at 3000 $\rm{\AA}$ for \ion{Mg}{2} \\
60 & LOG\_L\_BOL\_CIV                      & FLOAT64 & Bolometric AGN luminosity at 1350 $\rm{\AA}$ for \ion{C}{4} [$\log_{10}(L/\rm{erg\;s}^{-1})$] \\
61 & LOG\_L\_BOL\_CIV\_ERR                 & FLOAT64 & Uncertainty in the bolometric AGN luminosity at 1350 $\rm{\AA}$ for \ion{C}{4}\\
    \hline
    \multicolumn{4}{c}{ \bf{Eddington luminosities \& ratios}}\\
    \hline
62 & LOG\_L\_EDD\_H\_ALPHA                 & FLOAT64 & Eddington luminosity estimated from $M_{\rm{BH}}(\rm{H}\alpha)$ [$\log_{10}(L/\rm{erg\;s}^{-1})$] \\
63 & LOG\_L\_EDD\_H\_ALPHA\_ERR            & FLOAT64 & Uncertainty in the Eddington luminosity estimated from $M_{\rm{BH}}(\rm{H}\alpha)$\\
64 & LOG\_L\_EDD\_H\_BETA                  & FLOAT64 & Eddington luminosity estimated from $M_{\rm{BH}}(\rm{H}\beta)$ [$\log_{10}(L/\rm{erg\;s}^{-1})$] \\
65 & LOG\_L\_EDD\_H\_BETA\_ERR             & FLOAT64 & Uncertainty in the Eddington luminosity estimated from $M_{\rm{BH}}(\rm{H}\beta)$\\
66 & LOG\_L\_EDD\_MGII                     & FLOAT64 & Eddington luminosity estimated from $M_{\rm{BH}}(\textrm{\ion{Mg}{2}})$ [$\log_{10}(L/\rm{erg\;s}^{-1})$] \\
67 & LOG\_L\_EDD\_MGII\_ERR                & FLOAT64 & Uncertainty in the Eddington luminosity estimated from $M_{\rm{BH}}(\textrm{\ion{Mg}{2}})$\\
68 & LOG\_L\_EDD\_CIV                      & FLOAT64 & Eddington luminosity estimated from $M_{\rm{BH}}(\textrm{\ion{C}{4}})$ [$\log_{10}(L/\rm{erg\;s}^{-1})$] \\
69 & LOG\_L\_EDD\_CIV\_ERR                 & FLOAT64 & Uncertainty in the Eddington luminosity estimated from $M_{\rm{BH}}(\textrm{\ion{C}{4}})$\\
70 & EDD\_RATIO\_H\_ALPHA                  & FLOAT64 & Eddington Ratio estimated at 5100 $\rm{\AA}$ and $M_{\rm{BH}}(\rm{H}\alpha)$; $R_{\rm{Edd,H}\alpha}$ \\
71 & EDD\_RATIO\_H\_ALPHA\_ERR             & FLOAT64 & Uncertainty in $R_{\rm{Edd,H}\alpha}$\\
72 & EDD\_RATIO\_H\_BETA                   & FLOAT64 & Eddington Ratio estimated at 5100 $\rm{\AA}$ and $M_{\rm{BH}}(\rm{H}\beta)$; $R_{\rm{Edd,H}\beta}$ \\
73 & EDD\_RATIO\_H\_BETA\_ERR              & FLOAT64 & Uncertainty in $R_{\rm{Edd,H}\beta}$\\
74 & EDD\_RATIO\_MGII                      & FLOAT64 & Eddington Ratio estimated at 3000 $\rm{\AA}$ and $M_{\rm{BH}}(\textrm{\ion{Mg}{2}})$; $R_{\rm{Edd,MgII}}$ \\
75 & EDD\_RATIO\_MGII\_ERR                 & FLOAT64 & Uncertainty in $R_{\rm{Edd,MgII}}$\\
76 & EDD\_RATIO\_CIV                       & FLOAT64 & Eddington Ratio estimated at 1350 $\rm{\AA}$ and $M_{\rm{BH}}(\textrm{\ion{C}{4}})$; $R_{\rm{Edd,CIV}}$ \\
77 & EDD\_RATIO\_CIV\_ERR                  & FLOAT64 & Uncertainty in $R_{\rm{Edd,CIV}}$\\
    \hline
    \multicolumn{4}{c}{ \bf{Eigenvector 1 quantities}}\\
    \hline
78 & OPT\_FEII\_EW	                    	& FLOAT64 & Equivalent width of optical \ion{Fe}{2} from 4434 $\rm{\AA}$ to 4684 $\rm{\AA}$;~EW$_{\rm{FeII}}$~[$\rm{\AA}$] \\
79 & OPT\_FEII\_EW\_ERR	            	    & FLOAT64 & Uncertainty in EW$_{\rm{FeII}}$~[$\rm{\AA}$] \\
80 & R\_FEII	                            & FLOAT64 & Ratio of equivalent widths of \ion{Fe}{2} to braod H$\beta$;~$R(\textrm{\ion{Fe}{2}})$ \\
81 & R\_FEII\_ERR	                        & FLOAT64 & Uncertainty in $R(\textrm{\ion{Fe}{2}})$ \\
82 & R\_5007	                            & FLOAT64 & Ratio of equivalent widths of [\ion{O}{3}]$\lambda5007$ to braod H$\beta$;~$R(\lambda 5007)$ \\
83 & R\_5007\_ERR	                        & FLOAT64 & Uncertainty in $R(\lambda 5007)$ \\
84 & R\_4686	                            & FLOAT64 & Ratio of equivalent widths of \ion{He}{2}$\lambda4686$ to braod H$\beta$;~$R(\lambda 4686)$ \\
85 & R\_4686\_ERR	                        & FLOAT64 & Uncertainty in $R(\lambda 4686)$ \\
    \hline
    \multicolumn{4}{c}{ \bf{Stellar kinematics}}\\
    \hline
86 & STEL\_VEL\_CAT	                        & FLOAT64 & Stellar velocity measured from \ion{Ca}{0}T absorption [km s$^{-1}$]	\\
87 & STEL\_VEL\_CAT\_ERR	                & FLOAT64 & Uncertainty in the \ion{Ca}{0}T stellar velocity \\
88 & STEL\_DISP\_CAT	                    & FLOAT64 & Stellar dispersion measured from \ion{Ca}{0}T absorption [km s$^{-1}$]	\\
89 & STEL\_DISP\_CAT\_ERR	             	& FLOAT64 & Uncertainty in the \ion{Ca}{0}T stellar dispersion \\
90 & STEL\_VEL\_MG1B	                    & FLOAT64 & Stellar velocity measured from the \ion{Mg}{1}b absorption [km s$^{-1}$]	\\
91 & STEL\_VEL\_MG1B\_ERR	             	& FLOAT64 & Uncertainty in \ion{Mg}{1}b stellar velocity \\
92 & STEL\_DISP\_MG1B	                 	& FLOAT64 & Stellar dispersion measured from \ion{Mg}{1}b absorption [km s$^{-1}$]	\\
93 & STEL\_DISP\_MG1B\_ERR	             	& FLOAT64 & Uncertainty in \ion{Mg}{1}b stellar dispersion \\
94 & STEL\_VEL\_CAKH	                    & FLOAT64 & Stellar velocity measured from \ion{Ca}{0}H+K absorption [km s$^{-1}$] \\
95 & STEL\_VEL\_CAKH\_ERR	             	& FLOAT64 & Uncertainty in \ion{Ca}{0}H+K stellar velocity \\
96 & STEL\_DISP\_CAKH	                 	& FLOAT64 & Stellar dispersion measured from \ion{Ca}{0}H+K absorption [km s$^{-1}$] \\
97 & STEL\_DISP\_CAKH\_ERR	             	& FLOAT64 & Uncertainty in \ion{Ca}{0}H+K stellar dispersion \\
98 & NA\_OIII\_5007\_CORE\_DISP	            & FLOAT64 & Core [\ion{O}{3}]$\lambda5007$ dispersion as a proxy for $\sigma_*$ [km s$^{-1}$]	\\
99 & NA\_OIII\_5007\_CORE\_DISP\_ERR	    & FLOAT64 & Uncertainty in the core [\ion{O}{3}]$\lambda5007$ dispersion	\\
 \hline
 \multicolumn{4}{c}{ \bf{Emission line measurements}}\\
 \hline
100 & NA\_SII\_6732\_DISP	                & FLOAT64 & Dispersion of narrow [\ion{S}{2}]$\lambda6732$ [km s$^{-1}$]	\\
101 & NA\_SII\_6732\_DISP\_ERR	            & FLOAT64 & Uncertainty in narrow [\ion{S}{2}]$\lambda6732$ dispersion	\\
102 & NA\_SII\_6732\_FWHM	                & FLOAT64 & FWHM of narrow [\ion{S}{2}]$\lambda6732$ [km s$^{-1}$]	\\
103 & NA\_SII\_6732\_FWHM\_ERR	            & FLOAT64 & Uncertainty in narrow [\ion{S}{2}]$\lambda6732$ FWHM 	\\
104 & NA\_SII\_6732\_VOFF	                & FLOAT64 & Velocity of narrow [\ion{S}{2}]$\lambda6732$ [km s$^{-1}$]	\\
105 & NA\_SII\_6732\_VOFF\_ERR	            & FLOAT64 & Uncertainty in narrow [\ion{S}{2}]$\lambda6732$ velocity	\\
106	& NA\_SII\_6732\_EW	                 	& FLOAT64 & Equivalent width of narrow [\ion{S}{2}]$\lambda6732$ [$\rm{\AA}$] \\
107	& NA\_SII\_6732\_EW\_ERR	            & FLOAT64 & Uncertainty in narrow [\ion{S}{2}]$\lambda6732$ equivalent width \\
108	& NA\_SII\_6732\_LOG\_FLUX	         	& FLOAT64 & Flux of narrow [\ion{S}{2}]$\lambda6732$ [$\log_{10}(F/\rm{erg\;cm}^{-2}\;\rm{s}^{-1})$] \\
109	& NA\_SII\_6732\_LOG\_FLUX\_ERR	     	& FLOAT64 & Uncertainty in narrow [\ion{S}{2}]$\lambda6732$ flux \\
110	& NA\_SII\_6732\_LOG\_LUM	            & FLOAT64 & Luminosity of narrow [\ion{S}{2}]$\lambda6732$ [$\log_{10}(L/\rm{erg\;s}^{-1})$] \\
111	& NA\_SII\_6732\_LOG\_LUM\_ERR	     	& FLOAT64 & Uncertainty in narrow [\ion{S}{2}]$\lambda6732$ luminosity \\
112	& NA\_SII\_6732\_PROB	                & FLOAT64 & Detection probability of narrow [\ion{S}{2}]$\lambda6732$ \\
    \hline
113 & NA\_NII\_6585\_DISP	                & FLOAT64 & Dispersion of narrow [\ion{N}{2}]$\lambda6585$ [km s$^{-1}$]	\\
114 & NA\_NII\_6585\_DISP\_ERR	            & FLOAT64 & Uncertainty in narrow [\ion{N}{2}]$\lambda6585$ dispersion	\\
115 & NA\_NII\_6585\_FWHM	                & FLOAT64 & FWHM of narrow [\ion{N}{2}]$\lambda6585$ [km s$^{-1}$]	\\
116 & NA\_NII\_6585\_FWHM\_ERR	            & FLOAT64 & Uncertainty in narrow [\ion{N}{2}]$\lambda6585$ FWHM 	\\
117 & NA\_NII\_6585\_VOFF	                & FLOAT64 & Velocity of narrow [\ion{N}{2}]$\lambda6585$ [km s$^{-1}$]	\\
118 & NA\_NII\_6585\_VOFF\_ERR	            & FLOAT64 & Uncertainty in narrow [\ion{N}{2}]$\lambda6585$ velocity	\\
119	& NA\_NII\_6585\_EW	                 	& FLOAT64 & Equivalent width of narrow [\ion{N}{2}]$\lambda6585$ [$\rm{\AA}$] \\
120	& NA\_NII\_6585\_EW\_ERR	            & FLOAT64 & Uncertainty in narrow [\ion{N}{2}]$\lambda6585$ equivalent width \\
121	& NA\_NII\_6585\_LOG\_FLUX	         	& FLOAT64 & Flux of narrow [\ion{N}{2}]$\lambda6585$ [$\log_{10}(F/\rm{erg\;cm}^{-2}\;\rm{s}^{-1})$] \\
122	& NA\_NII\_6585\_LOG\_FLUX\_ERR	     	& FLOAT64 & Uncertainty in narrow [\ion{N}{2}]$\lambda6585$ flux \\
123	& NA\_NII\_6585\_LOG\_LUM	            & FLOAT64 & Luminosity of narrow [\ion{N}{2}]$\lambda6585$ [$\log_{10}(L/\rm{erg\;s}^{-1})$] \\
124	& NA\_NII\_6585\_LOG\_LUM\_ERR	     	& FLOAT64 & Uncertainty in narrow [\ion{N}{2}]$\lambda6585$ luminosity \\
125	& NA\_NII\_6585\_PROB	                & FLOAT64 & Detection probability of narrow [\ion{N}{2}]$\lambda6585$ \\
    \hline
126 & NA\_H\_ALPHA\_DISP                    & FLOAT64 & Dispersion of narrow H$\alpha$ [km s$^{-1}$]	\\
127 & NA\_H\_ALPHA\_DISP\_ERR               & FLOAT64 & Uncertainty in narrow H$\alpha$ dispersion	\\
128 & NA\_H\_ALPHA\_FWHM                    & FLOAT64 & FWHM of narrow H$\alpha$ [km s$^{-1}$]	\\
129 & NA\_H\_ALPHA\_FWHM\_ERR               & FLOAT64 & Uncertainty in narrow H$\alpha$ FWHM 	\\
130 & NA\_H\_ALPHA\_VOFF                    & FLOAT64 & Velocity of narrow H$\alpha$ [km s$^{-1}$]	\\
131 & NA\_H\_ALPHA\_VOFF\_ERR               & FLOAT64 & Uncertainty in narrow H$\alpha$ velocity	\\
132	& NA\_H\_ALPHA\_EW	                 	& FLOAT64 & Equivalent width of narrow H$\alpha$ [$\rm{\AA}$] \\
133	& NA\_H\_ALPHA\_EW\_ERR	             	& FLOAT64 & Uncertainty in narrow H$\alpha$ equivalent width \\
134	& NA\_H\_ALPHA\_LOG\_FLUX	            & FLOAT64 & Flux of narrow H$\alpha$ [$\log_{10}(F/\rm{erg\;cm}^{-2}\;\rm{s}^{-1})$] \\
135	& NA\_H\_ALPHA\_LOG\_FLUX\_ERR	     	& FLOAT64 & Uncertainty in narrow H$\alpha$ flux \\
136	& NA\_H\_ALPHA\_LOG\_LUM	            & FLOAT64 & Luminosity of narrow H$\alpha$ [$\log_{10}(L/\rm{erg\;s}^{-1})$] \\
137	& NA\_H\_ALPHA\_LOG\_LUM\_ERR	        & FLOAT64 & Uncertainty in narrow H$\alpha$ luminosity \\
138	& NA\_H\_ALPHA\_PROB	                & FLOAT64 & Detection probability of narrow H$\alpha$ \\
    \hline
139	& BR\_H\_ALPHA\_DISP	                & FLOAT64 & Dispersion of broad H$\alpha$ [km s$^{-1}$]	\\
140	& BR\_H\_ALPHA\_DISP\_ERR	            & FLOAT64 & Uncertainty in broad H$\alpha$ dispersion	\\
141	& BR\_H\_ALPHA\_FWHM	                & FLOAT64 & FWHM of broad H$\alpha$ [km s$^{-1}$]	\\
142	& BR\_H\_ALPHA\_FWHM\_ERR	            & FLOAT64 & Uncertainty in broad H$\alpha$ FWHM 	\\
143	& BR\_H\_ALPHA\_VOFF	                & FLOAT64 & Velocity of broad H$\alpha$ [km s$^{-1}$]	\\
144	& BR\_H\_ALPHA\_VOFF\_ERR	            & FLOAT64 & Uncertainty in broad H$\alpha$ velocity	\\
145	& BR\_H\_ALPHA\_EW	                 	& FLOAT64 & Equivalent width of broad H$\alpha$ [$\rm{\AA}$] \\
146	& BR\_H\_ALPHA\_EW\_ERR	             	& FLOAT64 & Uncertainty in broad H$\alpha$ equivalent width \\
147	& BR\_H\_ALPHA\_LOG\_FLUX	            & FLOAT64 & Flux of broad H$\alpha$ [$\log_{10}(F/\rm{erg\;cm}^{-2}\;\rm{s}^{-1})$] \\
148	& BR\_H\_ALPHA\_LOG\_FLUX\_ERR	     	& FLOAT64 & Uncertainty in broad H$\alpha$ flux \\
149	& BR\_H\_ALPHA\_LOG\_LUM	            & FLOAT64 & Luminosity of broad H$\alpha$ [$\log_{10}(L/\rm{erg\;s}^{-1})$] \\
150	& BR\_H\_ALPHA\_LOG\_LUM\_ERR	        & FLOAT64 & Uncertainty in broad H$\alpha$ luminosity \\
151	& BR\_H\_ALPHA\_PROB	                & FLOAT64 & broad H$\alpha$ detection probability \\
152	& BR\_H\_ALPHA\_QUALITY	                & FLOAT64 & broad H$\alpha$ fit quality \\
153	& BR\_H\_ALPHA\_RCHI2	                & FLOAT64 & broad H$\alpha$ reduced $\chi^2$ of fit\\
    \hline
154 & NA\_OI\_6302\_DISP                    & FLOAT64 & Dispersion of narrow [\ion{O}{1}]$\lambda6302$ [km s$^{-1}$]	\\
155 & NA\_OI\_6302\_DISP\_ERR               & FLOAT64 & Uncertainty in narrow [\ion{O}{1}]$\lambda6302$ dispersion	\\
156 & NA\_OI\_6302\_FWHM                    & FLOAT64 & FWHM of narrow [\ion{O}{1}]$\lambda6302$ [km s$^{-1}$]	\\
157 & NA\_OI\_6302\_FWHM\_ERR               & FLOAT64 & Uncertainty in narrow [\ion{O}{1}]$\lambda6302$ FWHM 	\\
158 & NA\_OI\_6302\_VOFF                    & FLOAT64 & Velocity of narrow [\ion{O}{1}]$\lambda6302$ [km s$^{-1}$]	\\
159 & NA\_OI\_6302\_VOFF\_ERR               & FLOAT64 & Uncertainty in narrow [\ion{O}{1}]$\lambda6302$ velocity	\\
160	& NA\_OI\_6302\_EW	                 	& FLOAT64 & Equivalent width of narrow [\ion{O}{1}]$\lambda6302$ [$\rm{\AA}$] \\
161	& NA\_OI\_6302\_EW\_ERR	             	& FLOAT64 & Uncertainty in narrow [\ion{O}{1}]$\lambda6302$ equivalent width \\
162	& NA\_OI\_6302\_LOG\_FLUX	            & FLOAT64 & Flux of narrow [\ion{O}{1}]$\lambda6302$ [$\log_{10}(F/\rm{erg\;cm}^{-2}\;\rm{s}^{-1})$] \\
163	& NA\_OI\_6302\_LOG\_FLUX\_ERR	     	& FLOAT64 & Uncertainty in narrow [\ion{O}{1}]$\lambda6302$ flux \\
164	& NA\_OI\_6302\_LOG\_LUM	            & FLOAT64 & Luminosity of narrow [\ion{O}{1}]$\lambda6302$ [$\log_{10}(L/\rm{erg\;s}^{-1})$] \\
165	& NA\_OI\_6302\_LOG\_LUM\_ERR	        & FLOAT64 & Uncertainty in narrow [\ion{O}{1}]$\lambda6302$ luminosity \\
166	& NA\_OI\_6302\_PROB	                & FLOAT64 & Detection probability of narrow [\ion{O}{1}]$\lambda6302$ \\
    \hline
167 & NA\_OIII\_5007\_DISP                  & FLOAT64 & Dispersion of narrow [\ion{O}{3}]$\lambda5007$ [km s$^{-1}$]	\\
168 & NA\_OIII\_5007\_DISP\_ERR             & FLOAT64 & Uncertainty in narrow [\ion{O}{3}]$\lambda5007$ dispersion	\\
169 & NA\_OIII\_5007\_FWHM                  & FLOAT64 & FWHM of narrow [\ion{O}{3}]$\lambda5007$ [km s$^{-1}$]	\\
170 & NA\_OIII\_5007\_FWHM\_ERR             & FLOAT64 & Uncertainty in narrow [\ion{O}{3}]$\lambda5007$ FWHM 	\\
171 & NA\_OIII\_5007\_VOFF                  & FLOAT64 & Velocity of narrow [\ion{O}{3}]$\lambda5007$ [km s$^{-1}$]	\\
172 & NA\_OIII\_5007\_VOFF\_ERR             & FLOAT64 & Uncertainty in narrow [\ion{O}{3}]$\lambda5007$ velocity	\\
173	& NA\_OIII\_5007\_EW	                & FLOAT64 & Equivalent width of narrow [\ion{O}{3}]$\lambda5007$ [$\rm{\AA}$] \\
174	& NA\_OIII\_5007\_EW\_ERR	            & FLOAT64 & Uncertainty in narrow [\ion{O}{3}]$\lambda5007$ equivalent width \\
175	& NA\_OIII\_5007\_LOG\_FLUX	         	& FLOAT64 & Flux of narrow [\ion{O}{3}]$\lambda5007$ [$\log_{10}(F/\rm{erg\;cm}^{-2}\;\rm{s}^{-1})$] \\
176	& NA\_OIII\_5007\_LOG\_FLUX\_ERR	    & FLOAT64 & Uncertainty in narrow [\ion{O}{3}]$\lambda5007$ flux \\
177	& NA\_OIII\_5007\_LOG\_LUM	         	& FLOAT64 & Luminosity of narrow [\ion{O}{3}]$\lambda5007$ [$\log_{10}(L/\rm{erg\;s}^{-1})$] \\
178	& NA\_OIII\_5007\_LOG\_LUM\_ERR	     	& FLOAT64 & Uncertainty in narrow [\ion{O}{3}]$\lambda5007$ luminosity \\
179	& NA\_OIII\_5007\_PROB	             	& FLOAT64 & Detection probability of narrow [\ion{O}{3}]$\lambda5007$ \\
    \hline
180 & NA\_H\_BETA\_DISP                     & FLOAT64 & Dispersion of narrow H$\beta$ [km s$^{-1}$]	\\
181 & NA\_H\_BETA\_DISP\_ERR                & FLOAT64 & Uncertainty in narrow H$\beta$ dispersion	\\
182 & NA\_H\_BETA\_FWHM                     & FLOAT64 & FWHM of narrow H$\beta$ [km s$^{-1}$]	\\
183 & NA\_H\_BETA\_FWHM\_ERR                & FLOAT64 & Uncertainty in narrow H$\beta$ FWHM 	\\
184 & NA\_H\_BETA\_VOFF                     & FLOAT64 & Velocity of narrow H$\beta$ [km s$^{-1}$]	\\
185 & NA\_H\_BETA\_VOFF\_ERR                & FLOAT64 & Uncertainty in narrow H$\beta$ velocity	\\
186	& NA\_H\_BETA\_EW	                    & FLOAT64 & Equivalent width of narrow H$\beta$ [$\rm{\AA}$] \\
187	& NA\_H\_BETA\_EW\_ERR	             	& FLOAT64 & Uncertainty in narrow H$\beta$ equivalent width \\
188	& NA\_H\_BETA\_LOG\_FLUX	            & FLOAT64 & Flux of narrow H$\beta$ [$\log_{10}(F/\rm{erg\;cm}^{-2}\;\rm{s}^{-1})$] \\
189	& NA\_H\_BETA\_LOG\_FLUX\_ERR	        & FLOAT64 & Uncertainty in narrow H$\beta$ flux \\
190	& NA\_H\_BETA\_LOG\_LUM	             	& FLOAT64 & Luminosity of narrow H$\beta$ [$\log_{10}(L/\rm{erg\;s}^{-1})$] \\
191	& NA\_H\_BETA\_LOG\_LUM\_ERR	        & FLOAT64 & Uncertainty in narrow H$\beta$ luminosity \\
192	& NA\_H\_BETA\_PROB	                 	& FLOAT64 & Detection probability of narrow H$\beta$ \\
    \hline
193	& BR\_H\_BETA\_DISP	                 	& FLOAT64 & Dispersion of broad H$\beta$ [km s$^{-1}$]	\\
194	& BR\_H\_BETA\_DISP\_ERR	            & FLOAT64 & Uncertainty in broad H$\beta$ dispersion	\\
195	& BR\_H\_BETA\_FWHM	                 	& FLOAT64 & FWHM of broad H$\beta$ [km s$^{-1}$]	\\
196	& BR\_H\_BETA\_FWHM\_ERR	            & FLOAT64 & Uncertainty in broad H$\beta$ FWHM 	\\
197	& BR\_H\_BETA\_VOFF	                 	& FLOAT64 & Velocity of broad H$\beta$ [km s$^{-1}$]	\\
198	& BR\_H\_BETA\_VOFF\_ERR	            & FLOAT64 & Uncertainty in broad H$\beta$ velocity	\\
199	& BR\_H\_BETA\_EW	                    & FLOAT64 & Equivalent width of broad H$\beta$ [$\rm{\AA}$] \\
200	& BR\_H\_BETA\_EW\_ERR	             	& FLOAT64 & Uncertainty in broad H$\beta$ equivalent width \\
201	& BR\_H\_BETA\_LOG\_FLUX	            & FLOAT64 & Flux of broad H$\beta$ [$\log_{10}(F/\rm{erg\;cm}^{-2}\;\rm{s}^{-1})$] \\
202	& BR\_H\_BETA\_LOG\_FLUX\_ERR	        & FLOAT64 & Uncertainty in broad H$\beta$ flux \\
203	& BR\_H\_BETA\_LOG\_LUM	             	& FLOAT64 & Luminosity of broad H$\beta$ [$\log_{10}(L/\rm{erg\;s}^{-1})$] \\
204	& BR\_H\_BETA\_LOG\_LUM\_ERR	        & FLOAT64 & Uncertainty in broad H$\beta$ luminosity \\
205	& BR\_H\_BETA\_PROB	                 	& FLOAT64 & broad H$\beta$ detection probability \\
206	& BR\_H\_BETA\_QUALITY	                & FLOAT64 & broad H$\beta$ fit quality\\
207	& BR\_H\_BETA\_RCHI2	                & FLOAT64 & broad H$\beta$ reduced $\chi^2$ of fit\\
    \hline
208	& BR\_MGII\_2799\_DISP	             	& FLOAT64 & Dispersion of \ion{Mg}{2}$\lambda2799$ [km s$^{-1}$]	\\
209	& BR\_MGII\_2799\_DISP\_ERR	         	& FLOAT64 & Uncertainty in \ion{Mg}{2}$\lambda2799$ dispersion	\\
210	& BR\_MGII\_2799\_FWHM	             	& FLOAT64 & FWHM of \ion{Mg}{2}$\lambda2799$ [km s$^{-1}$]	\\
211	& BR\_MGII\_2799\_FWHM\_ERR	         	& FLOAT64 & Uncertainty in \ion{Mg}{2}$\lambda2799$ FWHM 	\\
212	& BR\_MGII\_2799\_VOFF	             	& FLOAT64 & Velocity of \ion{Mg}{2}$\lambda2799$ [km s$^{-1}$]	\\
213	& BR\_MGII\_2799\_VOFF\_ERR	         	& FLOAT64 & Uncertainty in \ion{Mg}{2}$\lambda2799$ velocity	\\
214	& BR\_MGII\_2799\_EW	                & FLOAT64 & Equivalent width of \ion{Mg}{2}$\lambda2799$ [$\rm{\AA}$] \\
215	& BR\_MGII\_2799\_EW\_ERR	            & FLOAT64 & Uncertainty in \ion{Mg}{2}$\lambda2799$ equivalent width \\
216	& BR\_MGII\_2799\_LOG\_FLUX	         	& FLOAT64 & Flux of \ion{Mg}{2}$\lambda2799$ [$\log_{10}(F/\rm{erg\;cm}^{-2}\;\rm{s}^{-1})$] \\
217	& BR\_MGII\_2799\_LOG\_FLUX\_ERR	    & FLOAT64 & Uncertainty in \ion{Mg}{2}$\lambda2799$ flux \\
218	& BR\_MGII\_2799\_LOG\_LUM	         	& FLOAT64 & Luminosity of \ion{Mg}{2}$\lambda2799$ [$\log_{10}(L/\rm{erg\;s}^{-1})$] \\
219	& BR\_MGII\_2799\_LOG\_LUM\_ERR	     	& FLOAT64 & Uncertainty in \ion{Mg}{2}$\lambda2799$ luminosity \\
220	& BR\_MGII\_2799\_PROB	             	& FLOAT64 & \ion{Mg}{2}$\lambda2799$ detection probability \\
221	& BR\_MGII\_2799\_QUALITY	            & FLOAT64 & \ion{Mg}{2}$\lambda2799$ fit quality \\
222	& BR\_MGII\_2799\_RCHI2	             	& FLOAT64 & \ion{Mg}{2}$\lambda2799$ reduced $\chi^2$ of fit \\
    \hline
223	& BR\_CIV\_1549\_DISP	                & FLOAT64 & Dispersion of \ion{C}{4}$\lambda1549$ [km s$^{-1}$]	\\
224	& BR\_CIV\_1549\_DISP\_ERR	         	& FLOAT64 & Uncertainty in \ion{C}{4}$\lambda1549$ dispersion	\\
225	& BR\_CIV\_1549\_FWHM	                & FLOAT64 & FWHM of \ion{C}{4}$\lambda1549$ [km s$^{-1}$]	\\
226	& BR\_CIV\_1549\_FWHM\_ERR	         	& FLOAT64 & Uncertainty in \ion{C}{4}$\lambda1549$ FWHM 	\\
227	& BR\_CIV\_1549\_VOFF	                & FLOAT64 & Velocity of \ion{C}{4}$\lambda1549$ [km s$^{-1}$]	\\
228	& BR\_CIV\_1549\_VOFF\_ERR	         	& FLOAT64 & Uncertainty in \ion{C}{4}$\lambda1549$ velocity	\\
229	& BR\_CIV\_1549\_EW	                 	& FLOAT64 & Equivalent width of \ion{C}{4}$\lambda1549$ [$\rm{\AA}$] \\
230	& BR\_CIV\_1549\_EW\_ERR	            & FLOAT64 & Uncertainty in \ion{C}{4}$\lambda1549$ equivalent width \\
231	& BR\_CIV\_1549\_LOG\_FLUX	         	& FLOAT64 & Flux of \ion{C}{4}$\lambda1549$ [$\log_{10}(F/\rm{erg\;cm}^{-2}\;\rm{s}^{-1})$] \\
232	& BR\_CIV\_1549\_LOG\_FLUX\_ERR	     	& FLOAT64 & Uncertainty in \ion{C}{4}$\lambda1549$ flux \\
233	& BR\_CIV\_1549\_LOG\_LUM	            & FLOAT64 & Luminosity of \ion{C}{4}$\lambda1549$ [$\log_{10}(L/\rm{erg\;s}^{-1})$] \\
234	& BR\_CIV\_1549\_LOG\_LUM\_ERR	     	& FLOAT64 & Uncertainty in \ion{C}{4}$\lambda1549$ luminosity \\
235	& BR\_CIV\_1549\_PROB	                & FLOAT64 & \ion{C}{4}$\lambda1549$ detection probability \\
236	& BR\_CIV\_1549\_QUALITY	            & FLOAT64 & \ion{C}{4}$\lambda1549$ fit quality \\
237	& BR\_CIV\_1549\_RCHI2	                & FLOAT64 & \ion{C}{4}$\lambda1549$ reduced $\chi^2$ of fit \\
\enddata
\tablecomments{This table will be made available via the VizieR  (\url{https://vizier.u-strasbg.fr/viz-bin/VizieR-2}) and \url{https://crf.usno.navy.mil}.  The supplementary ``long'' table of all fitted emission lines and parameters not included in the ``short'' table will be made available via \url{https://crf.usno.navy.mil}.}
\end{deluxetable}
\end{longrotatetable}

\end{document}